\documentclass[pre, twocolumn, superscriptaddress, floatfix, nofootinbib]{revtex4-1}  

\usepackage[paperwidth=210mm,paperheight=297mm,centering,hmargin=2cm,vmargin=2.5cm]{geometry}

\usepackage[pdftex]{graphicx}
\usepackage{amsmath,amsfonts,amssymb}
\usepackage{MnSymbol}
\usepackage{dsfont}
\usepackage{enumitem}
\usepackage{csquotes}
\usepackage[normalem]{ulem}

\usepackage[left]{lineno}

\usepackage{hyperref}
\usepackage{xcolor}

\definecolor{bluemoi}{rgb}{0.25,0.50 ,0.75} 

\hypersetup{
  bookmarksopen=true,
  pdftitle=" ",
  pdfauthor=" ", 
  pdfsubject=" ", 
  pdftoolbar=true, 
  pdfmenubar=true, 
  pdfhighlight=/O, 
  colorlinks=true, 
  pdfpagemode=UseNone, 
  pdfpagelayout=SinglePage, 
  pdffitwindow=true, 
  linkcolor=bluemoi, 
  citecolor=bluemoi, 
  urlcolor=bluemoi 
}

\makeatletter
\renewcommand\@biblabel[1]{#1} 

\renewcommand\newblock{\hskip .11em\@plus.33em\@minus.07em}
\makeatother

\setcitestyle{authoryear,round}
\bibpunct{(}{)}{;}{a}{,}{}

\makeatletter
\newcommand{\removeperiod}{\@ifnextchar.{\@gobble}\relax}
\makeatother 

\makeatletter
\renewcommand{\figurename}{\sf \textbf{Figure}}
\renewcommand{\thefigure}{\arabic{figure}}
\renewcommand{\fnum@figure}{\sf\textbf{\figurename}~\textbf{\thefigure}}
\renewcommand{\tablename}{\sf\textbf{Table}}
\renewcommand{\thetable}{\arabic{table}}
\renewcommand{\fnum@table}{\sf\textbf{\tablename}~\textbf{\thetable}}
\makeatother

\begin{document}

\title{Assessing the effect of sample bias correction\\ in species distribution models} 

\author{Nicolas Dubos}
\thanks{Corresponding authors: nicolas.dubos@inrae.fr \& maxime.lenormand@inrae.fr}
\affiliation{TETIS, Univ Montpellier, AgroParisTech, Cirad, CNRS, INRAE, Montpellier, France}

\author{Cl{\'e}mentine Pr{\'e}au}
\affiliation{TETIS, Univ Montpellier, AgroParisTech, Cirad, CNRS, INRAE, Montpellier, France}

\author{Maxime Lenormand}
\thanks{Corresponding authors: nicolas.dubos@inrae.fr \& maxime.lenormand@inrae.fr}
\affiliation{TETIS, Univ Montpellier, AgroParisTech, Cirad, CNRS, INRAE, Montpellier, France}

\author{Guillaume Papuga}
\affiliation{AMAP, Univ Montpellier, CIRAD, CNRS, INRAE, IRD, Montpellier, France}

\author{Sophie Monsarrat}
\affiliation{Center for Biodiversity Dynamics in a Changing World (BIOCHANGE), Department of Biology, Aarhus University, Ny Munkegade 114, DK-8000 Aarhus C, Denmark}
\affiliation{Section for Ecoinformatics and Biodiversity, Department of Biology, Aarhus University, Ny Munkegade 114, DK-8000 Aarhus C, Denmark}

\author{Pierre Denelle}
\affiliation{Biodiversity, Macroecology \& Biogeography, University of G{\"o}ettingen, G{\"o}ttingen, Germany}

\author{Marine Le Louarn}
\affiliation{TETIS, Univ Montpellier, AgroParisTech, Cirad, CNRS, INRAE, Montpellier, France}

\author{Stien Heremans}
\affiliation{Research Institute for Nature and Forest (INBO), Brussels, Belgium}

\author{Roel May}
\affiliation{Norwegian Institute for Nature Research (NINA), P.O. Box 5685 Torgarden, NO-7485 Trondheim, Norway}

\author{Philip Roche}
\affiliation{INRAE, Aix Marseille Univ, RECOVER, Aix-en-Provence, France}

\author{Sandra Luque}
\affiliation{TETIS, Univ Montpellier, AgroParisTech, Cirad, CNRS, INRAE, Montpellier, France}

\begin{abstract}
\vspace*{0.1cm}
\noindent\textbf{ABSTRACT}
\vspace*{0.1cm}

\noindent\textbf{1.} Open-source biodiversity databases contain a large number of species occurrence records but are often spatially biased; which affects the reliability of species distribution models based on these records. Sample bias correction techniques require data filtering which comes at the cost of record numbers, or require considerable additional sampling effort. Since independent data is rarely available, assessment of the correction technique often relies solely  on performance metrics computed using subsets of the available -- biased -- data, which may prove misleading.

\vspace*{0.1cm}
\noindent\textbf{2.} Here, we assess the extent to which an acknowledged sample bias correction technique is likely to improve models' ability to predict species distributions in the absence of independent data. We assessed  variation in model predictions induced by the aforementioned correction and model stochasticity;  the variability between model replicates  related to a random component (pseudo-absences sets and cross-validation subsets). We present, then,  an index of the effect of correction relative to model stochasticity; the Relative Overlap Index (ROI). We investigated  whether the ROI better represented the effect of correction than classic performance metrics (Boyce index, cAUC, AUC and TSS) and absolute overlap metrics (Schoener's D, Pearson's and Spearman's correlation coefficients) when considering data related to 64 vertebrate species and 21 virtual species with a generated sample bias.

\vspace*{0.1cm}
\noindent\textbf{3.} When based on absolute overlaps and cross-validation performance metrics, we found that correction produced no significant effects. When considering its effect relative to model stochasticity, the effect of correction was strong for most species at one of the three sites. The use of virtual species enabled us to verify that the correction technique  improved both distribution  predictions and the biological relevance of the selected variables at the specific site, when these were not correlated with sample bias patterns.

\vspace*{0.1cm}
\noindent\textbf{4.} In the absence of additional independent data, the assessment of sample bias correction based on subsample data may be misleading. We propose to  investigate both  the biological relevance of environmental variables selected, and, the effect of sample bias correction based on  its effect relative to model stochasticity.

\vspace*{0.2cm}
\end{abstract}

\maketitle

\section*{Introduction}

While there is a growing demand for species data for the production of robust statistical models and evidence-based conservation actions,  the availability of standardised data remains limited. In recent years, the extensive development of  biodiversity databases has predominantly been  supported  by opportunistic, presence-only data collected by citizen science programs and naturalists associations. Despite its limitations, opportunistic, non-standardised data still constitute a promising avenue for improvement of  biodiversity assessments \citep{Mckinley2017}. Such data are often limited by the heterogeneity of their sources and spatial biases as a result of uneven sampling efforts \citep{Otegui2013,Beck2014,Bird2014,Johnston2020,Botella2021}. More specifically, sampling efforts may be biased by field accessibility such as the number of observations  influenced by the proximity to urban areas and roads,  often leading to spatial autocorrelation among observations \citep{Phillips2009,Stolar2015}. This may incur an environmental bias  and models tend to overestimate/underestimate environmental suitability in  zones with higher/lower density of occurrence data. This could prove to be problematic for studies that aim to provide guidelines for management \citep{Yackulic2013}. By accounting for spatial bias in  opportunistic data derived from heterogeneous sources, (e.g. citizen science, naturalist and expert associations) not only would increase the prospect of their potential use in ecological studies, it would also  enable the inclusion of a broader range of species in Species Distribution Models (SDMs).
SDMs are one of the most commonly used tools for  testing ecological hypotheses \citep{Anderson2009}, assessing  alien species invasion risks \citep{Bellard2013,Briscoe2019,Lanner2022}, forecasting the potential effect of environmental change \citep{Araujo2005}, and supporting conservation management efforts  \citep{Schwartz2012,Leroy2014,Mikolajczak2015,Dubos2021a}. Although presence-only biodiversity databases are frequently  used in SDMs, the adequacy of sample bias correction methods remains ambiguous \citep{Meynard2019,Johnston2020}.
Spatial sampling bias is a major factor affecting the predictive performance of SDMs \citep{Araujo2006,Barbet2012,Kramer2013,Meynard2019}. A number of procedures have been developed  to account for sampling bias, which include spatial filtering of presence points \citep{Edren2010,Boria2014,Matutini2021}, environmental filtering \citep{Varela2014,Gabor2020}, the combination of  presence-only and standardised presence-absence data \citep{Dorazio2014,Fithian2015,Koshkina2017} and the production of a similar sampling bias in non-presence background data/pseudo-absences \citep{Phillips2009}. However, presence points and environmental filtering consist in the removal of occurrence data, thereby inducing a loss of information and statistical power. This is particularly problematic when dealing with rare or poorly detected species \citep{Lobo2011,Kramer2013,Robinson2018,Vollering2019,Inman2021}. The most widely applicable method may therefore be the production of pseudo-absences that share the same bias as the presence data. In presence/background or presence/pseudo-absence models, a range of pseudo-absence selection techniques were recently developed which reduce the effect of sampling bias, improving model performance  without removing occurrence points (e.g. \citet{Senay2013,Fourcade2014,Hertzog2014,Iturbide2015}).  For instance, pseudo-absence selection based on sampling bias reference maps has been acknowledged as an efficient method to account for spatially biased occurrence data \citep{Phillips2009,Hertzog2014}. Reference maps, such as the target-group (TG) approach and accessibility maps, can be used to represent a sampling bias map that is specific to a given study area. This presents  a promising approach to improvement of the predictive performance of SDMs \citep{Ranc2017, Monsarrat2019}. The TG approach relies on the hypothesis that the study species share the same sampling pattern as the target group, whereas accessibility maps rely on the hypothesis that constraining features are identified (e.g. geographical barriers, social conflicts, long distances). Although neither mapping approach provides explicit information on sampling efforts they may prove  appropriate when species richness patterns are heterogeneous \citep{Ranc2017} or when treating data from heterogeneous sources with different sampling patterns \citep{Monsarrat2019}. Here we focus on a single sample bias correction technique (i.e. accessibility maps) to test a range of different methods. Accessibility maps do not require to subset the occurrence data and are therefore more appropriate for rare species. They are also more widely applicable than TG approach since TG requires information on sampling effort throughout an entire taxon.

\begin{figure*}
	\begin{center}
		\includegraphics[width=15cm]{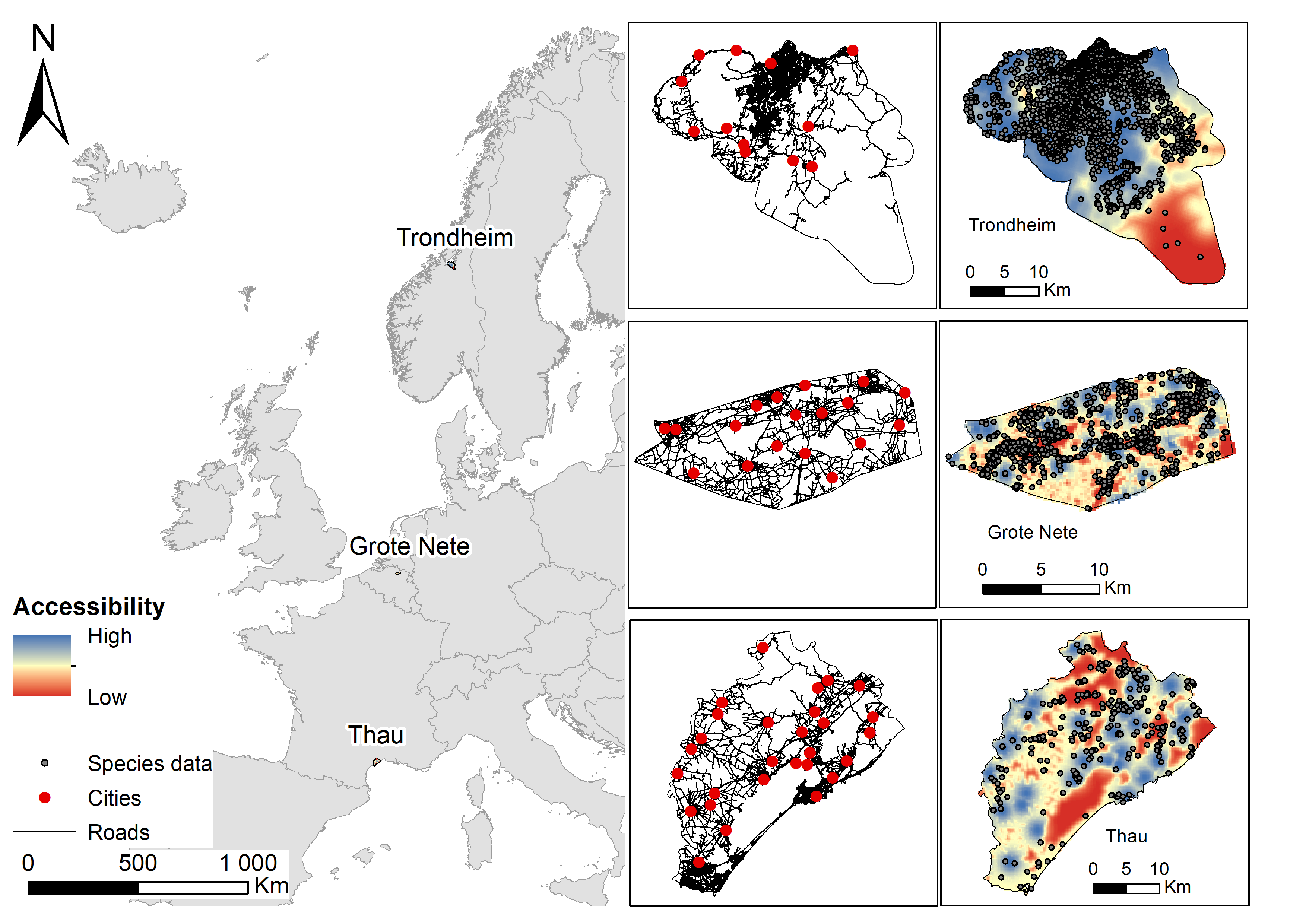}
		\caption{\sf \textbf{Location of the three study sites in Europe and accessibility maps for each site.} The accessibility index is inversely proportional to the Euclidean distance between cities and roads and represents the degree to which a geographic point is easily reached by an observer. Black dots represent occurrence records of all species pooled together (n = 46, 11 and 7 species in Trondheim, Grote Nete and Thau, respectively). \label{Fig1}}
	\end{center}
\end{figure*} 

The efficiency of a given sample bias correction technique is often measured by comparing the performance metrics of corrected and uncorrected models. In both corrected and uncorrected groups, performance metrics quantify the degree to which models built with a subset of the original data (i.e. training/calibration dataset) to accurately predict the remaining data (test/evaluation dataset). This process is commonly referred to as \enquote{cross-validation} (e.g. \citet{Senay2013,Boria2014}). The most common model performance metrics include  the Area Under the operating Curve (AUC), the True Skill Statistic (TSS), the Boyce index, and Similarity indices.  These metrics provide quantitative measurements of discrimination ability between models that are built with training data and those that utilise  the full dataset \citep{Fourcade2018}. The vast majority of studies test the efficiency of sample bias correction by performing internal cross-validation, (see point 3B of the Standard for SDMs on data sharing the same bias in \citet{Araujo2019}) an approach that has shown strong limitations when SDM is used to extrapolate predictions to a different time/region \citep{Araujo2005,Araujo2019,Beck2014,Hertzog2014,Fourcade2018}. Ideally, the improvement conferred by a correction technique should be evaluated with an independent, unbiased dataset \citep{Phillips2009,Hertzog2014,Norberg2019,Johnston2020}. Field validation (evaluations relative to independent standardised datasets) represents the best standard practice to assess models' ability to predict species distribution \citep{Araujo2019}. Nevertheless, field validation is labour-intensive  and sometimes unfeasible (e.g. taxonomically or geographically extensive scale study sites) and independent standardised datasets are rarely available \citep{Hao2019}. When relying on partitioned datasets, one possible method for assessing  sample bias correction techniques  is to select data subsets that are subject to different types of bias (e.g. \citet{Bean2012,Matutini2021}). However, this method may also prove  highly demanding in respect to spatio-temporal coverage and may not be feasible for noumerous  species \citep{Johnston2020}. A cost-effective method  proposed by \citet{Hijmans2012} to assess the potential efficiency of sample bias correction techniques uses  an AUC calibrated with a null geographic model. The efficiency of this method may vary through space and between species \citep{Hijmans2012}, calling for the characterisation of those sites and species. Virtual species can be used to assess bias correction techniques (e.g. \citet{Phillips2009,Fourcade2014,Varela2014,Ranc2017}), by simulating a sampling bias and producing performance metrics that are relative to a perfectly known distribution. The projection of a range of virtual species on multiple real regions may represent a cost-effective approach for assessing  whether a correction technique is likely to  improve the accuracy of SDMs, provided that virtual and real species are sufficiently comparable.
Species distribution models can be calibrated with a range of model parameters generated with a random component: (e.g. pseudo-absence selection, cross-validation subsets) inducing a stochasticity among models that are otherwise identical. These model parameters can be sources of uncertainty in model projections \citep{Buisson2010,Thibaud2014}. Sample bias correction should induce variation in the predicted values and subsequently in species range projections, but may prove  negligible if the variation is of the same magnitude as the sources of uncertainty. Therefore, the effect of sample bias correction may be assessed on the basis of its effect between corrected and uncorrected modalities relative to intra-modality variation.

We present the Relative Overlap Index, which informs the extent of spatial similarity between corrected and uncorrected models relative to the variability between model replicates. We hypothesised that  sample bias correction  improves model predictions if its relative effect is stronger than that of the remaining model input parameters. We tested this assumption using virtual species with a generated sample bias, modelled at the same sites and with the same range of model parameters as the real species. The aim of this study  was (1) to assess the effect of a sampling bias correction technique on distribution projections over a range of terrestrial vertebrate species (n = 64) in three contrasting regions and in the absence of independent data. We measured the effect of correction by computing the degree of overlap between uncorrected and corrected models. We predicted that the effect of correction on projections differs between sites and species. We further evaluated (2) whether the effect of correction could be assessed with a range of validation metrics. We also tested whether this effect could  be represented by an index of overlap between correction modalities relative to intra-modality variation. We tested (3) whether sample bias correction actually improved model predictions by using virtual species with a simulated sample bias. Finally, we provided recommendations for the assessment of sample bias correction when independent data are unavailable.

\section*{Materials and Methods}

\subsection*{Study sites and accessibility}

We focused on three regions located in the vicinity  of Thau (southern France), Grote Nete (northern Belgium) and Trondheim (central Norway). These sites were characterised by contrasting distributions of roads and cities. While there was an altogether  homogeneous distribution of roads and towns  in Thau and  Grote Nete,   there was a strong gradient in road and town density in Trondheim.  We produced accessibility maps for each site (Figure \ref{Fig1}) by computing an Accessibility Index (AI). Accessibility Indices represent the degree to which a geographic point is easily reached by an observer and is context specific \citep{Monsarrat2019}. In this instance,  we have used occurrence data from heterogeneous sources. For that reason, we assumed that accessibility was -- for the most part -- dependent  upon distance from cities and  roads (e.g. \citet{Sicacha2020}). The Accessibility Index was computed as follows:
\begin{equation}
\displaystyle AI_{i}=\displaystyle \frac{1}{2}\left(e^{\displaystyle -\frac{1}{2}\left(\frac{dist_C}{\sigma_C}\right)^2}+e^{\displaystyle -\frac{1}{2}\left(\frac{dist_R}{\sigma_R}\right)^2} \right)  
\label{AI}
\end{equation}
where $AI_i$ is the accessibility index at pixel $i$, $dist_C$ is the Euclidean distance from the closest city centre ($\geq$ 200 inhabitants), $dist_R$ is the Euclidean distance from the closest primary and secondary road. $\sigma_C$ and $\sigma_R$ are the standard deviations of the distances distributions to the nearest city and road, respectively.

\subsection*{Environmental data}

We used land use variables retrieved from Corine Land Cover habitat classes, a European biophysical dataset derived from remote sensing. At the local scale, land  use variables are more relevant to species distribution models than climatic predictors at the local scale \citep{Soberon2009,Ficetola2014}. For each $200 \times 200 \mbox{ m}^2$ pixel, we measured the distance from the nearest habitat features using 8 habitat classes: artificial surface, forest edge, intensive farmland, non-intensive farmland, scrubland/herbaceous areas, coastal areas, water courses, water bodies. We also computed the proportion of a given habitat type within a range of buffer zones around occurrence points (200, 500 and 1,000 meters). This corresponded to species habitat use at the landscape scale in accordance to previous studies (e.g. in reptiles, amphibians and bats; \citet{Jeliazkov2014,Azam2016}). In birds, the landscape may be influential at larger scales (e.g. 5,000 meters; \citet{Dubos2018}). Given the scale of our study sites, the use of larger buffer zones would result in a lack of variability in environmental conditions at occurrence points, so we limited the extent of our buffer zones to 1,000 meters.

\begin{figure}[!h]
	\begin{center}
		\includegraphics[width=9cm]{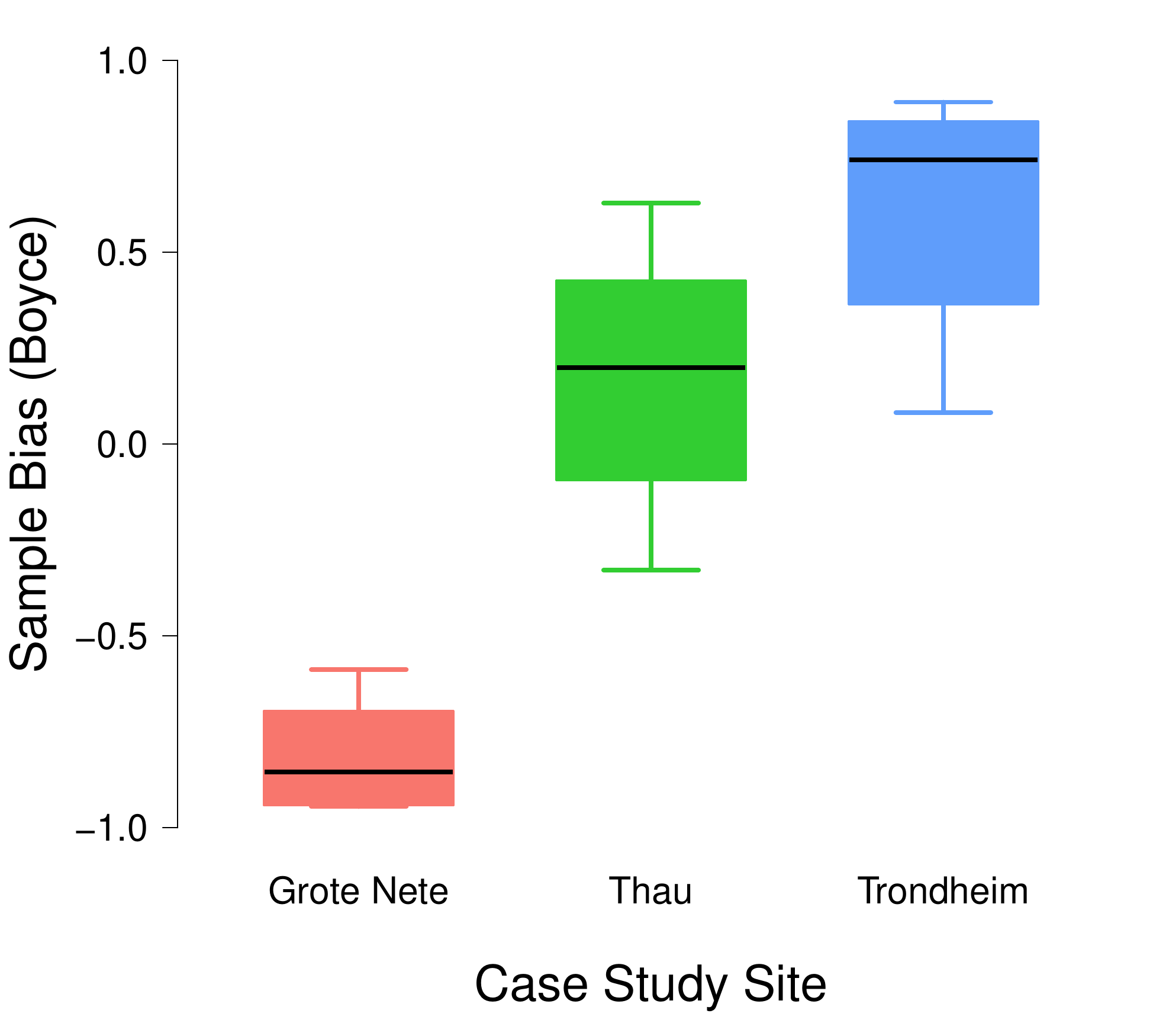}
		\caption{\sf \textbf{Boxplots of measured sample bias in the spatial distribution of the 64 real species according to the case study site.} The bias is assessed with Boyce indices measuring how accurately occurrence data are predicted by the Accessibility Index. Each boxplot is composed of the first decile, the first quartile, the median, the third quartile and the ninth decile. \label{Fig2}}
	\end{center}
\end{figure}

\begin{figure*}
	\begin{center}
		\includegraphics[width=14cm]{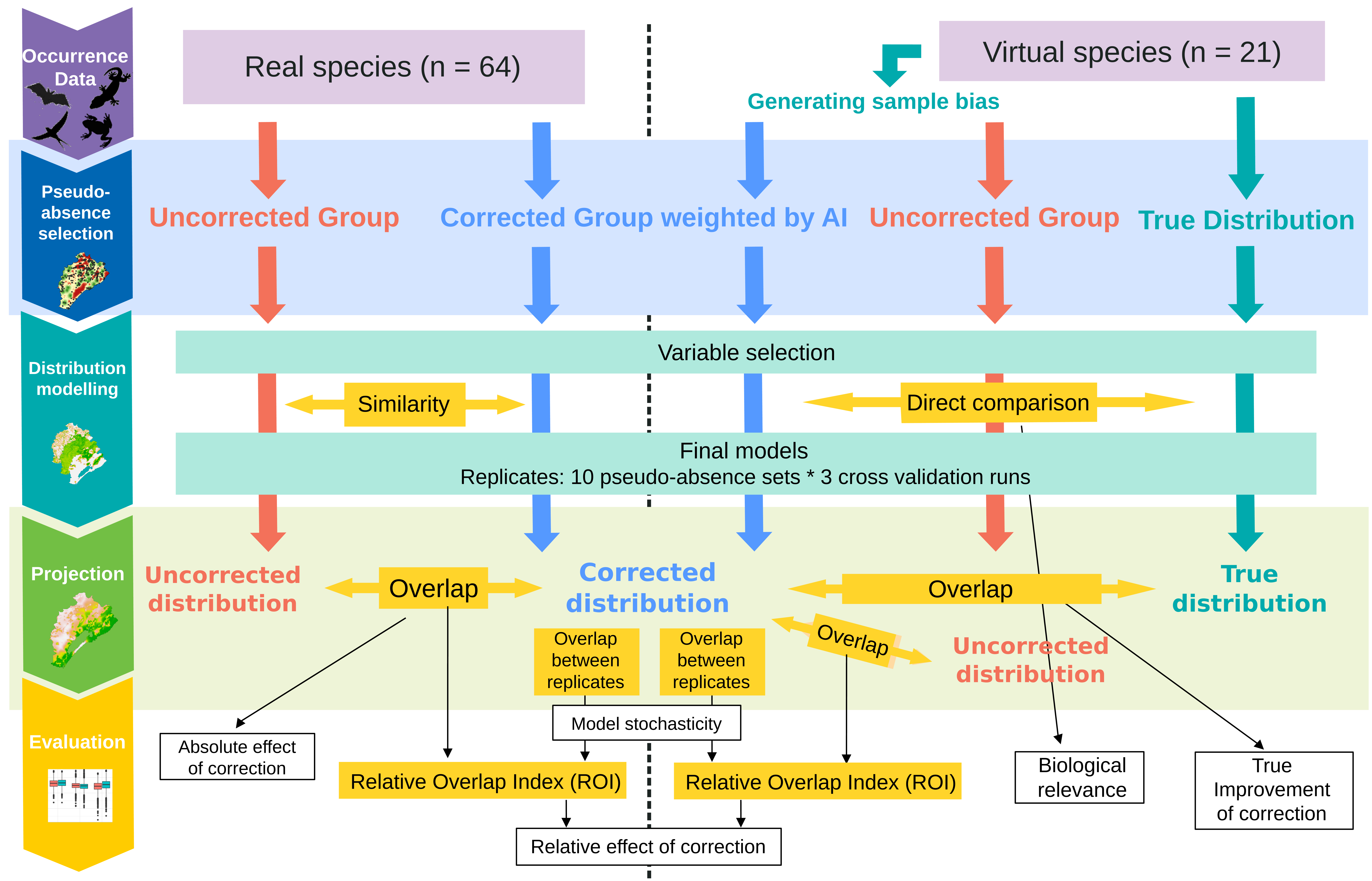}
		\caption{\sf \textbf{Steps of the methodology developed to study the effect of sample bias correction applied on the species distribution modelling of 64 real and 21 virtual species.} \label{Fig3}}
\end{center}
\end{figure*} 

\subsection*{Occurrence records}

At each of the three sites, we  used occurrence records obtained from biodiversity databases (\textit{Ligue pour la Protection des Oiseaux}\footnote[1]{\url{https://www.lpo.fr}, last accessed 17/02/2021}, the \textit{Artsobservasjoner}\footnote[2]{\url{https://www.artsobservasjoner.no}, last accessed 14/06/2018} and \textit{Natuurpunt Studie Association}\footnote[3]{\url{https://www.natuurpunt.be/afdelingen/natuurpunt-studie}, last accessed 17/02/2021}) for 79 terrestrial vertebrate species (58 birds, 10 mammals, 6 amphibians and 5 reptiles). We selected one occurrence point per  $200 \times 200 \mbox{ m}^2$ pixel (i.e. data thinning/resampling at the resolution of our environmental variables). Whilst this process is used as a rule of thumb to limit sampling bias driven by multiple observations within the pixel, but it does not account for sampling bias driven by aggregated observations in the surrounding pixels and at larger scales. After data thinning species with fewer than 10 occurrence points per site  were discarded (n = 15). The level of sample bias can be estimated using the Boyce index, which is usually used as an evaluation metric for presence-only data to assess the extent to which a spatial layer correctly predicts presence points. More details about the species name, sample size (after thinning/resampling) and sample bias are available in Table S1 in Appendix.

In our case, the spatial layer pertains to the accessibility map (Figure \ref{Fig1}). As  observed in Figure \ref{Fig2}, the spatial bias due to accessibility in occurrence data was negative in Grote Nete (average Boyce index = -0.75), slightly positive in Thau (average Boyce index = 0.16) and positive in Trondheim (average Boyce index = 0.59).

\subsection*{Distribution modelling}

We built species distribution models using the \textit{biomod2} R package \citep{Thuiller2009} and an ensemble of eight modelling techniques: generalised linear modelling (GLM), generalised boosting modelling (GBM), classification tree analysis (CTA), artificial neural network (ANN), surface range envelop (SRE), flexible discriminant analysis (FDA), general additive modelling (GAM) and random forest (RF). The modelling procedure included (1) a method for pseudo-absence selection, (2) an environmental variable selection process, (3) a final model calibration, and (4) a model evaluation process as summarised in Figure \ref{Fig3}.\\

\noindent\textbf{Pseudo-absence selection.} For each group, (uncorrected and corrected groups as defined below) we included 10 different sets of generated pseudo-absences equal in number to that of presence points \citep{Meynard2019}. For models which did not account for field accessibility, (and subsequent sampling bias) hereafter referred as the \enquote{uncorrected group}, we randomly and evenly selected a number of pseudo-absences in the study area equal to the number of occurrence data within the background \citep{Barbet2012,Liu2019}. For models which accounted for sampling bias, we randomly selected pseudo-absences with a sampling probability weighted by AIs after excluding presence pixels. This enabled pseudo-absences to share the same bias as presence points in accordance with the original concept proposed in \citet{Phillips2009}. Thus for species that were negatively biased by accessibility, i.e. more commonly found in inaccessible areas, we weighted sampling probability using negative AIs.\\

\noindent\textbf{Variable selection.} For both correction modalities, we selected one variable per group of inter-correlated variables to avoid collinearity, treating each site separately (Pearson's r $\geq$ 0.7). For each individual species, \citep{Hawkins2017} we assessed the relative importance of each variable (calculated as the Pearson's coefficient between initial model predictions and model predictions made when the assessed variable is randomly permuted) with 10 permutations. The final set of variables included in the final models were those with a relative importance of $\geq$ 0.05 across at least 50\% of model runs \citep{Bellard2016}.\\

\noindent\textbf{Final models.} We used the eight aforementioned modelling techniques with 10 sets of pseudo absence, and 3 runs of calibration over 80\% of the data (20\% for evaluation).

\subsection*{Effect of sample bias correction}

\noindent\textbf{Effect on model predictions.} We measured the \enquote{absolute} effect of sample bias correction using indices of similarity and correlation coefficients between uncorrected and corrected predictions. We computed the Schoener's D as a measure of projection overlap (computed with the \textit{ENMTool} R package \citep{Warren2010,Rodder2011}), the Pearson's correlation coefficient \citep{Li2013} and the Spearman's rank coefficient \citep{Phillips2009}. Schoener's D was computed as follows:
\begin{equation}
D(p_x,p_y)=1-\frac{1}{2} \sum_i |p_{x_i} - p_{y_i}| 
\label{D}
\end{equation}
For each species, modelling technique, cross-validation run, and pseudo-absence run individually, $p_{x_i}$ and $p_{y_i}$ are the normalised suitability scores for uncorrected $x$ and corrected $y$ prediction in grid cell $i$. This corresponds to the comparison of 30 corrected projections and 30 uncorrected projections (10 pseudo-absence datasets and 3 cross-validation subsets). This therefore represents  $8\cdot(3\cdot10)^2=7,200$ values for each species and overlap metrics. It is important to note that some predictions failed, particularly for the GBM modelling technique, and were therefore not included in the computation of the overlap metrics. See Table S2 and S3 in Appendix for more details.

For comparison, we also assessed model performance using four classical evaluation metrics based on cross-validation data subsets, namely: the Boyce index (computed with the \textit{ecospat} R package \citep{Cola2017}), the true skill statistic (TSS), the area under the relative operating characteristic curve (AUC), and a calibrated AUC (cAUC). The cAUC was computed following \citet{Hijmans2012} and calibrated on the AUC of a null geographic model. The null geographic model was computed with the \textit{geoDist} function of the \textit{dismo} R package \citep{Hijmans2015}. The Boyce index, a reliability metric, indicates the extent to which a spatial layer correctly predicts presence points. The TSS, AUC and cAUC are discrimination metrics that indicate the ability to distinguish between occupied and unoccupied sites. We also used a one-sided Student t-test in order to evaluate whether or not the mean corrected performance was significantly greater than the mean uncorrected performance for a given performance metric, species, and modelling technique. The p-value was computed with the \textit{rquery.t.test} function of the \textit{ContDataQC} R package \citep{Leppo2021}.\\

\noindent\textbf{Effect relative to model stochasticity.} For each species and modelling technique, we assessed the extent to which the correction technique affected predictions relative to the sources of stochastic variation between models of the corrected group (i.e. cross-validation runs and pseudo-absence set runs for each modelling technique). Model stochasticity was quantified using the aforementioned overlap metrics (Schoener's D, Pearson's and Spearman's coefficient) between all pairwise combinations of model projections for the each of the 64 species individually (10 pseudo-absence datasets, 3 cross validation subsets and 8 modelling techniques, resulting in $8\cdot\frac{(3\cdot10)^2-(3\cdot10)}{2}=3,480$ values for each species and overlap metric). We present the Relative Overlap Index (ROI), an index of mean overlap between predictions of the uncorrected and the corrected groups, relative to the average overlap between pairwise model projections of the corrected group. The two overlap components of the ROI can be assessed either with similarity metrics or correlation coefficients. When based on Schoener's D, the ROI was computed as follows.
\begin{equation}
ROI=\frac{\bar{D}_{0}-\bar{D}}{\bar{D}_{0}}
\label{ROI}
\end{equation}
Where $\bar{D}_{0}$ is the mean overlap between model runs of the corrected group, $\bar{D}$ is the mean overlap between runs of the uncorrected and the corrected groups. It is important to note that the ROI is always computed for a given species and modelling technique, where $\bar{D}_{0}$ is based on $\frac{(3\cdot10)^2-(3\cdot10)}{2}=435$ overlaps between model runs, and $\bar{D}$ is based on $(3\cdot10)^2=900$ overlaps between corrected and uncorrected models runs. A value close to 0 represents a perfect match between predictions, i.e. no effect of sample bias correction. The overlaps between uncorrected and corrected groups tend to be significantly smaller than the overlaps between runs when the ROI approaches 1 (i.e. strong effect of sample bias correction). A negative value can sometimes be obtained when the sample size is small, meaning that model stochasticity is of higher magnitude than the sample bias correction. In this case, we can thereby conclude that there is no effect of sample bias correction. The formula is similar when based on Pearson's and Spearman's rank coefficient, but values were transformed in order to range between 0 and 1 by adding 1 and dividing by 2. 

We also used a one-sided Student t-test  to evaluate whether or not $\bar{D}_{0}$ was significantly greater than $\bar{D}$ for a given overlap metric, species, and modelling technique. The p-value was computed with the \textit{rquery.t.test} function of the \textit{ContDataQC} R package \citep{Leppo2021}.\\

\noindent\textbf{Effect on variable selection.} We estimated the degree of similarity in the selected variables between the uncorrected and corrected groups. We used the Jaccard index \citep{Jaccard1912} only when considering whether or not the variable was selected. We also computed the Bray-Curtis index \citep{Bray1957} when accounting for variable importance.\\

\begin{figure*}
	\begin{center}
		\includegraphics[width=\linewidth]{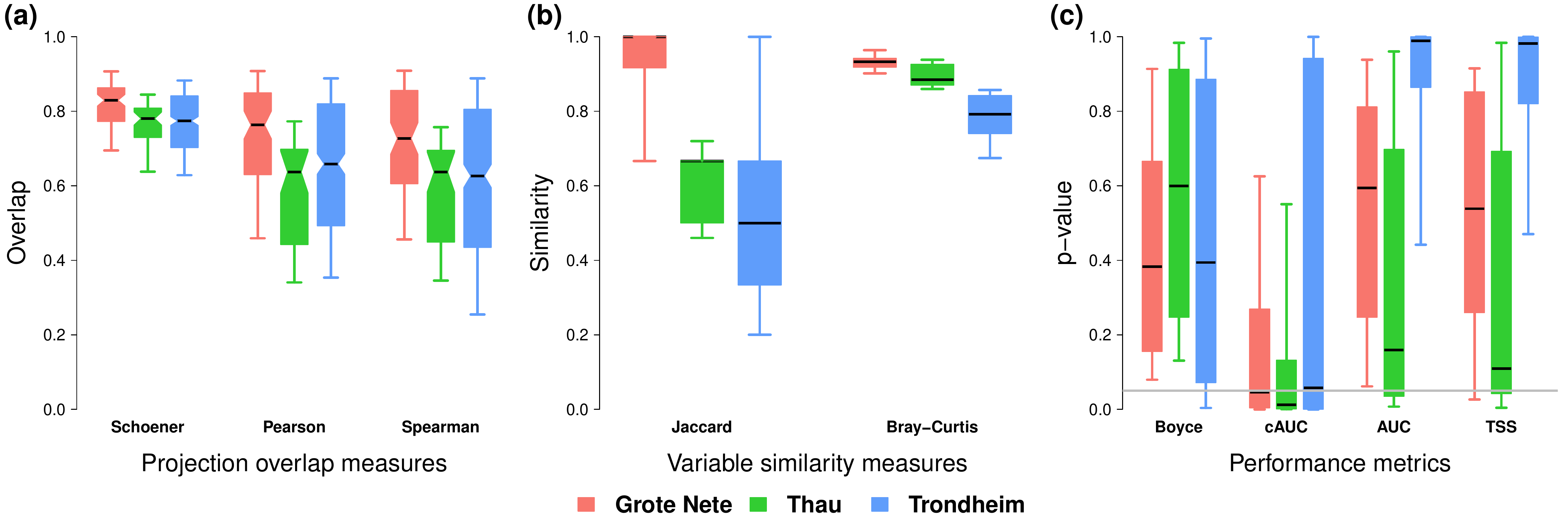}
		\caption{\sf \textbf{Site-specific variation in the effect of sample bias correction.} (a) Boxplots of the three overlap measures between corrected and uncorrected predictions. Each boxplot is composed of the average overlap between 30 corrected and 30 uncorrected projections obtained for each species and each modelling technique. (b) Boxplots of the variable selection similarity (one value per species). (c) Boxplots of the p-value of a one-sided Student t-test to evaluate whether the mean performance of corrected group is greater than the mean performance of corrected group. The grey horizontal line represents the 0.05 significance threshold. Each boxplot is composed of the first decile, the first quartile, the median, the third quartile and the ninth decile. \label{Fig4}}
\end{center}
\end{figure*}

\subsection*{Testing whether changes correspond to actual improvements using virtual species}

For each of the three study sites, we generated a set of 7 virtual species, with different ecological niches, generated at each site (21 species in total) using the \textit{Virtualspecies} R package \citep{Leroy2016}. The probability of their presence was generated according to a relationship with a single environmental variable (parameters and shape of the relationships are provided in Table \ref{Tab1}). We produced presence-absence maps using a probability threshold of 0.6 and sampled 300 presence points.  For species associated with the grassland index and the proportion of herbaceous areas within 500 meters, for which there were fewer than 300 pixels, we respectively selected 100 and 80 presence points. We then produced a spatially biased sample for the 21 virtual species by weighting the probability of selection using the Accessibility Index. More details about the species name, sample size, and sample bias are available in Table S4 in Appendix.

For the generated species, we used the same distribution modelling framework and metrics to assess the effect of sample bias correction as those used for the real species.  It is again worth noting that some predictions failed (particularly for the GBM modelling technique) and were therefore excluded from the computation of the overlap metrics (see Tables S5 and S6 in Appendix for more details).\\

\noindent\textbf{Did the correction effect correspond to an improvement?} The virtual species generated with simulated sample bias enabled us to test whether the correction technique actually improved the models' ability to predict the \enquote{true} distribution \citep{Meynard2012}. Using the  Schoener's D overlap, the Pearson's correlation coefficient and the root-mean-square error (RMSE) to quantify the degree to which model predictions were improved by the correction technique, we compared the predicted probability of occurrence of the corrected and uncorrected groups with the \enquote{true} probability of occurrence.

We relied on a one-sided Student t-test to evaluate whether or not the overlap with the \enquote{true} distribution was significantly greater for the corrected group than the uncorrected group for a given overlap metric, species, and modelling technique. The p-value was computed with the \textit{rquery.t.test} function of the \textit{ContDataQC} R package \citep{Leppo2021}.

We used the following procedure to assess Schoener's D and classic performance metrics’ (Boyce, cAUC, AUC and TSS) ability to assess the effect of correction. We took the overlap with the \enquote{true} distribution as reference for dividing each species and modelling technique into two groups (effect/no effect). This was based upon the value of the significance threshold $\delta$ determined by the one-sided Student t-test p-value (lower than $\delta$/higher than $\delta$). We followed the same process with the Schoener'D overlap (p-value associated with the comparison between $\bar{D}_{0}$ and $\bar{D}$), the Boyce, cAUC, AUC and TSS (p-values associated with the comparison of performance metrics for the corrected and uncorrected projections). We then built five tables of confusion to see how the five metrics classified the species for each modelling technique when compared with the reference for a given significance threshold $\delta$. To measure the proportion of species and modelling techniques classified into the same category (effect and no effect) we computed \enquote{accuracy}  for a given significance threshold $\delta$.\\ 

\noindent\textbf{Did the correction improve the biological relevance of variable selection?} We determined whether variable selection was relevant by comparing the variables used to generate the virtual species with those selected before and after correction.

\begin{figure*}
	\begin{center}
		\includegraphics[width=\linewidth]{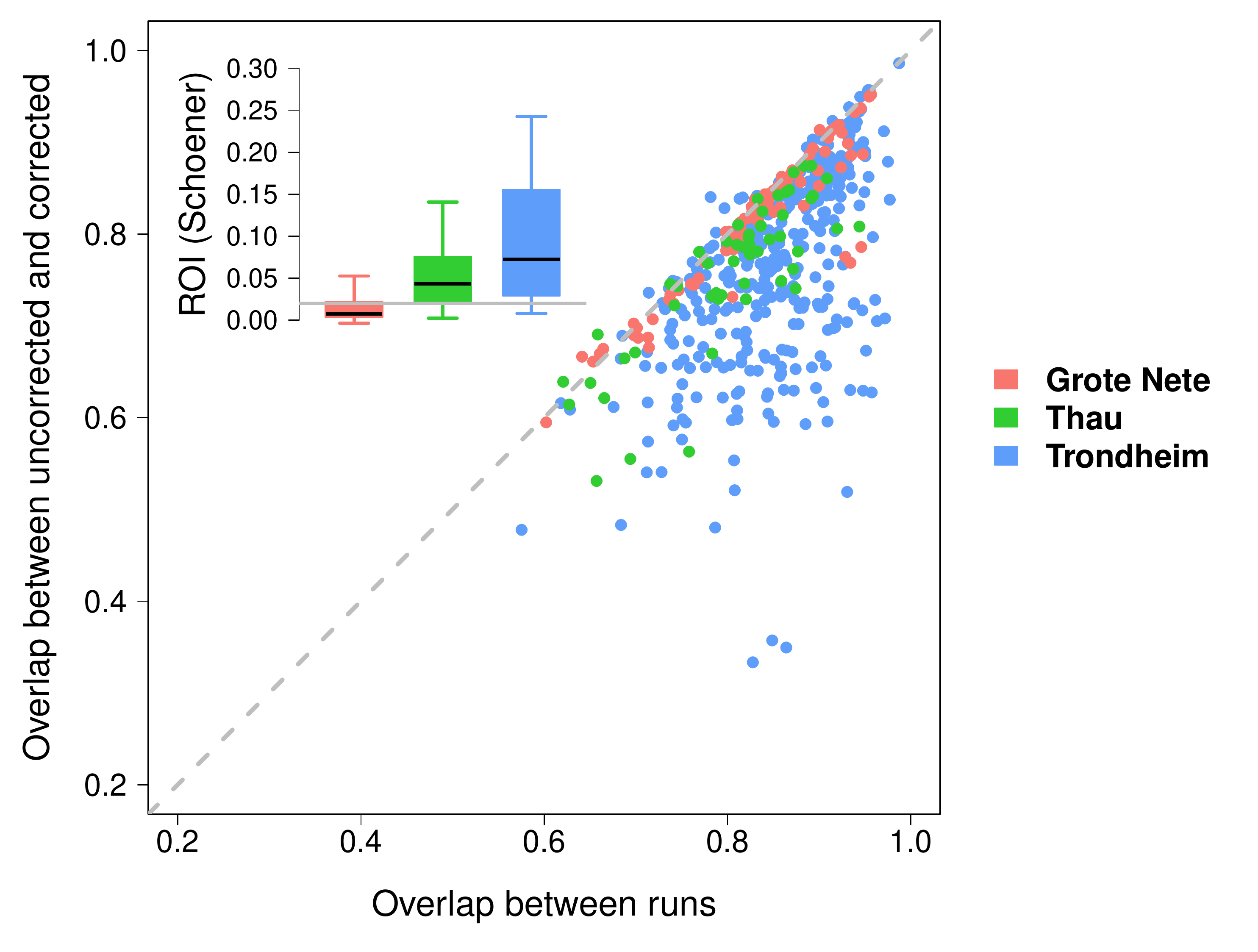}
		\caption{\sf \textbf{Site-specific variation in the relative effect of sample bias correction.} Relationship between the effect of sample bias correction (Schoener's D overlap between uncorrected and corrected predictions $\bar{D}$) and the effect of model stochasticity (Schoener's D overlap between model runs $\bar{D}_{0}$). Points located below the y = x line represent models for which the relative effect of sample bias correction exceeds that of model stochasticity. The inset shows the boxplots of the Relative Overlap Index (ROI) according to the case study site. The horizontal grey line represents the ROI threshold value 0.02 associated with the one-sided Student t-test significance threshold 0.05 (see Figure S29 in Appendix for more details). Each boxplot is composed of the first decile, the lower hinge, the median, the upper hinge and the ninth decile. \label{Fig5}}
\end{center}
\end{figure*} 

\section*{Results}

\subsection*{Effect of sample bias correction}

\noindent\textbf{Effect on model predictions.} The effect of correction was largely consistent between sites as shown by the overlap between projections built from uncorrected and corrected pseudo-absences, plotted in Figure \ref{Fig4}a. Model projections shared about 80\% of information (Schoener's D) common to uncorrected and corrected models. Maps of the projection are available in Appendix (Figures S1 to S17). Generally, there were no significant differences between Boyce, AUC, and TSS indices between groups (Figure \ref{Fig4}c). Only the cAUC showed a significant effect of correction for a majority of species and modelling techniques over the three sites, as indicated by the low median p-value (Figure \ref{Fig4}c).\\

\noindent\textbf{Effect on variable selection.} At Thau and Trondheim, we found important differences in variable selection (Jaccard indices) and variable importance (Bray-Curtis indices) between corrected and uncorrected models (Figure \ref{Fig4}b).

\begin{figure*}
	\begin{center}
		\includegraphics[width=\linewidth]{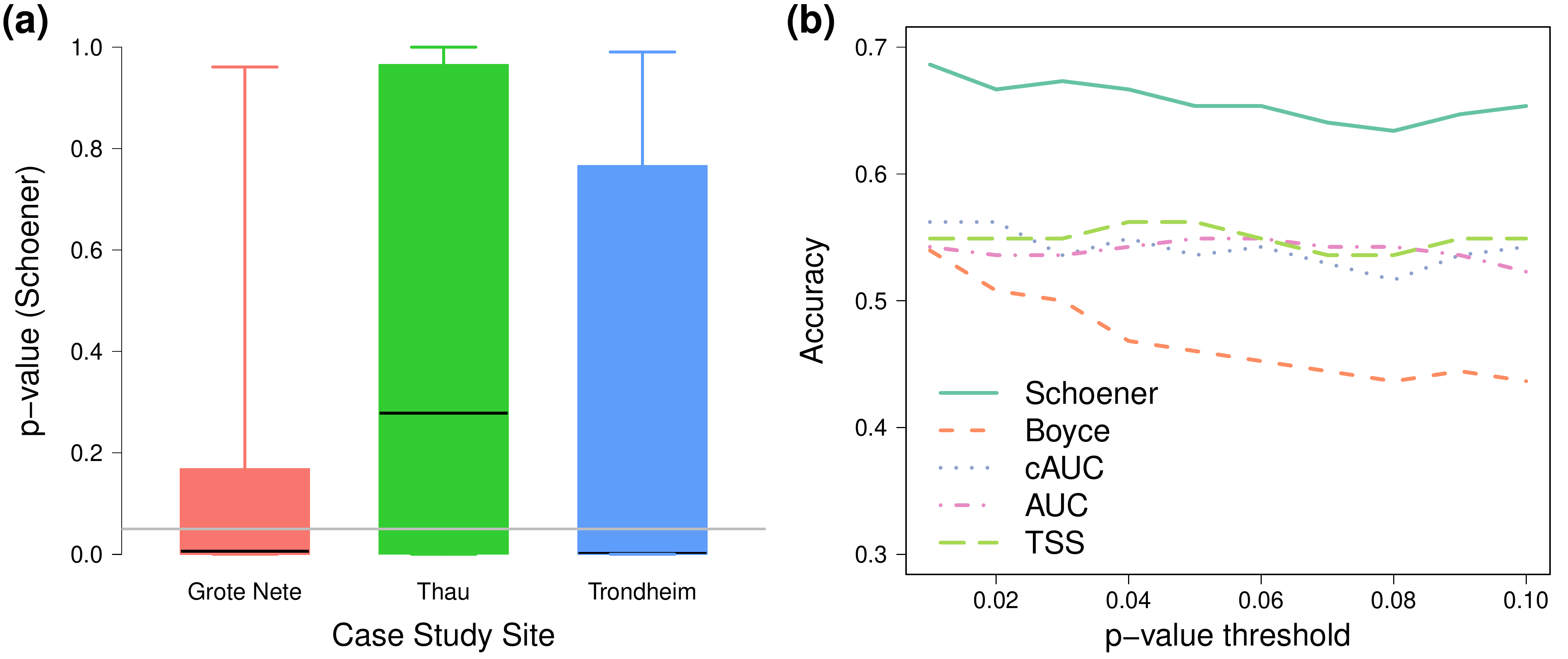}
		\caption{\sf \textbf{Comparison with the \enquote{true} probability of occurrence for the uncorrected and the corrected groups at three sites (virtual species).} (a) Boxplots of the p-value obtained from a one-sided Student t-test evaluating whether the overlap between Schoener's D and the \enquote{true} distribution is significantly greater for the corrected group than the uncorrected group, for a given species and modelling technique. Similar plots obtained with the Pearson's coefficient and the RMSE are available in Figure S30 in Appendix. (b) Evolution of the accuracy values as a function of the significance threshold $\delta$. The five tables of confusion pertaining to accuracy values obtained within a significance threshold $\delta = 0.05$ are available in Table S7 in Appendix. \label{Fig6}}
\end{center}
\end{figure*}

\subsection*{Relative effect of correction}

The effect of correction was high compared to model stochasticity in Trondheim, and to a lesser extent in Thau. This is shown by the lower overlaps between treatments (i.e. corrected and uncorrected groups) compared to overlaps between model replicates (Figure \ref{Fig5}). At the remaining site, the effect of correction was of similar magnitude to that of model replicates. This result remained consistent when using Pearson's and Spearman's coefficients as a measure of overlap (Figure S28 in Appendix).

A direct relationship exists between the ROI and the one-sided Student t-test p-value used to evaluate whether the overlap between runs ($\bar{D}_{0}$) was significantly greater than the overlap between corrected and uncorrected projections ($\bar{D}$). Indeed, we can show that all the p-values associated with ROI were strictly higher than 0.02 and lower than 0.05 (Figure S29 in Appendix). The majority of ROI values in Thau and Trondheim were higher than 0.02 (as shown in the inset of Figure \ref{Fig5}).

\subsection*{Did the correction actually improve predictions?}

The correction remarkably improved distribution predictions in Trondheim and Grote Nete for the vast majority of species and modelling techniques (Figure \ref{Fig6}a). The results were less impressive in Thau with less than 42\% of the species and modelling techniques showing an improvement. Maps of the projection are available in Appendix (Figures S18 to S23).

In terms of \enquote{accuracy} with regards to the classification of species and modelling techniques into two groups (effect/no effect), the results obtained using Schoener's D (p-value associated with the comparison between $\bar{D}_{0}$ and $\bar{D}$) are more in line with the classification (considering the \enquote{true} distribution) than those obtained using the classic performance metrics (Boyce, cAUC, AUC and TSS). This may be observed in Figure \ref{Fig6}b. 

\subsection*{Did the correction improve the biological relevance of the selected variables?}

Sample bias correction improved the biological relevance for 17 virtual species (Table \ref{Tab1}). The improvement was characterised by an increase in relative importance for the relevant variable (i.e. used to generate the species; n = 9), or by an increase in the importance of the relevant variable and a decrease the importance of irrelevant variables (n = 8). For two such instances, the improvement was mitigated by an increase in the relative importance of irrelevant variables.

\begin{table*}
	\caption{\textbf{Variables and response functions used to generate virtual species, and variables selected during the modelling process for the uncorrected and the corrected group.\footnote{We have provided the distribution family and the parameters of the response function. The variables used to generate the species for which the relative importance increased after correction, and variables which were not used for which the relative importance decreased are shown in bold.}}}
	\label{Tab1}
	\vspace*{0.5cm}
	\centering 
	\includegraphics[width=\linewidth]{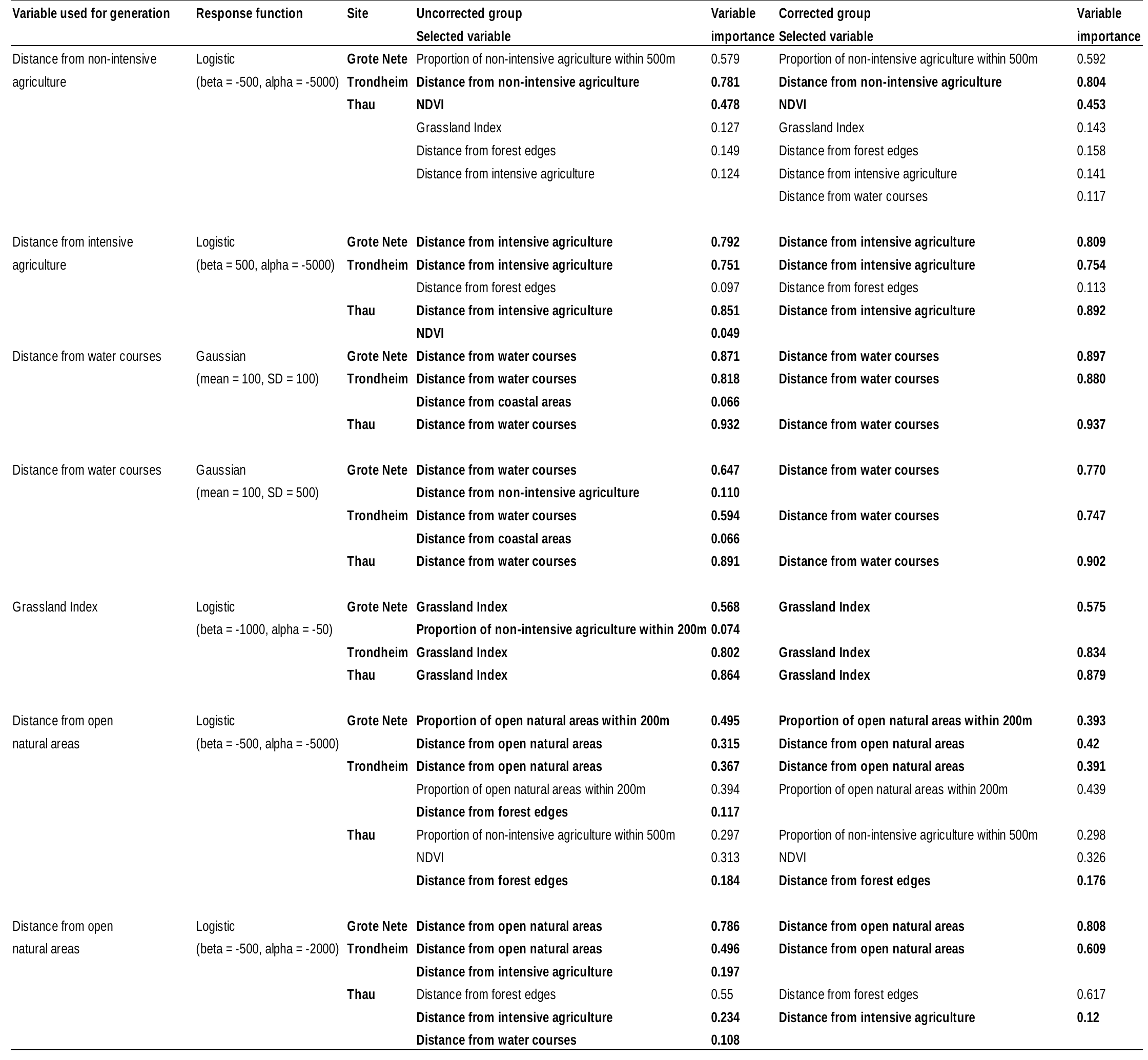}
\end{table*}

\section*{Discussion}

The efficiency of sample bias correction based on accessibility varies between sites and species. The effect of correction depends upon the landscape complexity at each site, and on the degree of spatial bias in the occurrence data and predictor variables. In absence of independent data, the effect of the correction cannot be assessed with classic evaluation metrics (difference in performance, or absolute overlap between corrected and uncorrected projections). The sample generated bias in virtual species enabled us to verify that the Relative Overlap Index represented better the effect of correction than alternative approaches.

\subsection*{Effect of sample bias correction}

In this study, we found that the impact of sample bias correction primarily depended upon the study site and, to a lesser extent, on species. Sample bias correction primarily affected species distribution predictions in Trondheim and, to a lesser extent, in Thau. The fact that the impact of correction between sites differed may be explained by variations in the spatial scale and landscape complexity between sites. The accessibility index was based upon site-specific characteristicsand reflects a relative measure of distance. The less accessible zones were geographically closer to the more accessible ones at smaller spatial scales, and the absolute distance may prove unproblematic in terms of accessibility. This suggests that sampling bias correction, when based on accessibility, is less necessary when studying areas with homogeneous distributions of roads and towns at smaller spatial scales. 

Additionally, accessibility maps may not reflect sample bias equally for all species and sites. Accessibility maps can be built on the basis of topography, land use, and property, to better represent species or site-specific sample bias. In any case, the comparisons with the \enquote{true}probability of occurrence of virtual species suggested that the correction was likely to improve model predictions at each site for a high number of species and modelling techniques (Figure \ref{Fig6}a).

We also investigated whether the effect of correction differed between sites and according to the sample bias, sample size and the modelling technique. Methodological details, results and related discussion are available in Appendix. The effect of correction on predictions was greater when sample bias was positive but remained highly heterogeneous, even among species with strong sample biases. Variation in ROI value tends to decrease with increasing sample size. Species with low sample sizes (fewer than 100 occurrences after thinning/resampling) cover a very wide range of ROI values from negative value (close to -0.1) to very high ROI value (up to 0.6). This is presumably because the model stochasticity may be higher as a result of lower accuracy \citep{Stockwell2002}. More details are available in Appendix.

\subsection*{Biological relevance of variable selection}

Variable selection differed the most after correction at Trondheim, where correction was most effective and accessibility gradients were clearest. For the most part, models of the uncorrected group selected additional variables that were not biologically meaningful, presumably because they were locally correlated with the accessibility index (intensive agriculture, water courses and distance to the coast in Trondheim; Figures S1 to S17 in Appendix). In Trondheim, models tended to explain the absence of occurrence data in the inaccessible area by the environmental variable which was most represented there. Virtual species indicated that the correction increased the relative importance of biologically meaningful variables, since they were often used to generate species distribution. This result is supported by the virtual species analysis, wherein the effect of correction decreased when the species habitat variable was correlated with accessibility (e.g. distance from intensive agriculture in Trondheim correlated with the AI, Pearson's r = 0.80). This result underlines that the biological relevance of the variable selected before and after correction needs to be carefully investigated, in accordance with the recommendations given in  \citet{Hijmans2012}and \citet{Fourcade2018}.

When the geographic sampling bias translates into a bias in the environmental predictors, the benefit conferred by sample bias correction with accessibility maps may be limited – as is consistent with the target-group approach \citep{Ranc2017}. In other terms, when the distribution of environmental variables matches that of accessibility, issues related to sample biases cannot be corrected with accessibility maps. In our study regions, this may be the case for species that are most impacted by urbanisation \citep{Geslin2013} and intensive agriculture \citep{Jeliazkov2016,Olivier2020}.

\subsection*{Correction effect relative to model stochasticity}

In the absence of independent, standardised data, the performance of SDMs and correction methods cannot be properly assessed. Here, we propose to measure the effect of correction relative to the within-model stochasticity (between runs of varying input parameters, individually, for each modelling technique) to inform the potential benefit conferred by correction. We show that the Relative Overlap Index was in better agreement than the classic cross-validation performance metrics when concerning changes between corrected and uncorrected predictions of virtual species. The ROI also yielded better results with regards to changes in variable selection and the relative importance across sites and species. The use of this index may be generalisable to species for which habitat is not restricted to the same section of the accessibility gradient. This metric can be used to indicate whether species distribution models are likely to be improved by sample bias correction.

\subsection*{Cross-validation metrics}

The performance metrics based on cross-validation failed to detect an improvement in species range predictions, even for cAUC. This differs from the findings of \citet{Hertzog2014}, which assessed the performance of a variety of bias correction techniques based on the model's ability to predict the range of a dung beetle (Coprophagous Scarabaeidae). This study found a striking difference between the evaluation of partitioned datasets and field validation. In their study, cAUC was in agreement with field validation. In our case, the cAUC showed an effect of correction on real species but not virtual ones with perfectly known distribution and generated bias. This calls into question the reliability of cAUC to properly identify a change after correction. As specified by \citet{Hijmans2012}, a null geographic model may not be relevant when a species occurs in a single continuous range. Perhaps this may explain why no improvement of correction was detected in some cases. Another recently developed evaluation method considers the accumulation curve of occurrences within the area predicted as suitable as well as the amount of uninformative niche space predicted \citep{Jimenez2020}. This method is appropriate when absence data are unavailable and is similar to our approach in that the evaluation is relative to a random component. It is recommended that multiple metrics describing various aspects of model performance be reported, so as  to improve the understanding and transparency of SDMs \citep{Araujo2019}. Our results suggest that the performance of sample bias correction should be assessed using the Relative Overlap Index, and that alternative metrics may be misleading. The ROI is to be interpreted as the degree of change in spatial predictions whilst accounting for variation between model replicates.

\subsection*{Generalisation and limitations}

Our analyses included a wide range of species (n = 64) of various taxa (four classes of vertebrates) with varying responses to environmental predictors. We generated a set of virtual species with contrasting ecological preferences (e.g. dependence on water courses or open, natural areas) and varying degrees of specialisation commonly found in amphibians, bats, and birds \citep{Godet2015,Jeliazkov2014,Dubos2021}. We also attempted to represent the diversity of species sensitivity to anthropogenic disturbance by including species with a negative response to intensive agriculture \citep{Chiron2014}. However, our study does not encompass the entirety of species’ distribution, nor does it account for climatic predictors, which may reduce the scope of our conclusions. The correction method best improved our predictions at the largest site which suggests, that our correction method, based on accessibility, may prove efficient at larger geographic scales where terrain accessibility is a clear issue. Aside from terrain accessibility, the efficiency of a correction technique may also differ depending upon the type of bias. Further studies should be designed at broader taxonomic and geographic scales and should assess potential differences between various correction techniques.

\subsection*{Increasing the potential use of biodiversity databases}

The limited availability of high-quality data is still a major hindrance to effective decision-making, despite the increasing availability of open-source, publicly available data and EU policies to promote open access and data sharing. Accounting for sample bias is challenging, especially for rare species and those for which the distribution range is small with subsequently fewer occurrence points. Pseudo-absence selection, weighted by accessibility maps, enables us to account for sample bias without the use of filtering techniques that reduce the amount of data available. Sample bias correction is also viable for when concerning broadly distributed species in the event that  occurrence data are spatially biased – provided that the effect of correction is assessed \citep{Hertzog2014}. When standardised data are unavailable, effectively assessing the efficiency of sample bias correction techniques remains challenging. However, the relative measure of its effect and the use of virtual species could prove to be critical for the increased inclusion of large, heterogeneous, biased datasets in species distribution models and biodiversity assessments.
\section*{Acknowledgements}

This study was partly supported by the IMAGINE project (ERANET BIODIVERSA). Natuurpunt Studie Association provided the data for Grote Nete, Region in Belgium, LPO (\textit{Ligue pour la Protection des Oiseaux}) Occitanie, provided the data for Thau Region, the French site. Data from Norway are part of GBIF. A special thank goes to Boris Leroy for helpful discussions and wise advices. 

\section*{Author contributions}

ND, ML and SL designed the analyses. ND and ML performed the analyses. CP built the environmental variables and the map. GP, SM and PD aided in interpreting the results and worked on the manuscript. MLL and FR, SH and RM helped to organize data collected in France, Belgium and Norway, respectively. ND led the writing of the manuscript. All authors read, commented and validated the final version of the manuscript.

\section*{Data Accessibility Statement}

All relevant data and code are available online (\url{https://gitlab.com/maximelenormand/sample-bias-correction-sdms}).

\bibliographystyle{myapalike}
\bibliography{Accessibility}

\onecolumngrid

\makeatletter
\renewcommand{\fnum@figure}{\sf\textbf{\figurename~\textbf{S}\textbf{\thefigure}}}
\renewcommand{\fnum@table}{\sf\textbf{\tablename~\textbf{S}\textbf{\thetable}}}
\makeatother

\setcounter{figure}{0}
\setcounter{table}{0}
\setcounter{equation}{0}

\newpage
\clearpage
\newpage
\section*{Appendix}

\subsection*{Ensemble projection maps}

\begin{figure}[!h]
	\begin{center}
		\includegraphics[width=16cm]{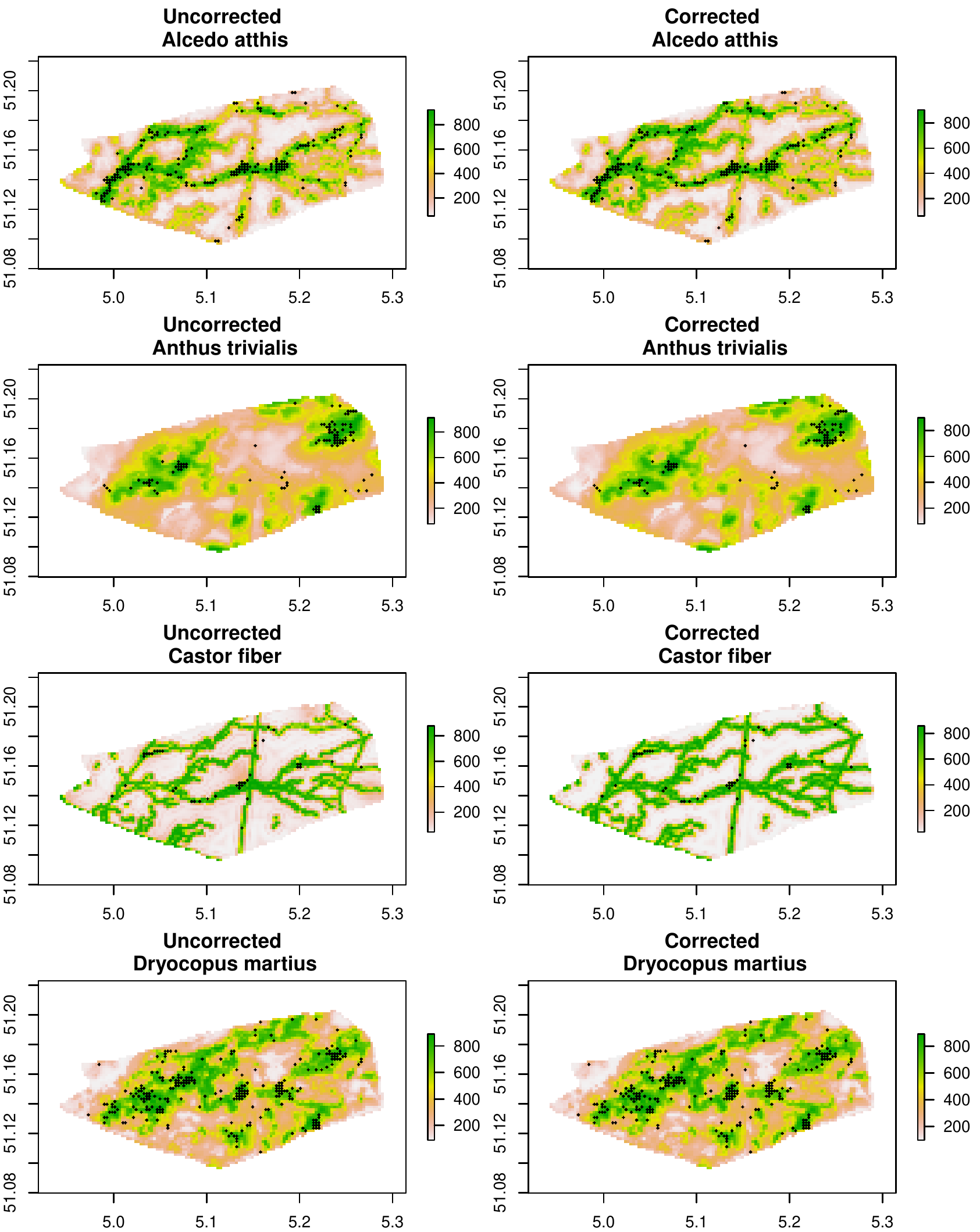}
		\caption{\sf \textbf{Species projections in Grote Nete (uncorrected versus corrected group).} From top to bottom, \textit{Alcedo atthis}, \textit{Anthus trivialis}, \textit{Castor fiber} and \textit{Dryocopus martius}. 			
			\label{FigS1}}
	\end{center}
\end{figure}

\begin{figure}[!h]
	\begin{center}
		\includegraphics[width=\linewidth]{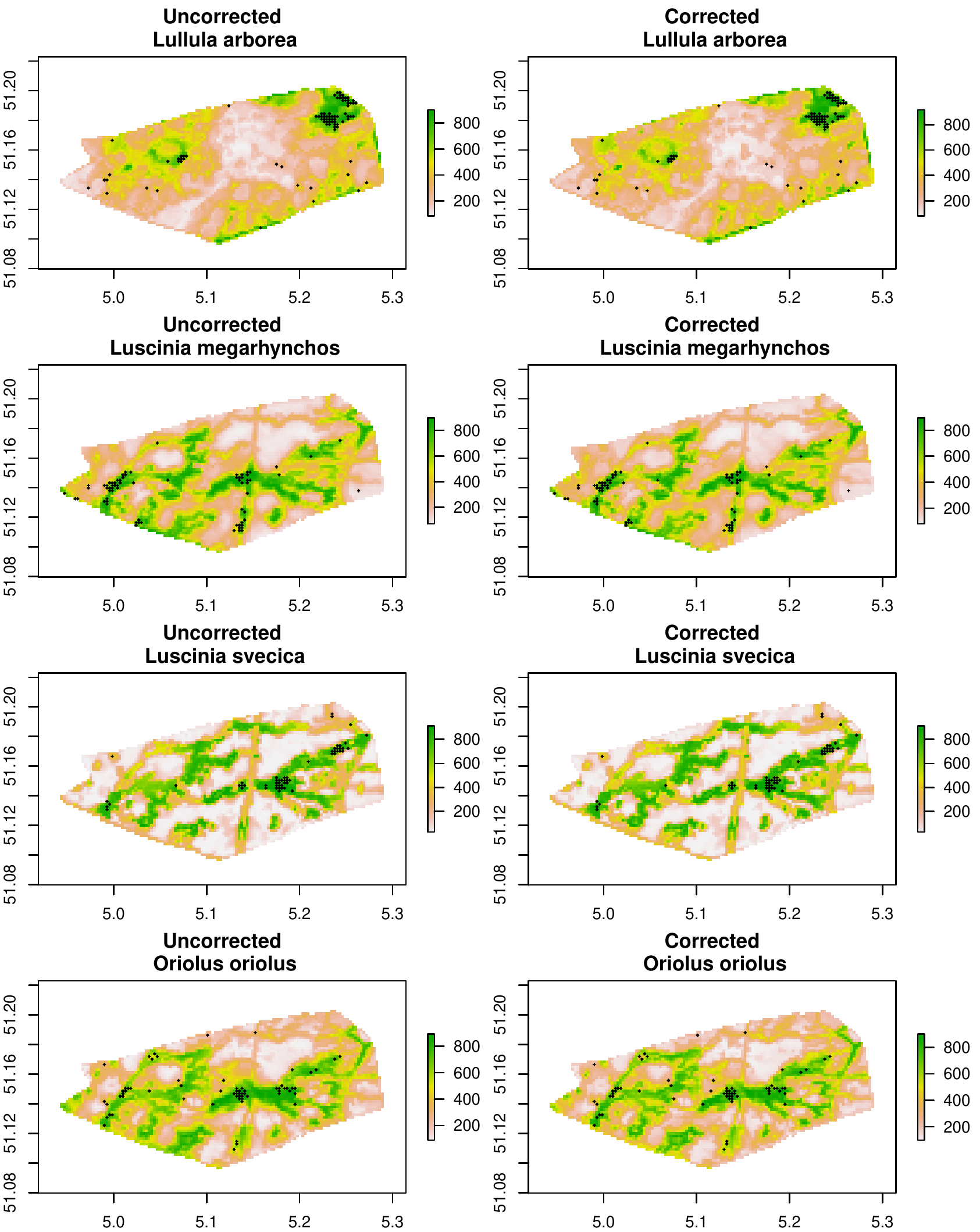}
		\caption{\sf \textbf{Species projections in Grote Nete (uncorrected versus corrected group).} From top to bottom, \textit{Lullula arborea}, \textit{Luscinia megarhynchos}, \textit{Luscinia svecica} and \textit{Oriolus oriolus}. 			
			\label{FigS2}}
	\end{center}
\end{figure}

\begin{figure}[!h]
	\begin{center}
		\includegraphics[width=\linewidth]{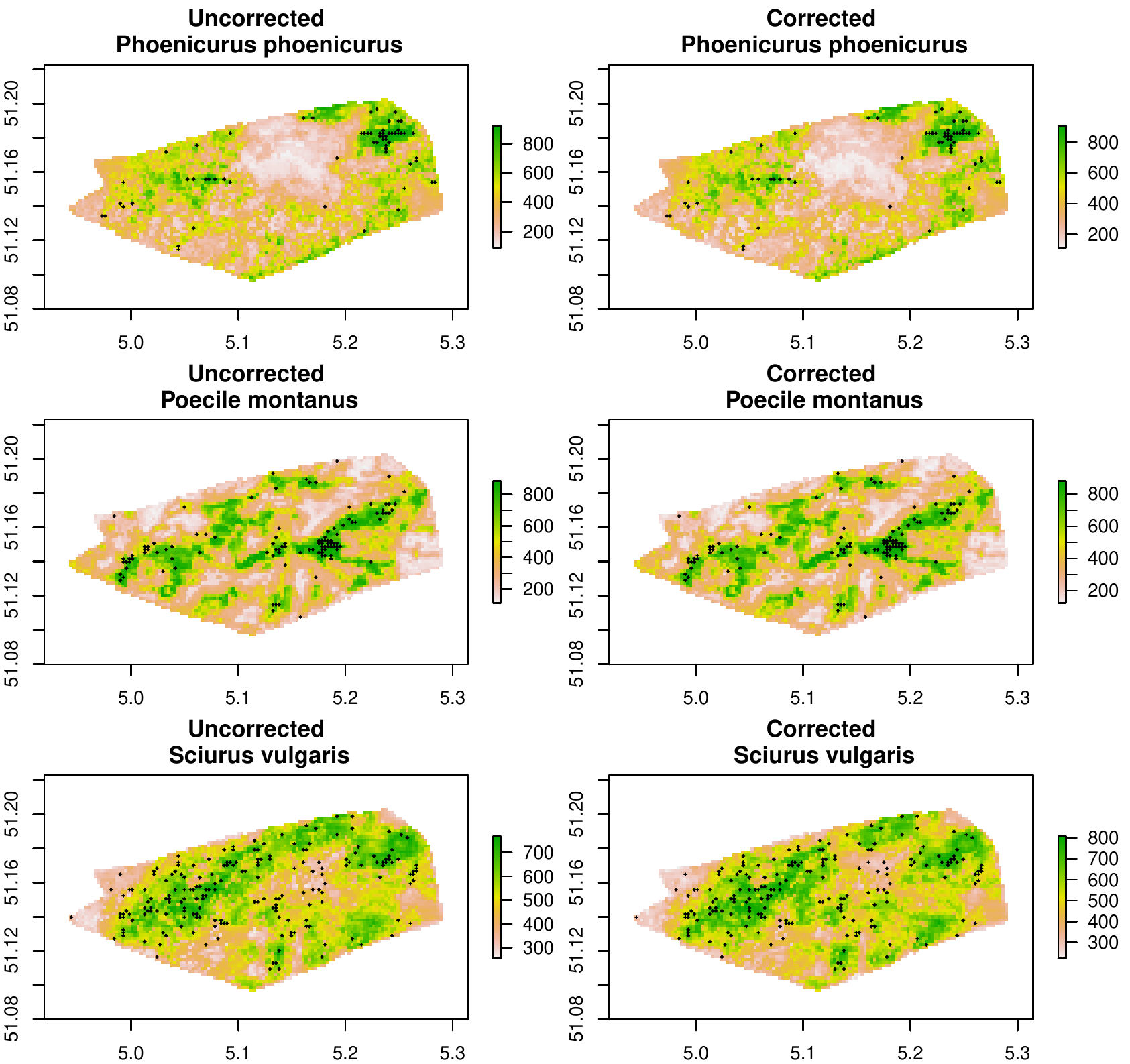}
		\caption{\sf \textbf{Species projections in Grote Nete (uncorrected versus corrected group).} From top to bottom, \textit{Phoenicurus hoenicurus}, \textit{Poecile montanus} and \textit{Sciurus vulgaris}. 			
			\label{FigS3}}
	\end{center}
\end{figure}

\begin{figure}[!h]
	\begin{center}
		\includegraphics[width=\linewidth]{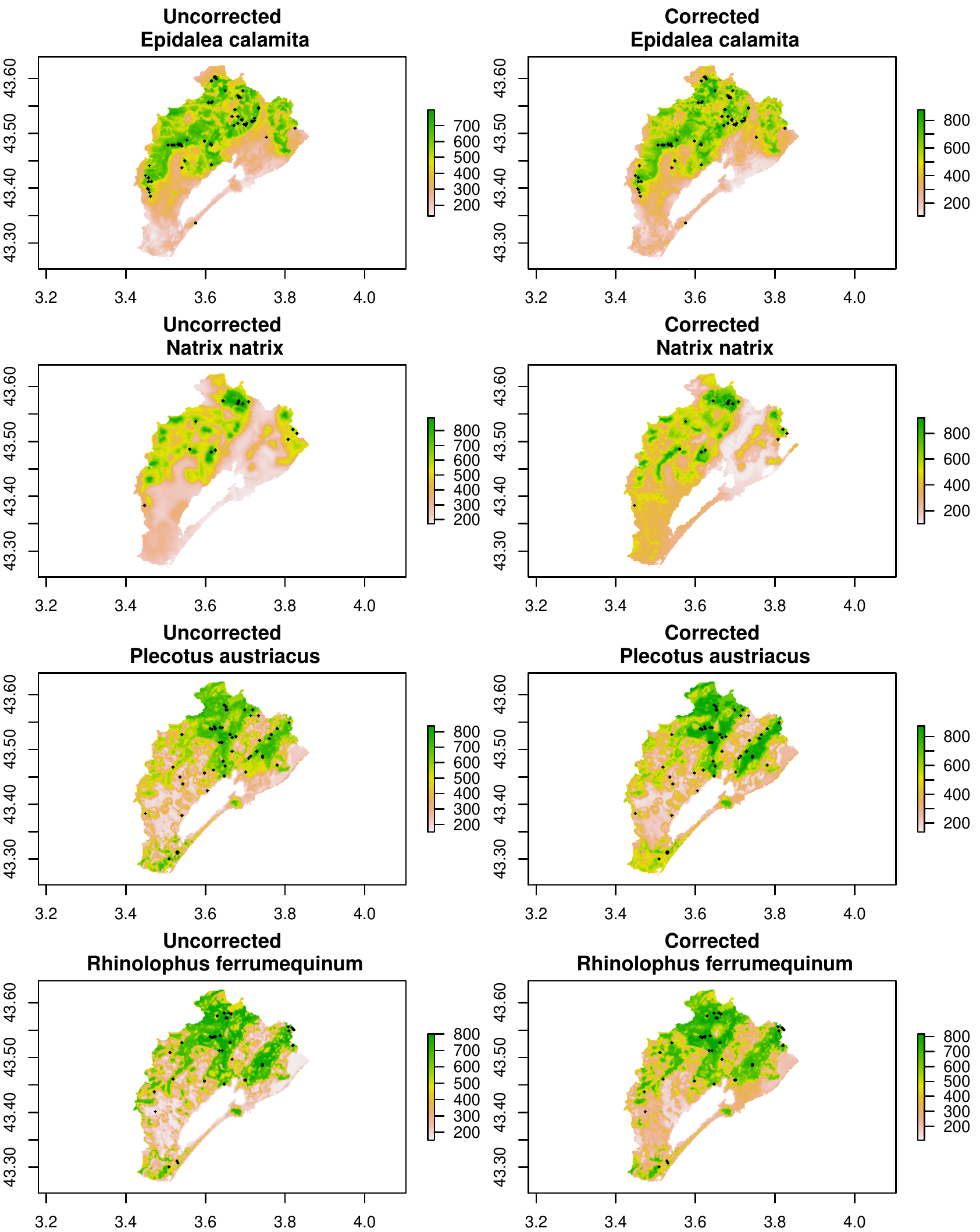}
		\caption{\sf \textbf{Species projections in Thau (uncorrected versus corrected group).} From top to bottom, \textit{Epidalea calamita}, \textit{Natrix natrix}, \textit{Plecotus austriacus} and \textit{Rhinolophus ferrumequinum}. 			
			\label{FigS4}}
	\end{center}
\end{figure}

\begin{figure}[!h]
	\begin{center}
		\includegraphics[width=\linewidth]{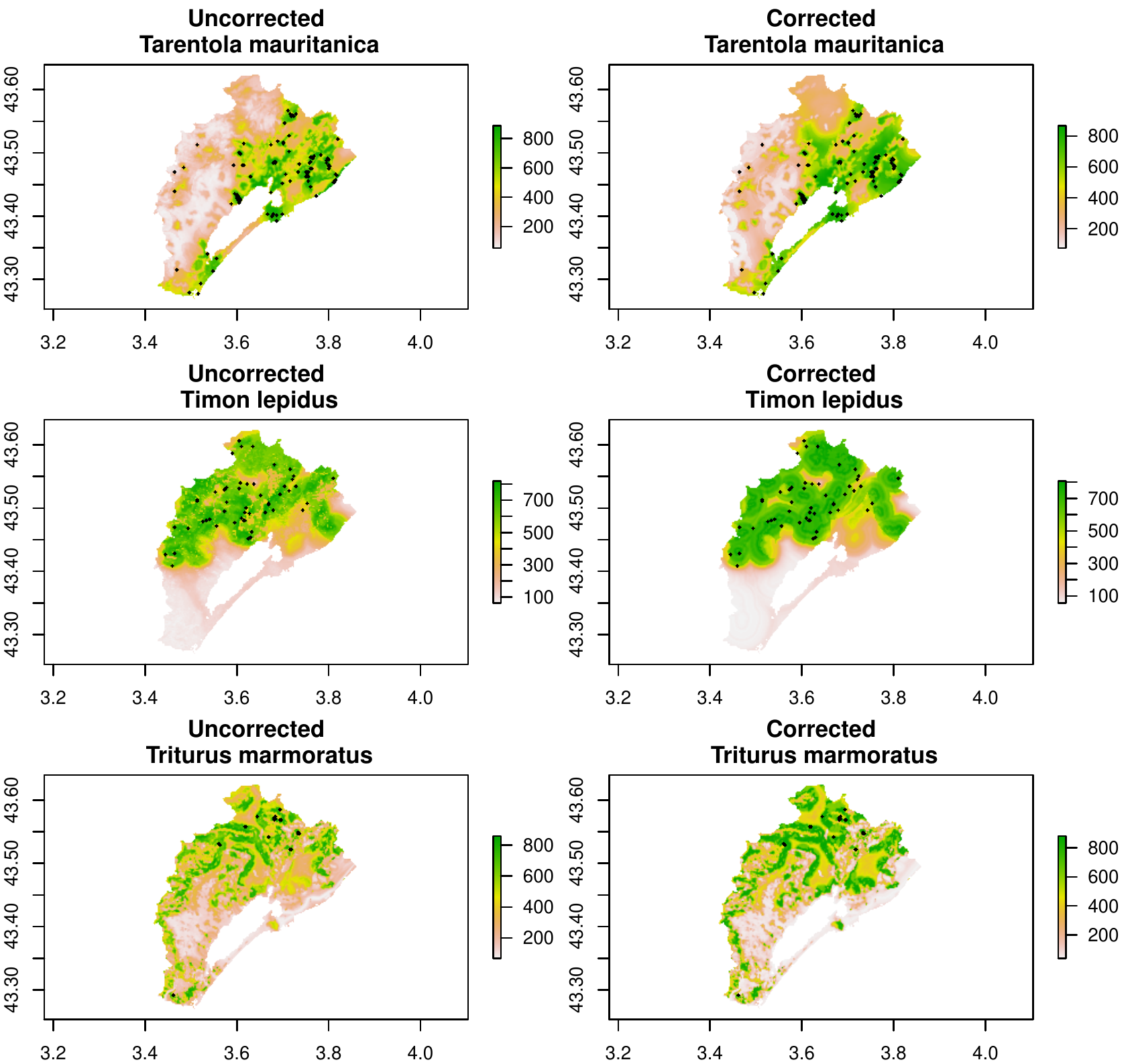}
		\caption{\sf \textbf{Species projections in Thau (uncorrected versus corrected group).} From top to bottom, \textit{Tarentola mauritanica}, \textit{Timon lepidus} and \textit{Triturus marmoratus}. 			
			\label{FigS5}}
	\end{center}
\end{figure}

\begin{figure}[!h]
	\begin{center}
		\includegraphics[width=\linewidth]{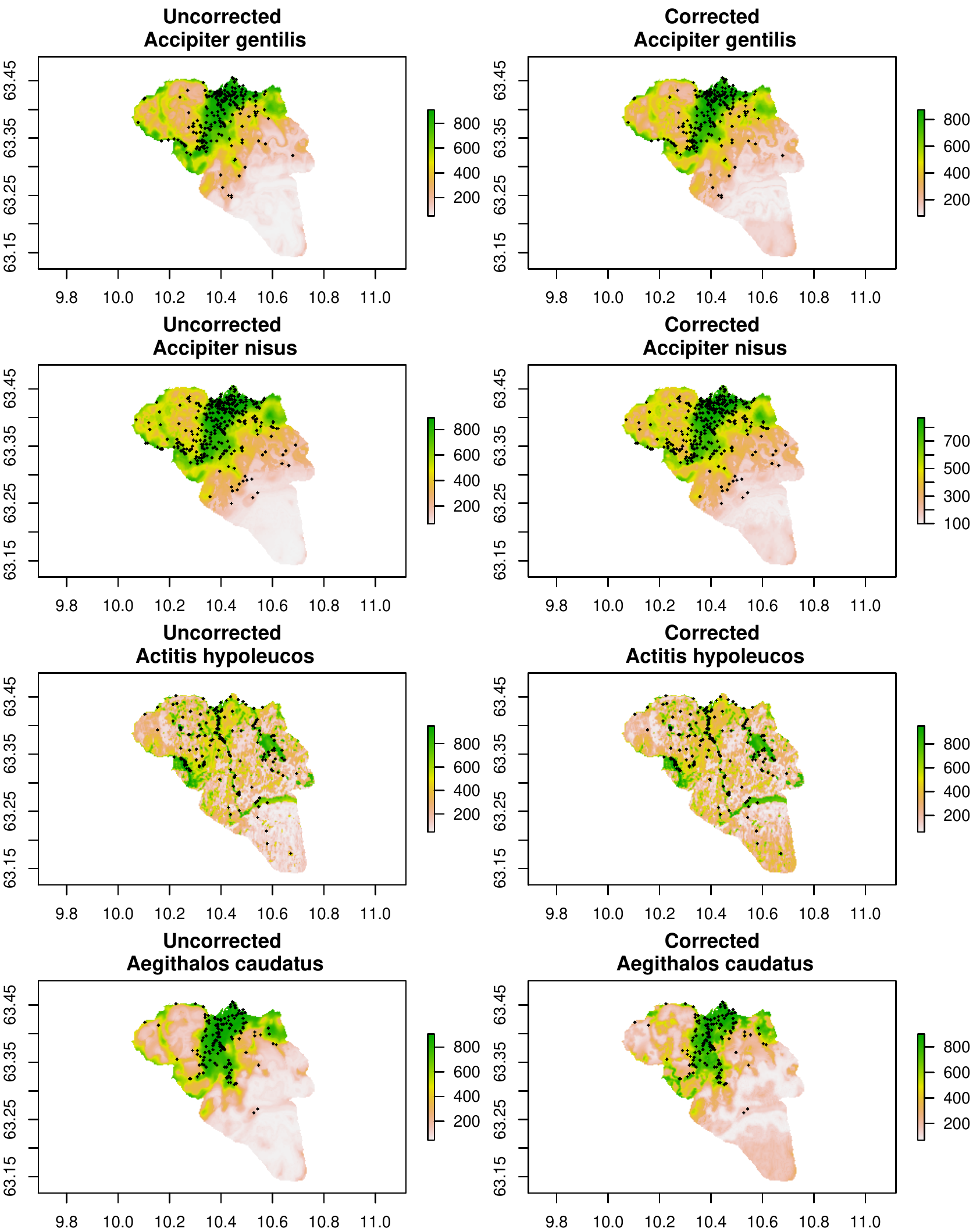}
		\caption{\sf \textbf{Species projections in Trondheim (uncorrected versus corrected group).} From top to bottom, \textit{Accipiter gentilis}, \textit{Accipiter nisus}, \textit{Actitis hypoleucos} and \textit{Aegithalos caudatus}. 			
			\label{FigS6}}
	\end{center}
\end{figure}

\begin{figure}[!h]
	\begin{center}
		\includegraphics[width=\linewidth]{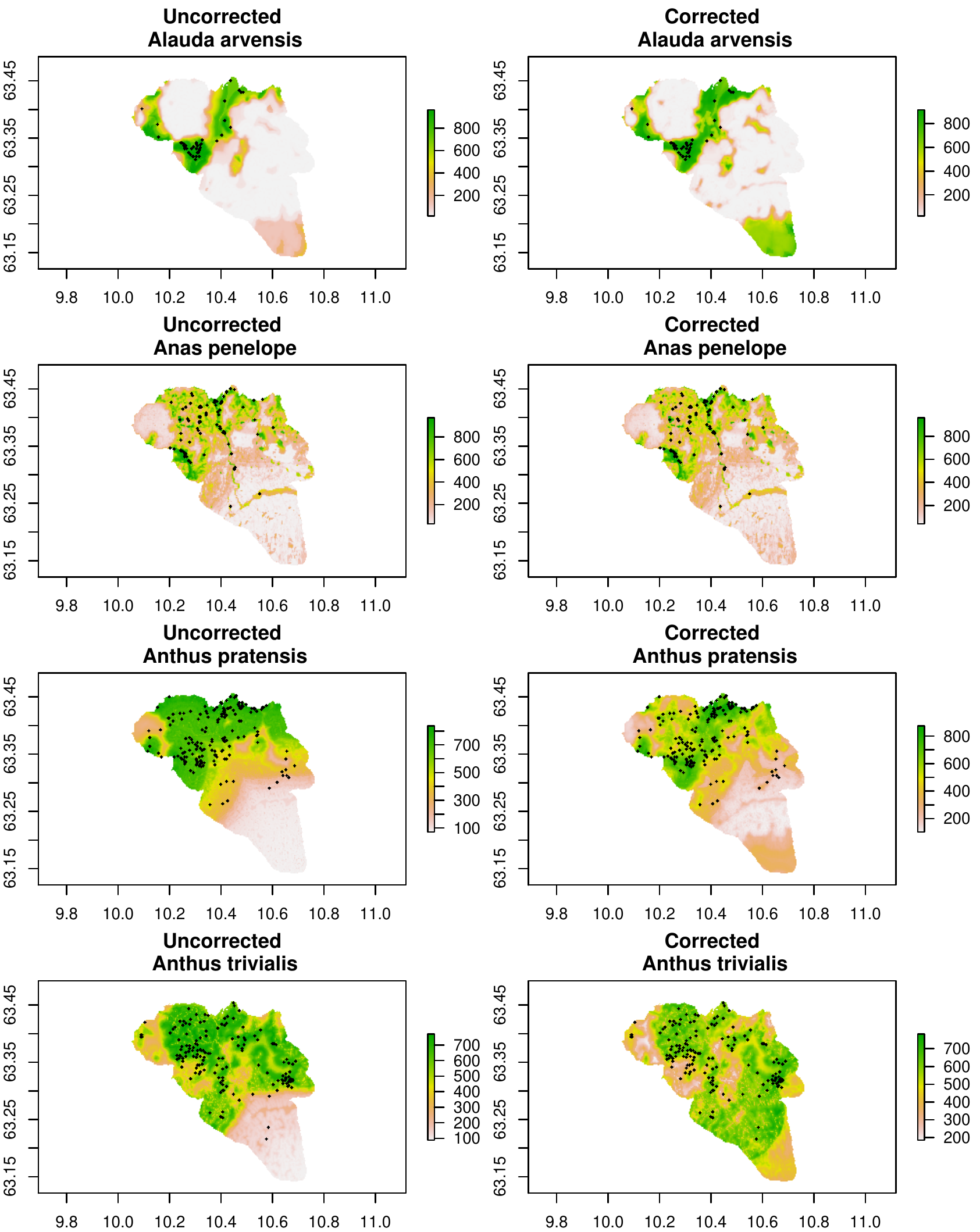}
		\caption{\sf \textbf{Species projections in Trondheim (random versus corrected group).} From top to bottom, \textit{Alauda arvensis}, \textit{Anas acuta},  \textit{Anthus pratensis} and \textit{Anthus trivialis}. 		
			\label{FigS7}}
	\end{center}
\end{figure}

\begin{figure}[!h]
	\begin{center}
		\includegraphics[width=\linewidth]{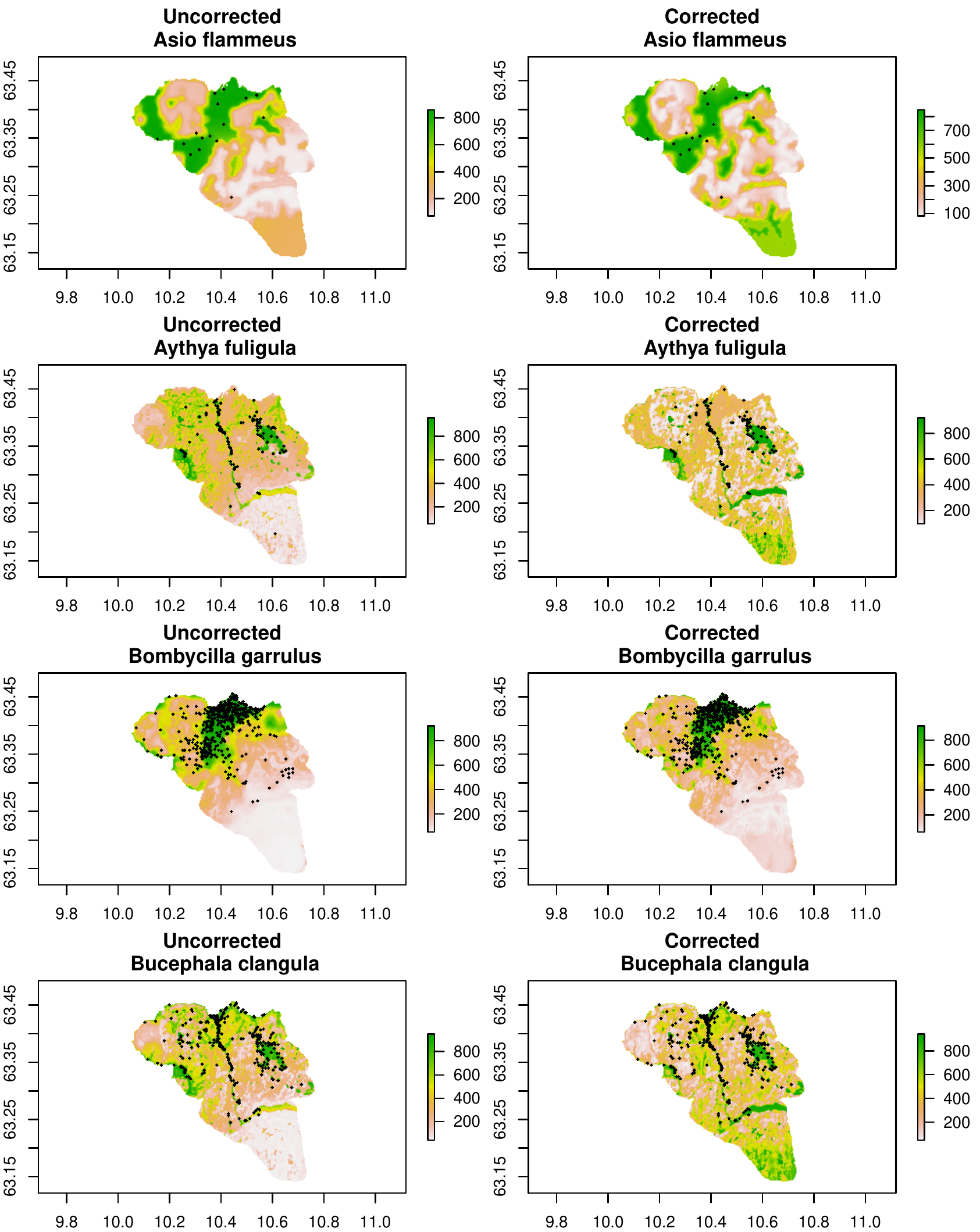}
		\caption{\sf \textbf{Species projections in Trondheim (uncorrected versus corrected group).} From top to bottom,\textit{Asio flammeus}, \textit{Aythya fuligula}, \textit{Bombycilla garrulus} and \textit{Bucephala clangula}. 		
			\label{FigS8}}
	\end{center}
\end{figure}

\begin{figure}[!h]
	\begin{center}
		\includegraphics[width=\linewidth]{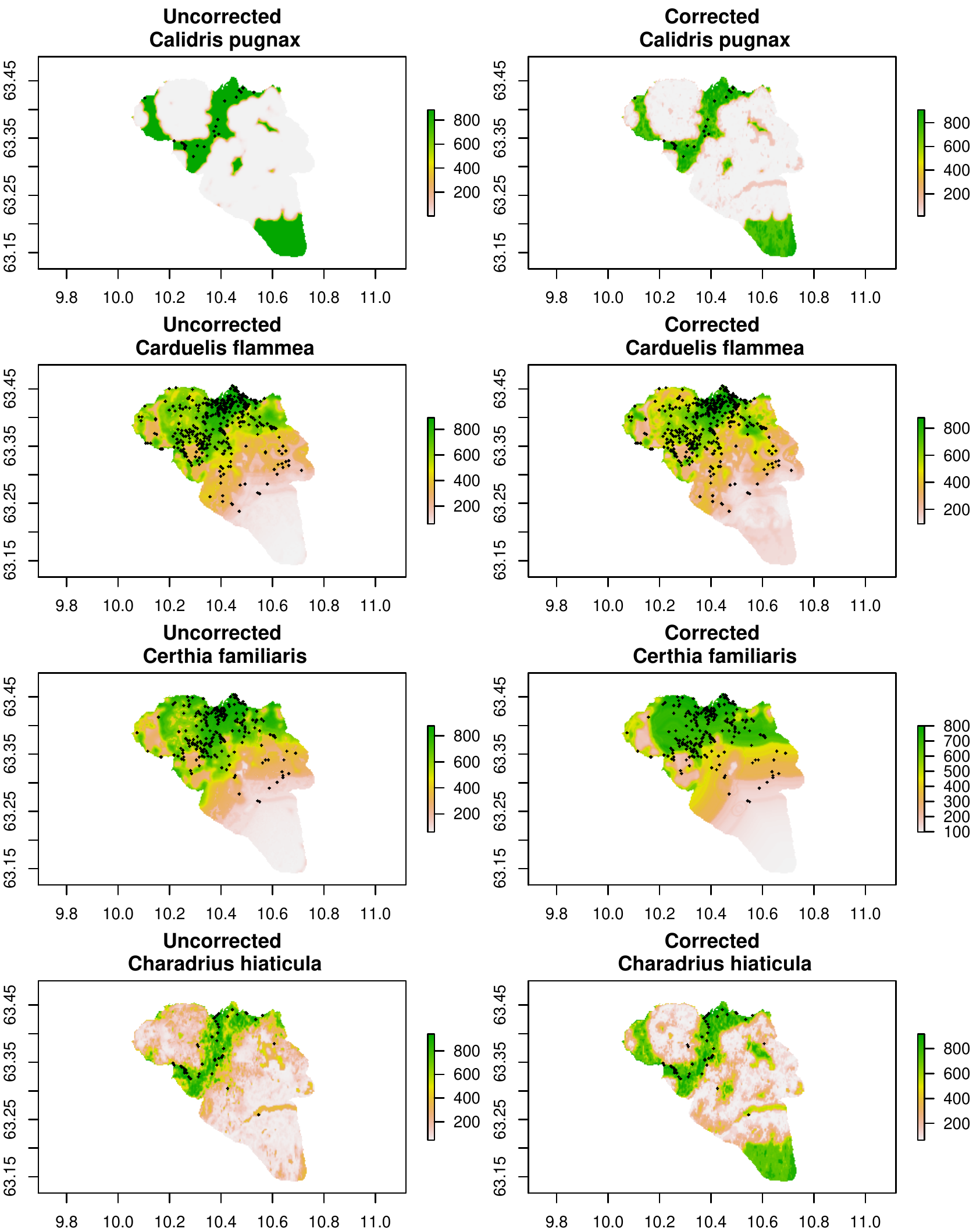}
		\caption{\sf \textbf{Species projections in Trondheim (uncorrected versus corrected group).} From top to bottom, \textit{Calidris pugnax}, \textit{Carduelis flammea}, \textit{Certhia familiaris} and \textit{Charadrius hiaticula}. 		
			\label{FigS9}}
	\end{center}
\end{figure}

\begin{figure}[!h]
	\begin{center}
		\includegraphics[width=\linewidth]{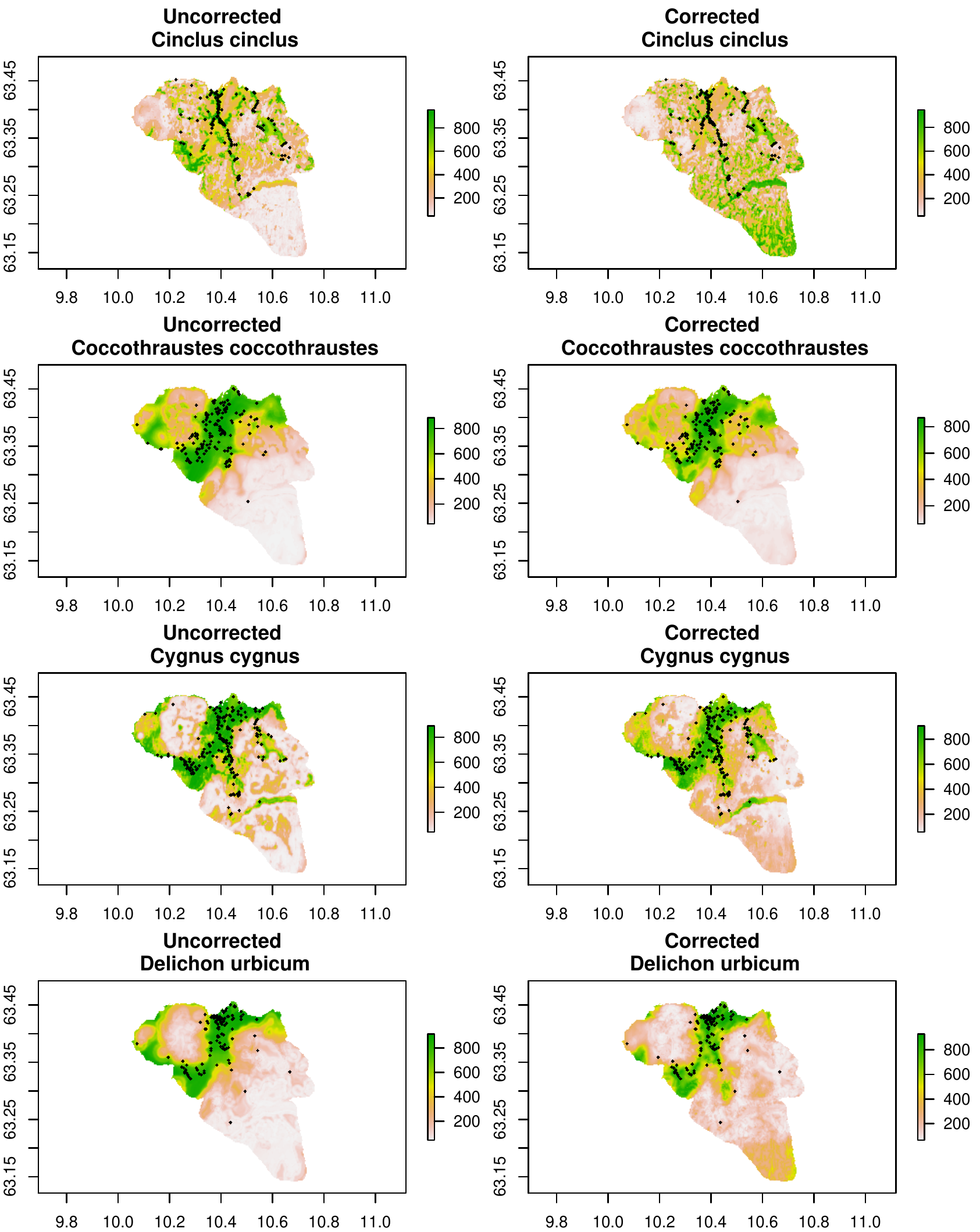}
		\caption{\sf \textbf{Species projections in Trondheim (uncorrected versus corrected group).} From top to bottom, \textit{Cinclus cinclus}, \textit{Coccothraustes coccothraustes}, \textit{Cygnus cygnus} and \textit{Delichon urbicum}. 		
			\label{FigS10}}
	\end{center}
\end{figure}

\begin{figure}[!h]
	\begin{center}
		\includegraphics[width=\linewidth]{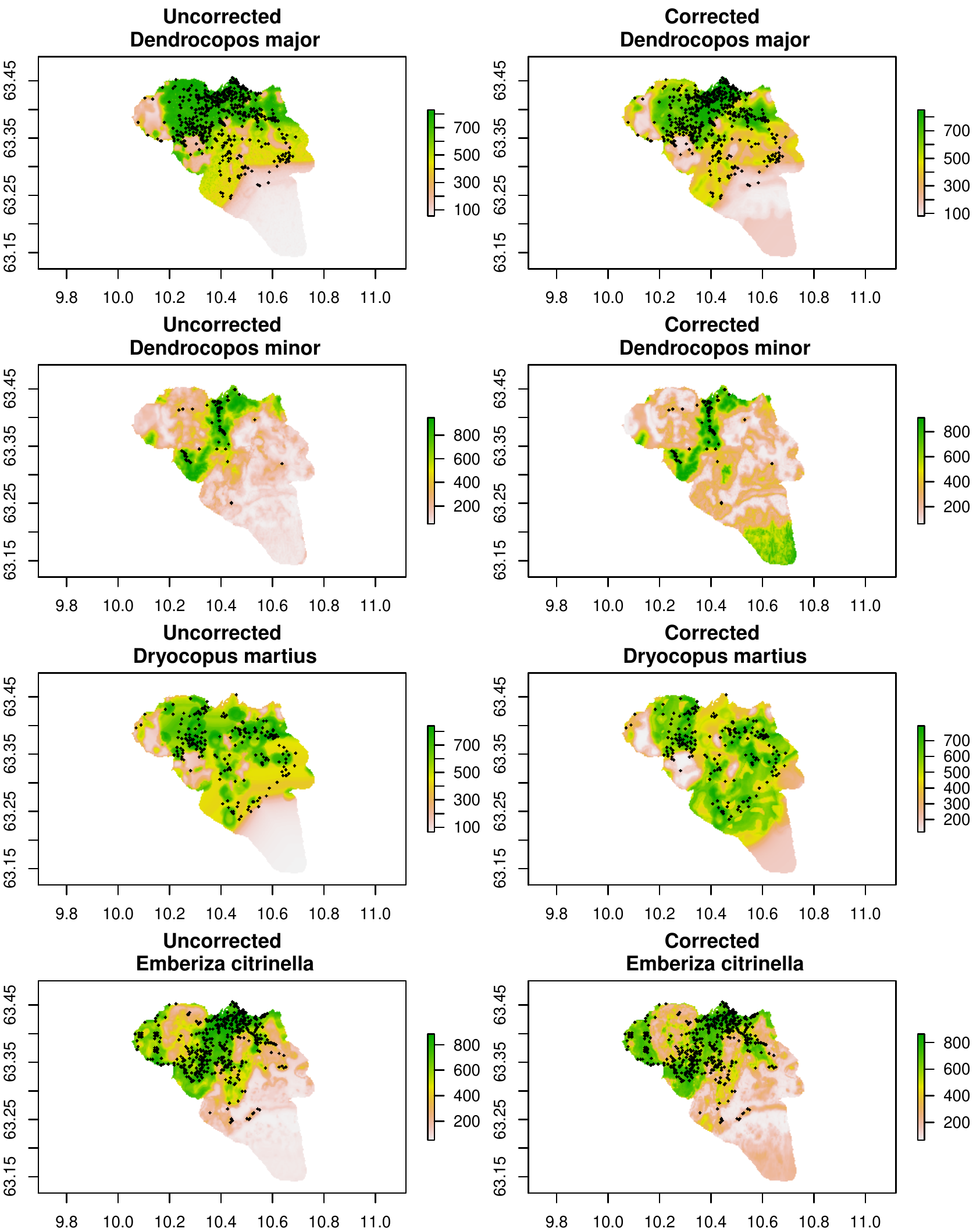}
		\caption{\sf \textbf{Species projections in Trondheim (uncorrected versus corrected group).} From top to bottom, \textit{Dendrocopos major}, \textit{Dendrocopos minor}, \textit{Dryocopus martius} and \textit{Emberiza citrinella}. 		
			\label{FigS11}}
	\end{center}
\end{figure}

\begin{figure}[!h]
	\begin{center}
		\includegraphics[width=\linewidth]{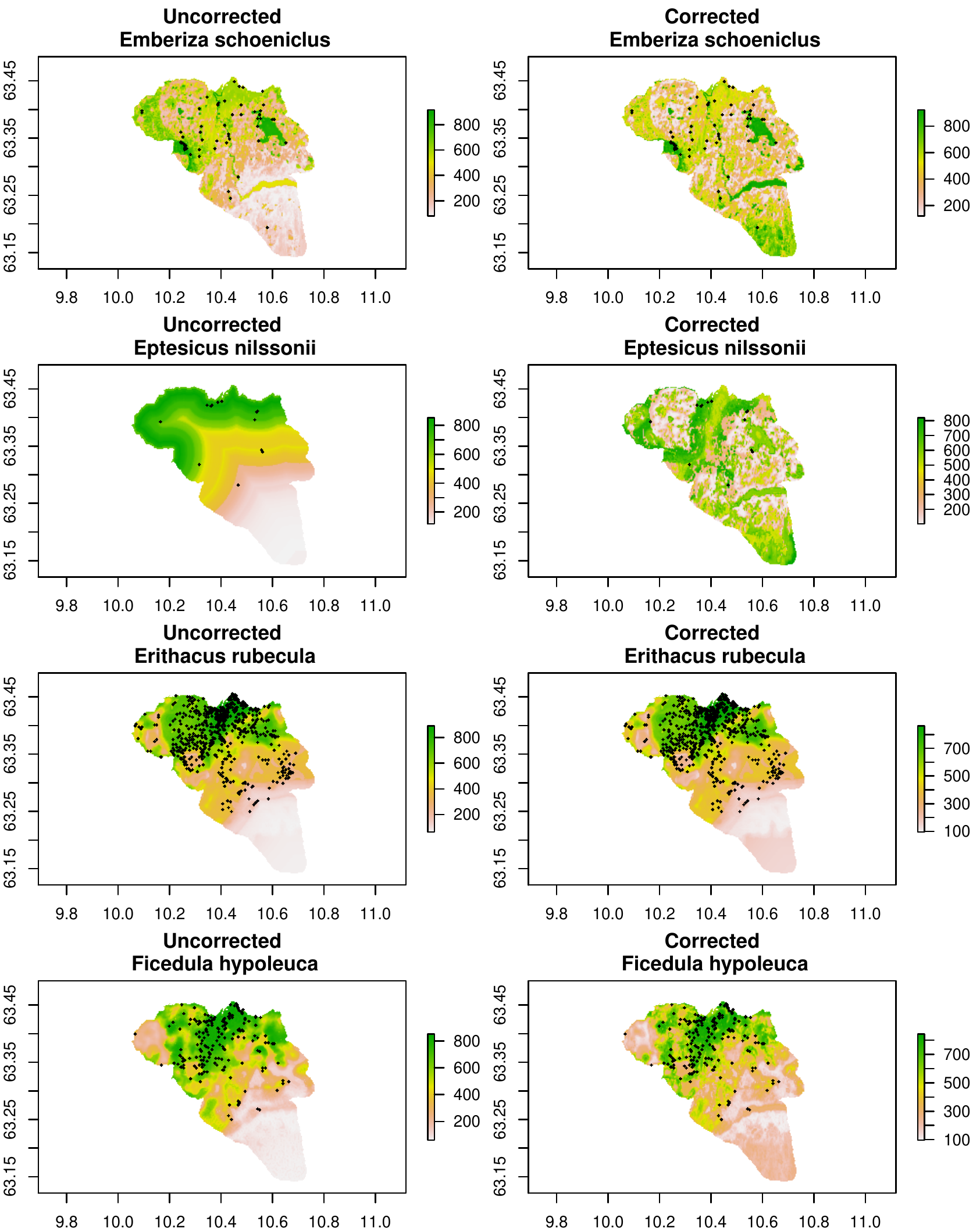}
		\caption{\sf \textbf{Species projections in Trondheim (uncorrected versus corrected group).} From top to bottom, \textit{Emberiza schoeniclus}, \textit{Eptesicus nilssonii}, \textit{Erithacus rubecula} and \textit{Ficedula hypoleuca}.		
			\label{FigS12}}
	\end{center}
\end{figure}

\begin{figure}[!h]
	\begin{center}
		\includegraphics[width=\linewidth]{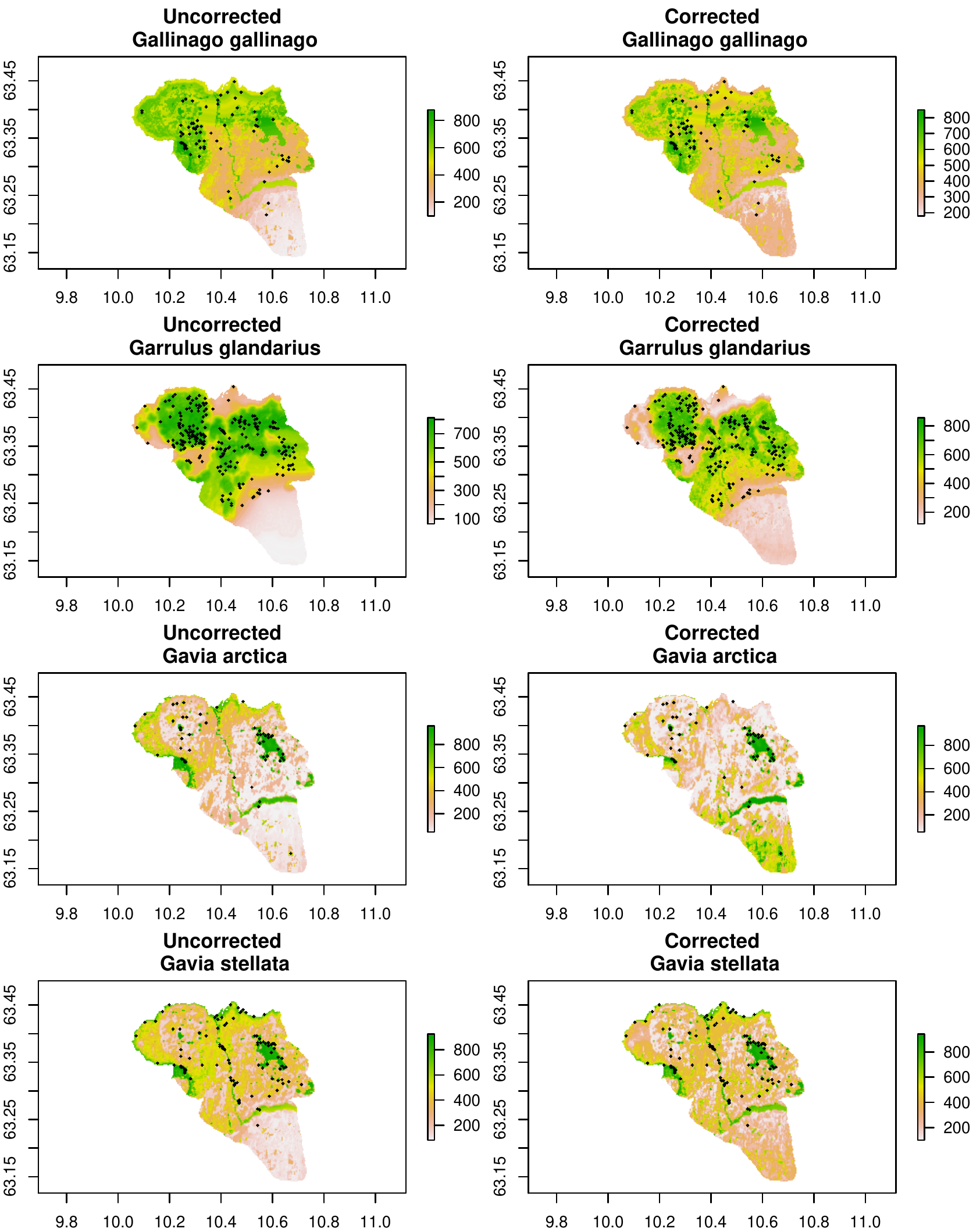}
		\caption{\sf \textbf{Species projections in Trondheim (uncorrected versus corrected group).} From top to bottom, \textit{Gallinago gallinago}, \textit{Garrulus glandarius}, \textit{Gavia arctica} and \textit{Gavia stellata}. 		
			\label{FigS13}}
	\end{center}
\end{figure}

\begin{figure}[!h]
	\begin{center}
		\includegraphics[width=\linewidth]{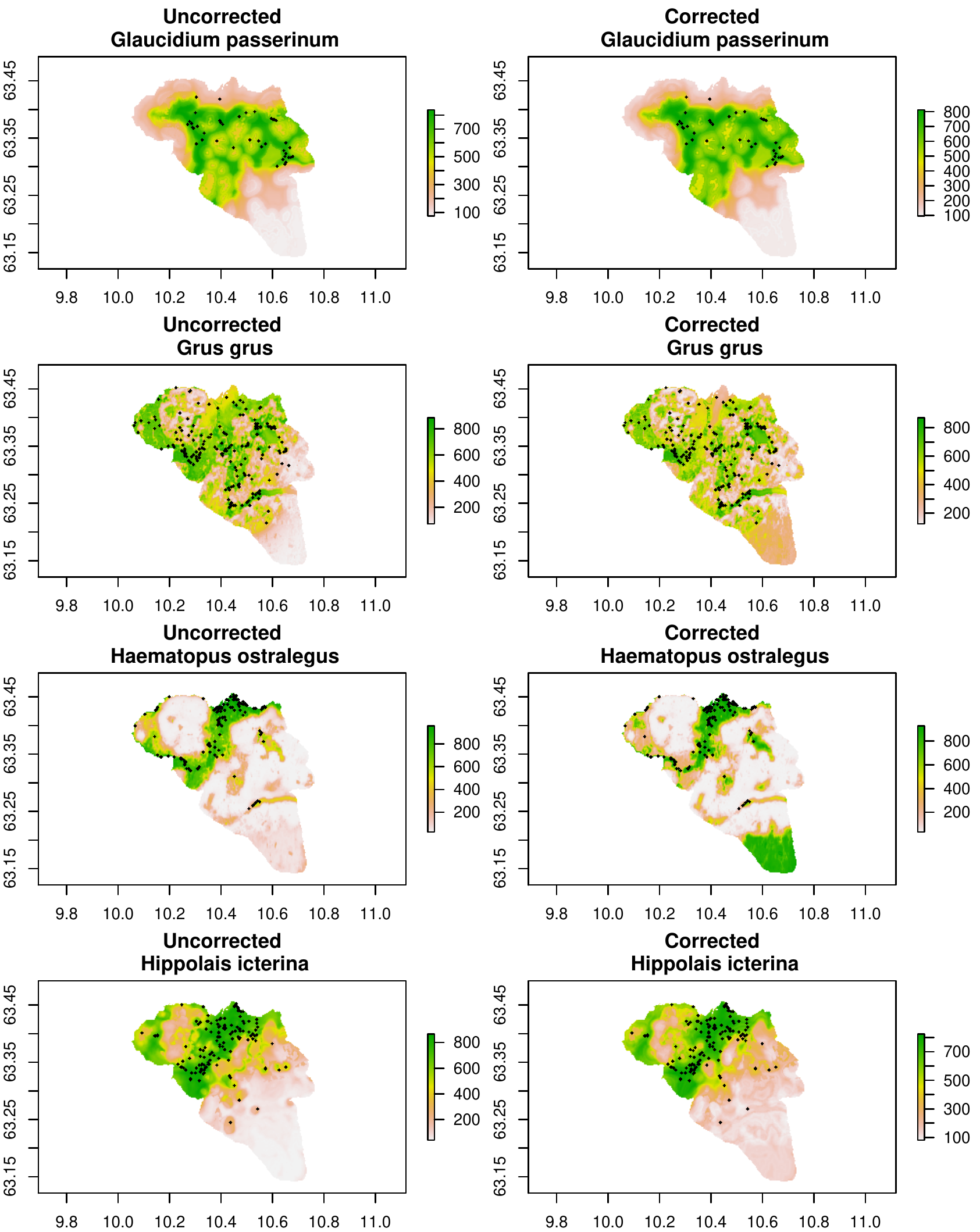}
		\caption{\sf \textbf{Species projections in Trondheim (uncorrected versus corrected group).} From top to bottom, \textit{Glaucidium passerinum}, \textit{Grus grus}, \textit{Haematopus ostralegus} and \textit{Hippolais icterina}. 	
			\label{FigS14}}
	\end{center}
\end{figure}

\begin{figure}[!h]
	\begin{center}
		\includegraphics[width=\linewidth]{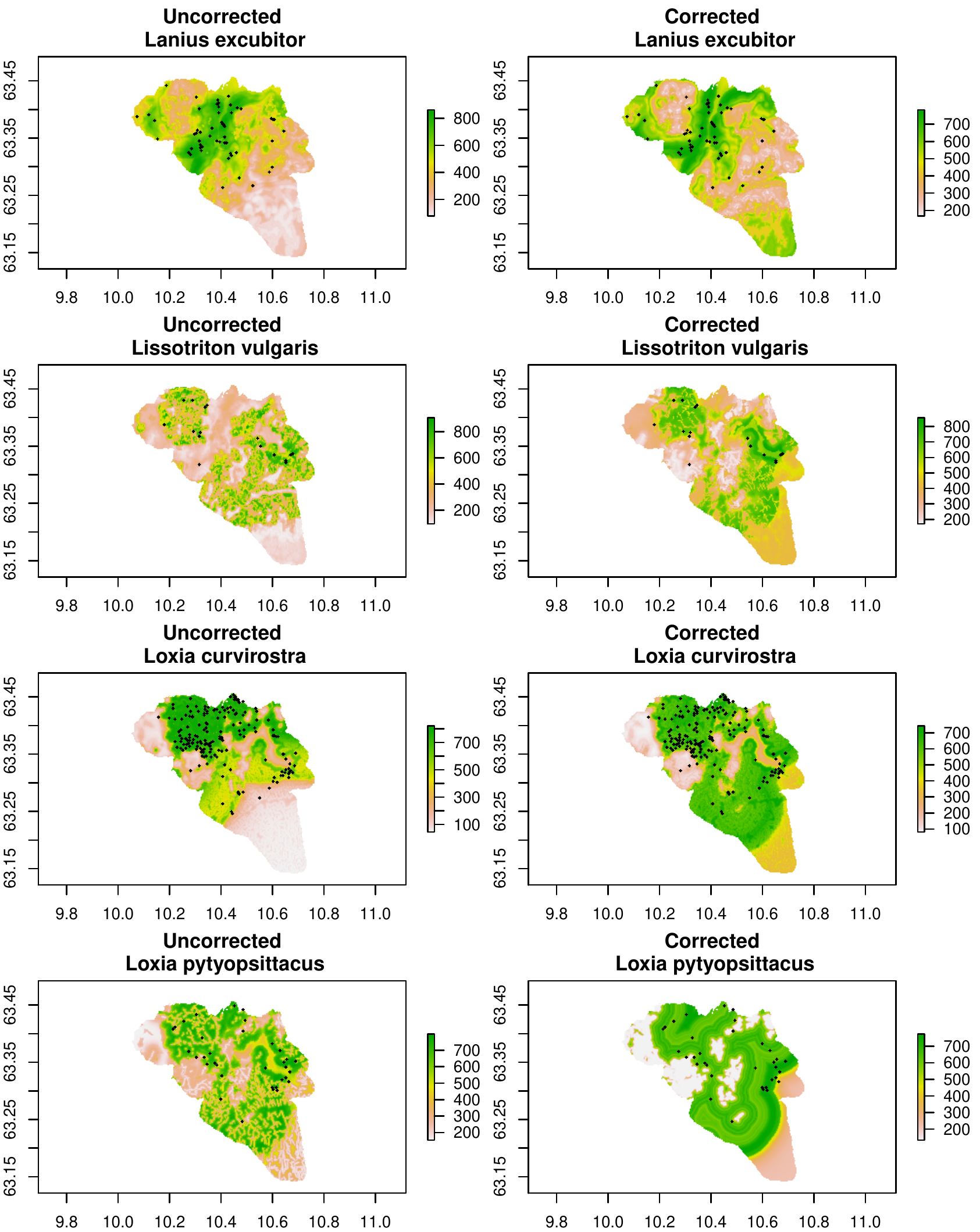}
		\caption{\sf \textbf{Species projections in Trondheim (uncorrected versus corrected group).} From top to bottom, \textit{Lanius excubitor}, \textit{Lissotriton vulgaris}, \textit{Loxia curvirostra} and \textit{Loxia pytyopsittacus}.
			\label{FigS15}}
	\end{center}
\end{figure}

\begin{figure}[!h]
	\begin{center}
		\includegraphics[width=\linewidth]{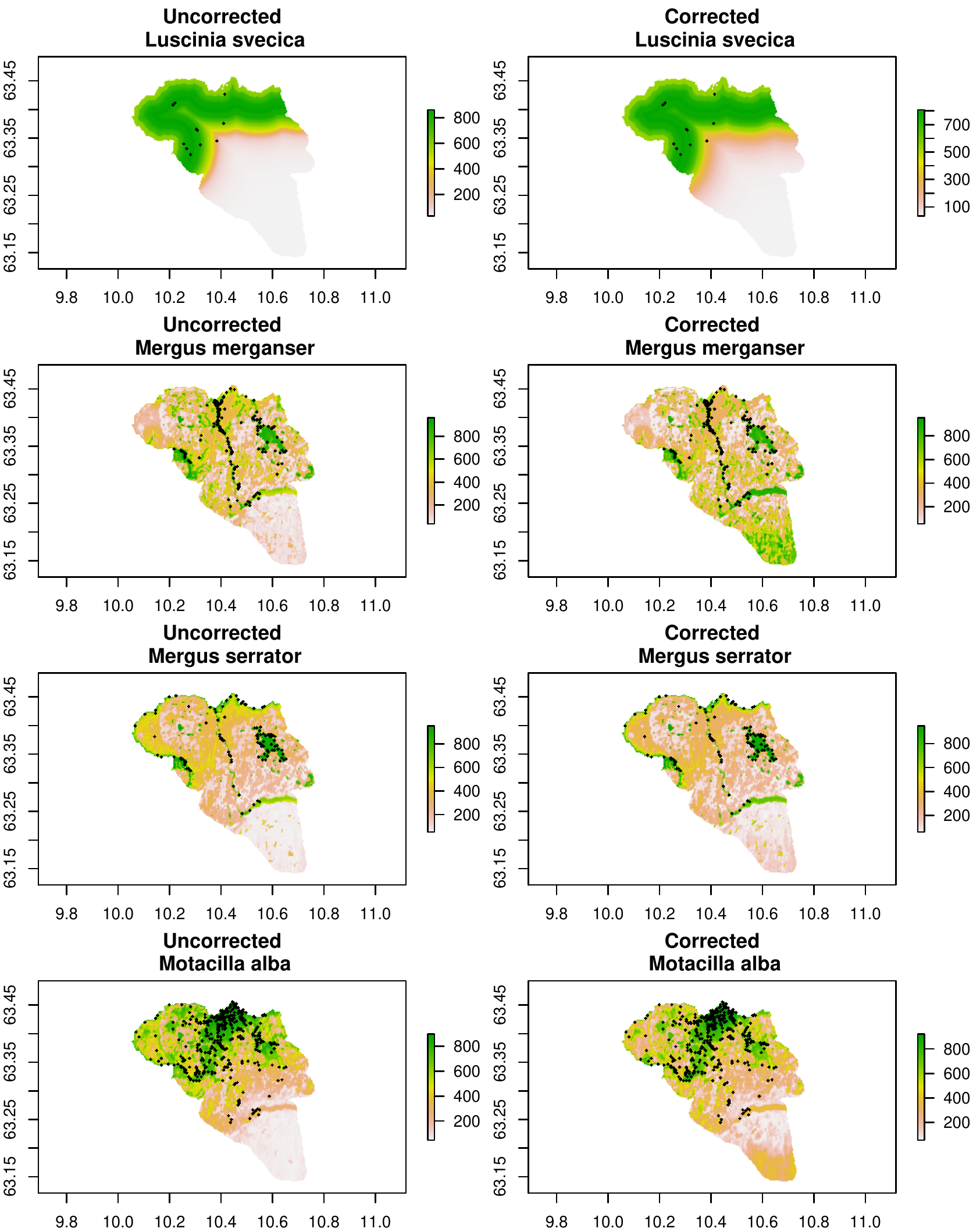}
		\caption{\sf \textbf{Species projections in Trondheim (uncorrected versus corrected group).} From top to bottom, \textit{Luscinia svecica}, \textit{Mergus merganser}, \textit{Mergus serrator} and \textit{Motacilla alba}. 
			\label{FigS16}}
	\end{center}
\end{figure}

\begin{figure}[!h]
	\begin{center}
		\includegraphics[width=\linewidth]{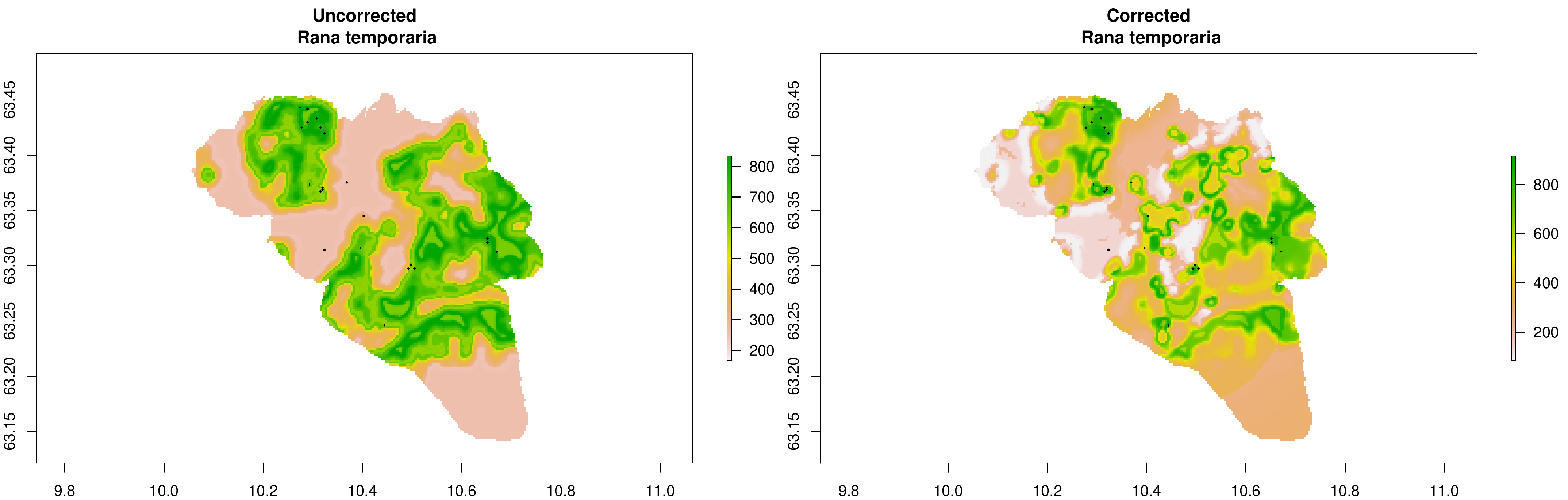}
		\caption{\sf \textbf{Species projections of \textit{Rana temporaria} in Trondheim (uncorrected versus corrected group).} 
			\label{FigS17}}
	\end{center}
\end{figure}

\begin{figure}[!h]
	\begin{center}
		\includegraphics[width=\linewidth]{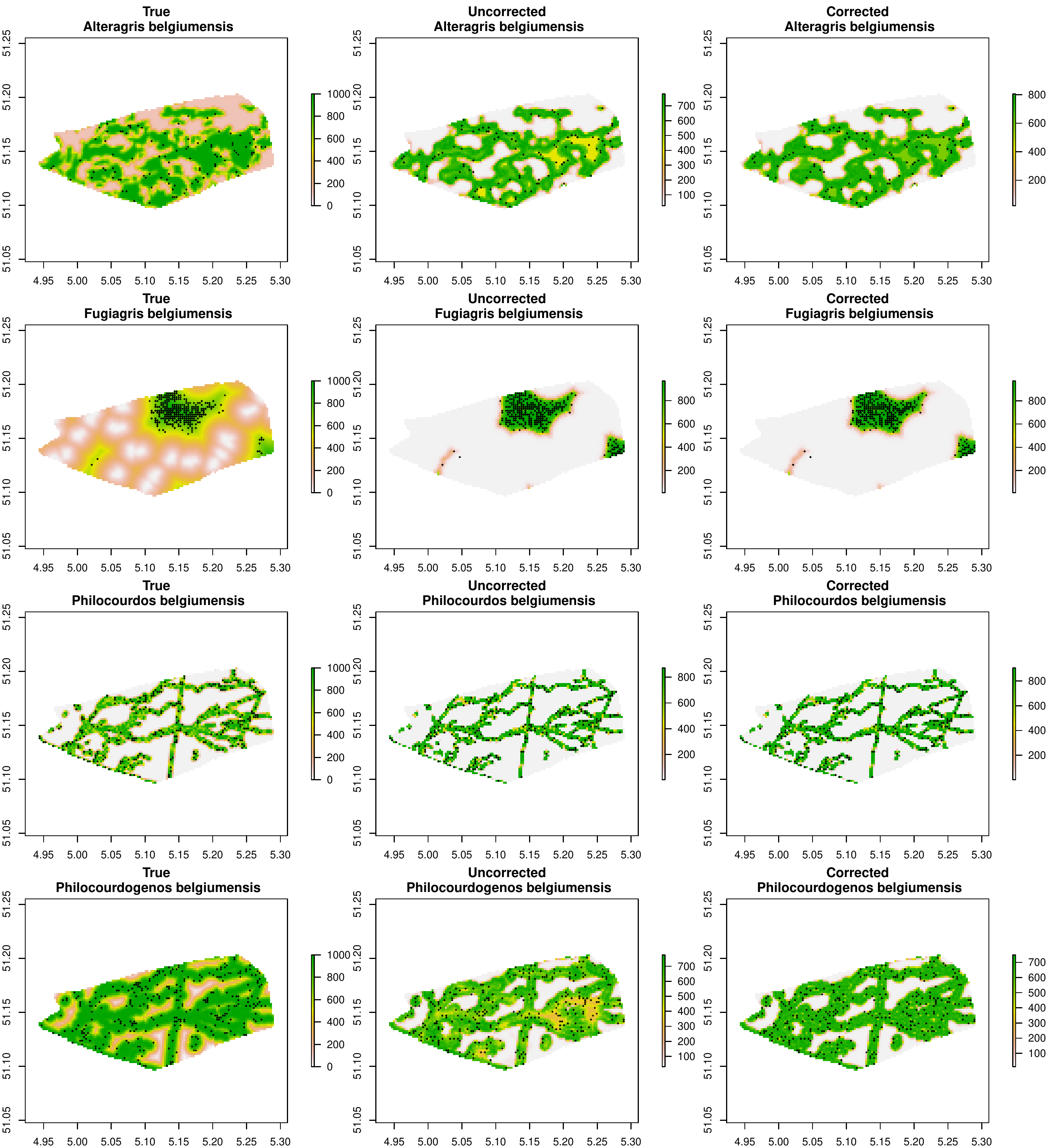}
		\caption{\sf \textbf{Virtual species projections in Grote Note (true, uncorrected and corrected group).} From top to bottom, \textit{Alteragris belgiumensis}, \textit{Fugiagris belgiumensis}, \textit{Philocourdos belgiumensis} and \textit{Philocourdogenos belgiumensis}.
			\label{FigS18}}
	\end{center}
\end{figure}

\begin{figure}[!h]
	\begin{center}
		\includegraphics[width=\linewidth]{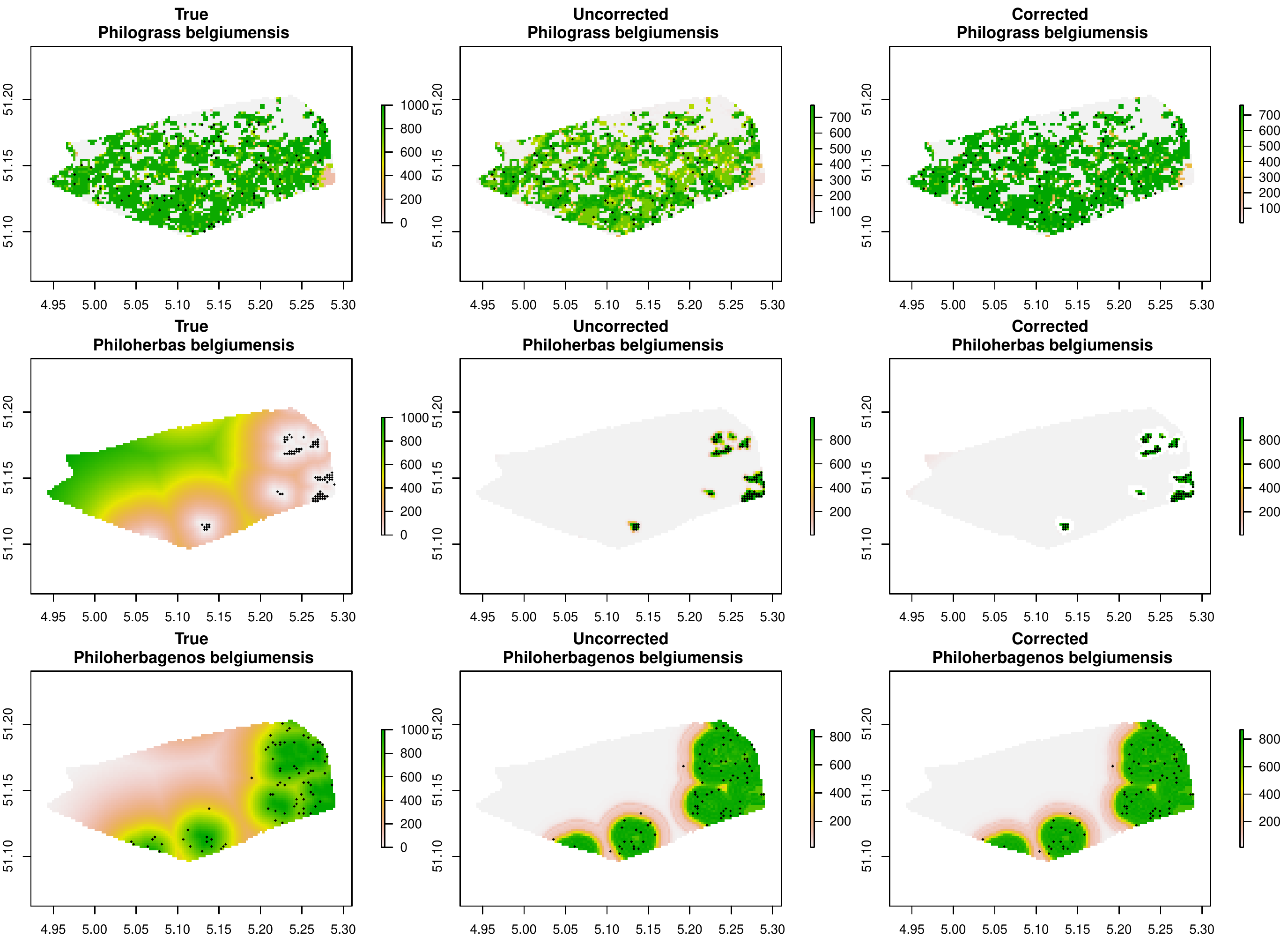}
		\caption{\sf \textbf{Virtual species projections in Grote Note (true, uncorrected and corrected group).} From top to bottom, \textit{Philograss belgiumensis}, \textit{Philoherbas belgiumensis} and \textit{Philoherbagenos belgiumensis}.
			\label{FigS19}}
	\end{center}
\end{figure}

\begin{figure}[!h]
	\begin{center}
		\includegraphics[width=\linewidth]{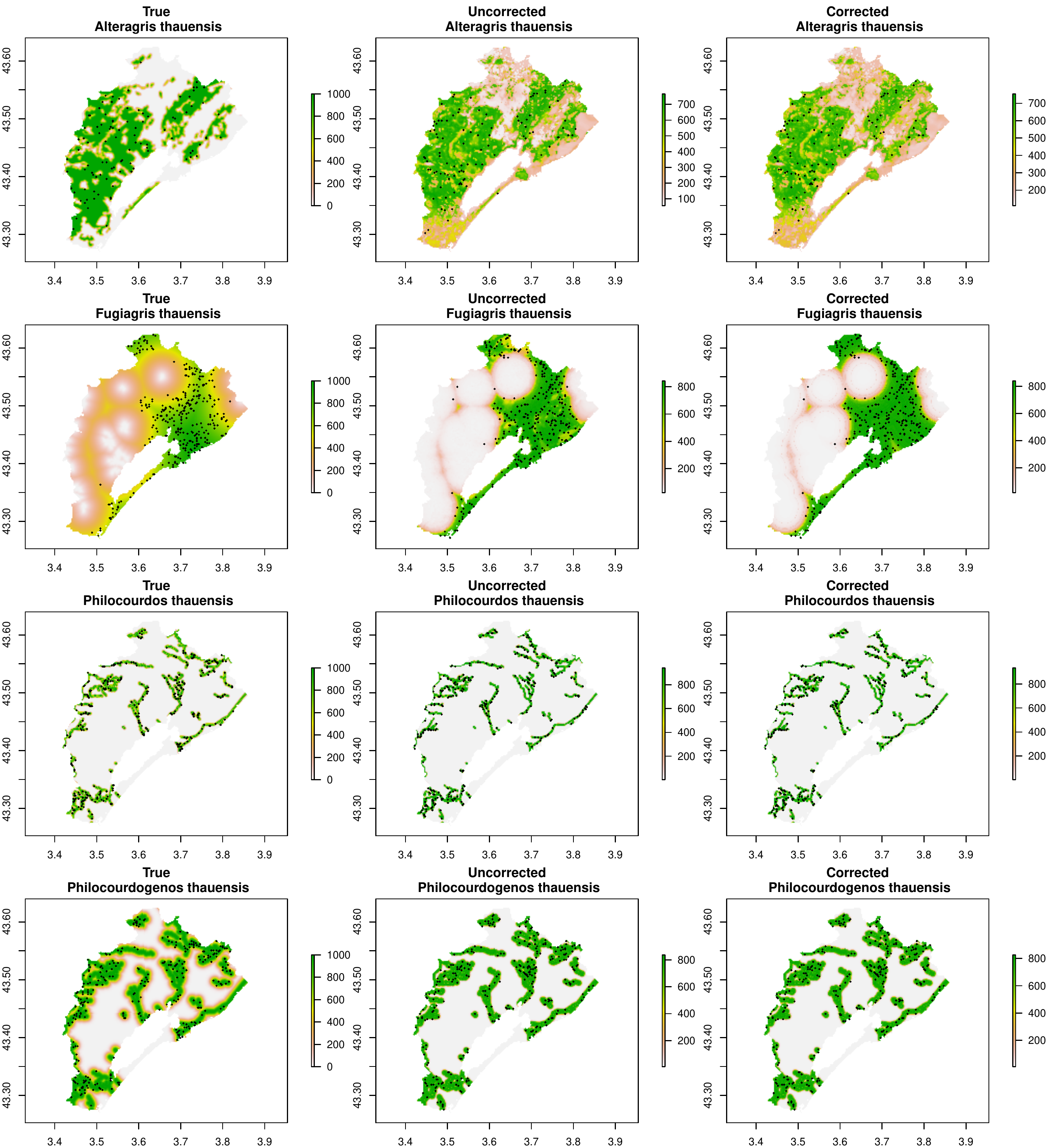}
		\caption{\sf \textbf{Virtual species projections in Thau (true, uncorrected and corrected group).} From top to bottom, \textit{Alteragris thauensis}, \textit{Fugiagris thauensis}, \textit{Philocourdos thauensis} and \textit{Philocourdogenos thauensis}.
			\label{FigS20}}
	\end{center}
\end{figure}

\begin{figure}[!h]
	\begin{center}
		\includegraphics[width=\linewidth]{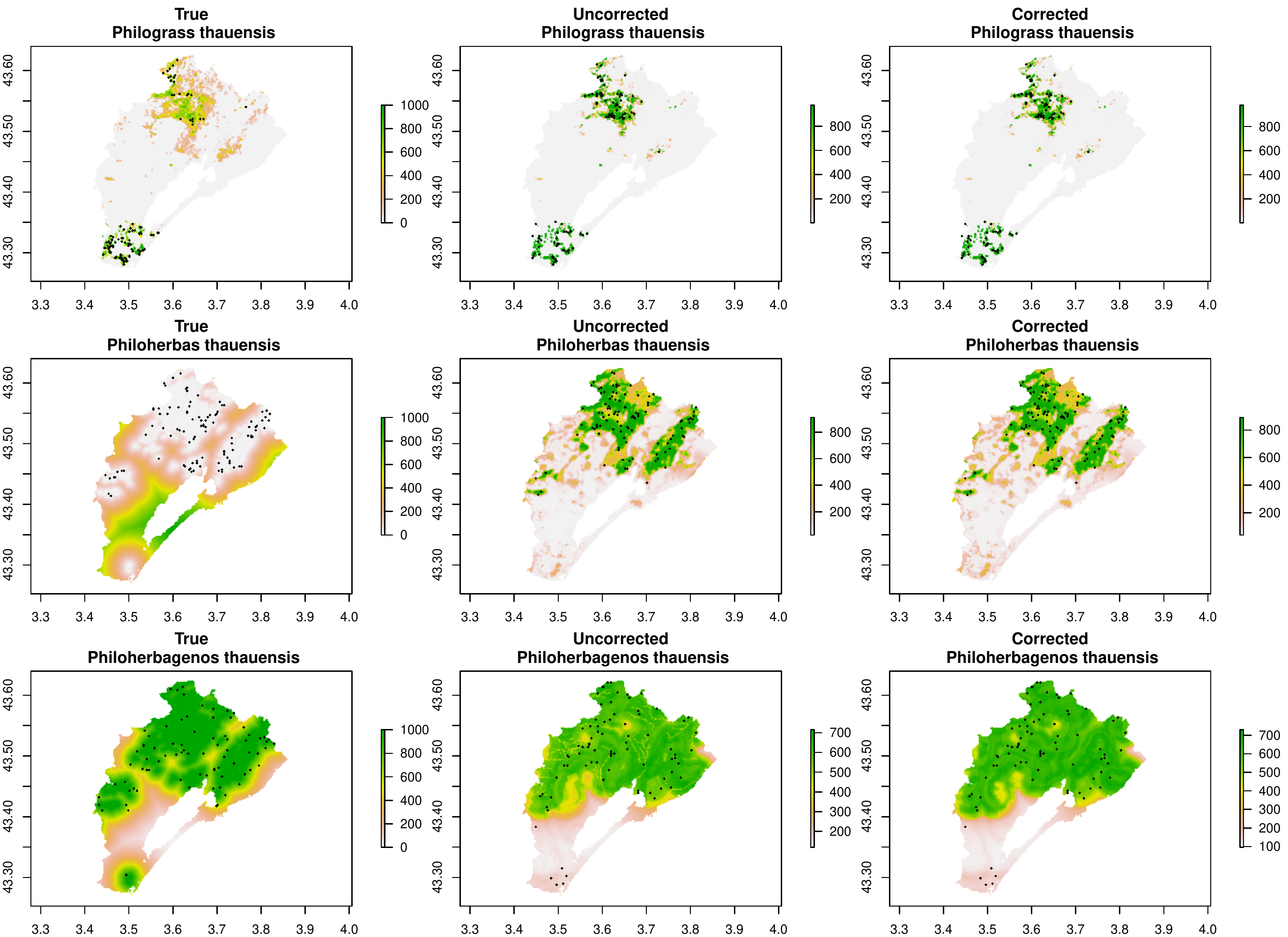}
		\caption{\sf \textbf{Virtual species projections in Thau (true, uncorrected and corrected group).} From top to bottom, \textit{Philograss thauensis}, \textit{Philoherbas thauensis} and \textit{Philoherbagenos thauensis}.
			\label{FigS21}}
	\end{center}
\end{figure}

\begin{figure}[!h]
	\begin{center}
		\includegraphics[width=\linewidth]{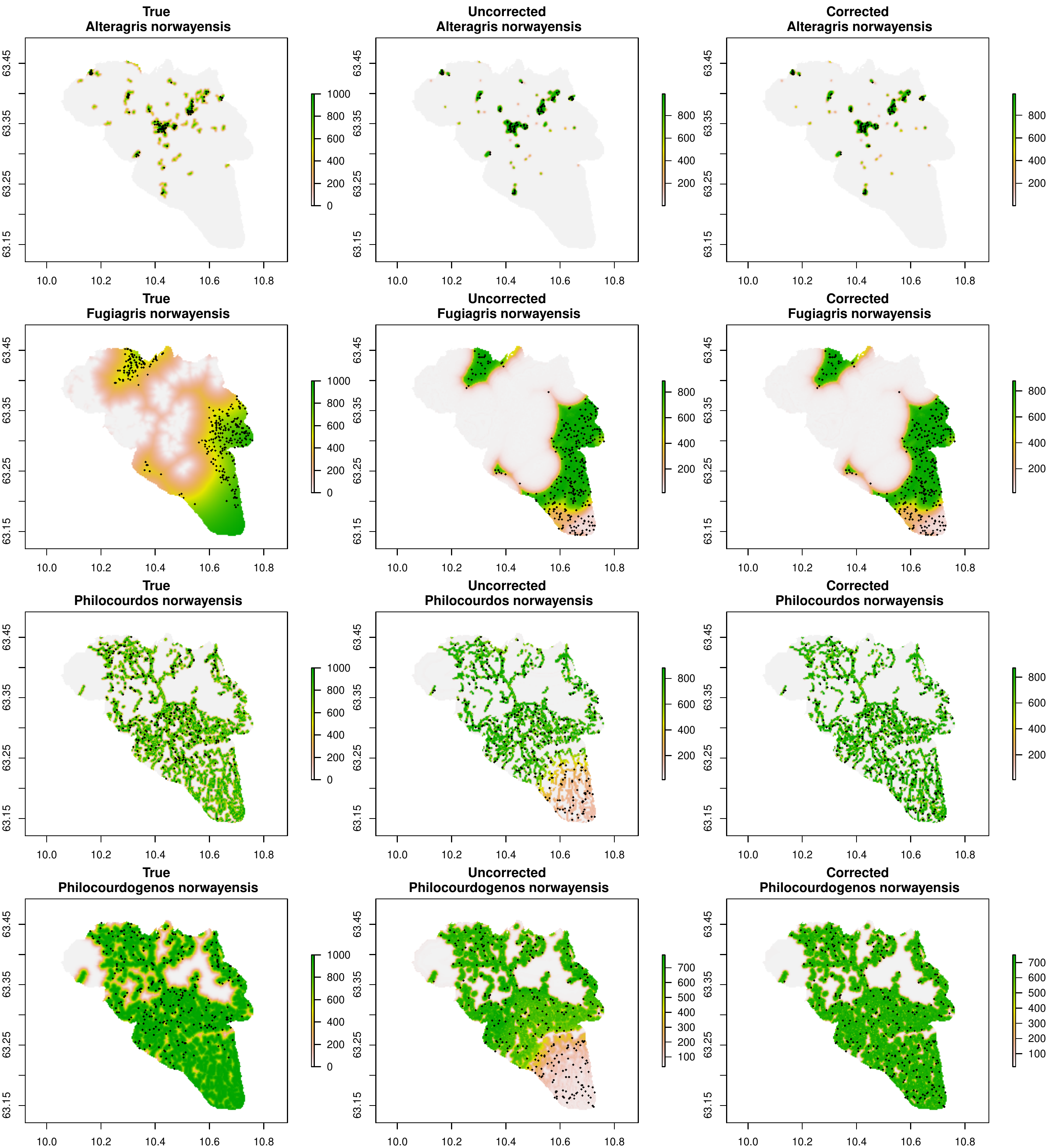}
		\caption{\sf \textbf{Virtual species projections in Trondheim (true, uncorrected and corrected group).} From top to bottom, \textit{Alteragris norwayensis}, \textit{Fugiagris norwayensis}, \textit{Philocourdos norwayensis} and \textit{Philocourdogenos norwayensis}
			\label{FigS22}}
	\end{center}
\end{figure}

\begin{figure}[!h]
	\begin{center}
		\includegraphics[width=\linewidth]{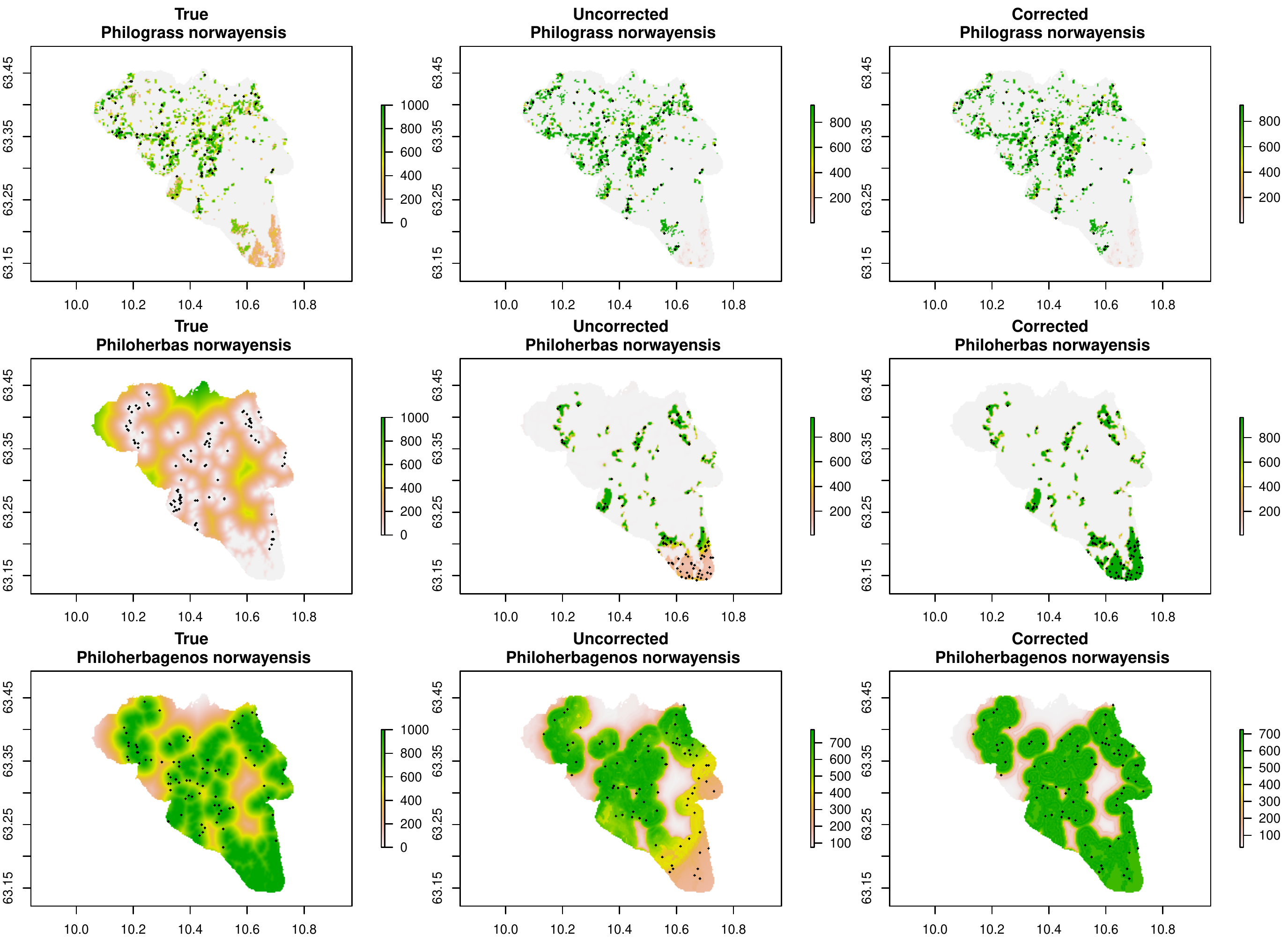}
		\caption{\sf \textbf{Virtual species projections in Trondheim (true, uncorrected and corrected group).} From top to bottom, \textit{Philograss norwayensis}, \textit{Philoherbas norwayensis} and \textit{Philoherbagenos norwayensis}.
			\label{FigS23}}
	\end{center}
\end{figure}

\newpage
\clearpage
\newpage
\subsection*{Factors influencing the effect of correction}

We investigated whether the effect of correction differed between sites and according to the sample bias,inferred from Boyce indices computed with species occurrences and accessibility maps, and whether the effect differed depending upon sample size (number of occurrence points after data thinning/resampling at the resolution of our environmental variables). The latter two factors were slightly correlated (Figure S24).

\begin{figure}[!h]
	\begin{center}
		\includegraphics[width=13cm]{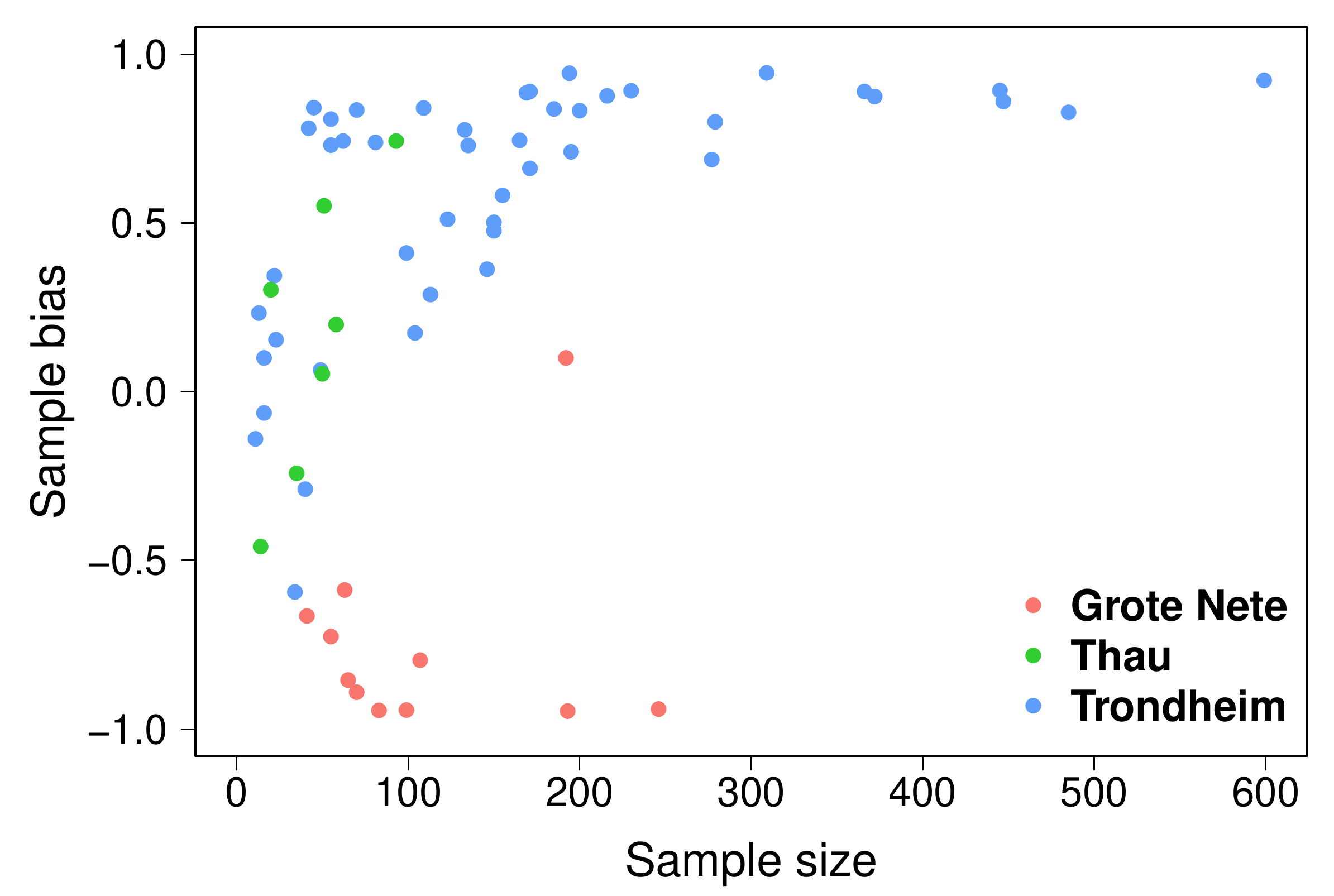}
		\caption{\sf \textbf{Relationship between the sample size and the sample bias according to the case study site.} \label{FigS24}}
	\end{center}
\end{figure}

As observed in Figure S25 the Relative Overlap Index (ROI) tends to globally decrease with sample size and to increase with sample bias. There is an overrepresentation of high and negative ROI values for the small sample size (fewer than 100). The rare species with a high sample size (greater than 300) tend to be characterized by low ROI values (lower than 0.15). It is important to note that, for species with a sample size greater than 100, the difference between the mean Schoener's D overlap between corrected and uncorrected groups and the mean Schoener's D overlap between corrected model runs is almost always significant (Figure S26). Regarding the sample bias, the ROI tends to be higher for species with a positive sample bias than for species with a negative sample bias.

\begin{figure}[!h]
	\begin{center}
		\includegraphics[width=\linewidth]{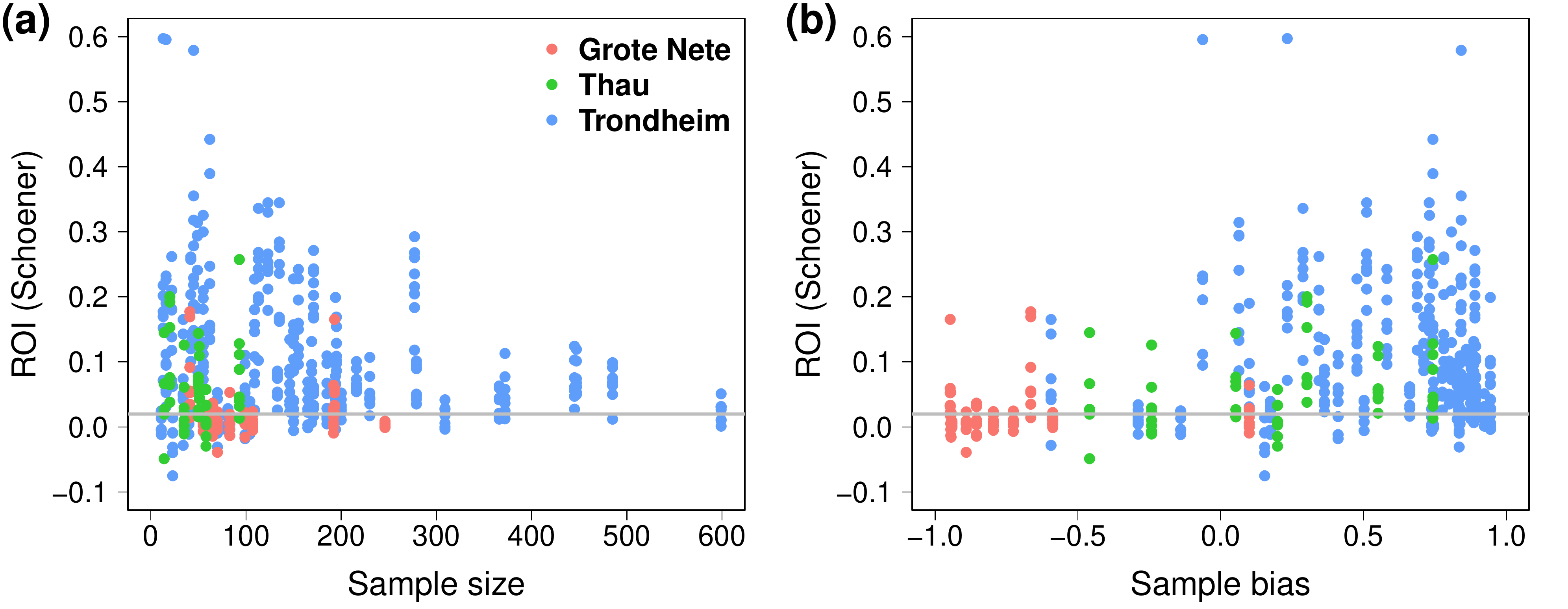}
		\caption{\sf \textbf{Evolution of the Relative Overlap Index (ROI) as a function of (a) the sample size (after thinning/resampling) and (b) the sample bias according to the study site for 64 species and for each modelling technique.} The horizontal grey line represents the ROI threshold value 0.02 associated with the one-sided Student t-test significance threshold 0.05 (see Figure \ref{FigS26} in Appendix for more details). \label{FigS25}}
	\end{center}
\end{figure}

\begin{figure}[!h]
	\begin{center}
		\includegraphics[width=\linewidth]{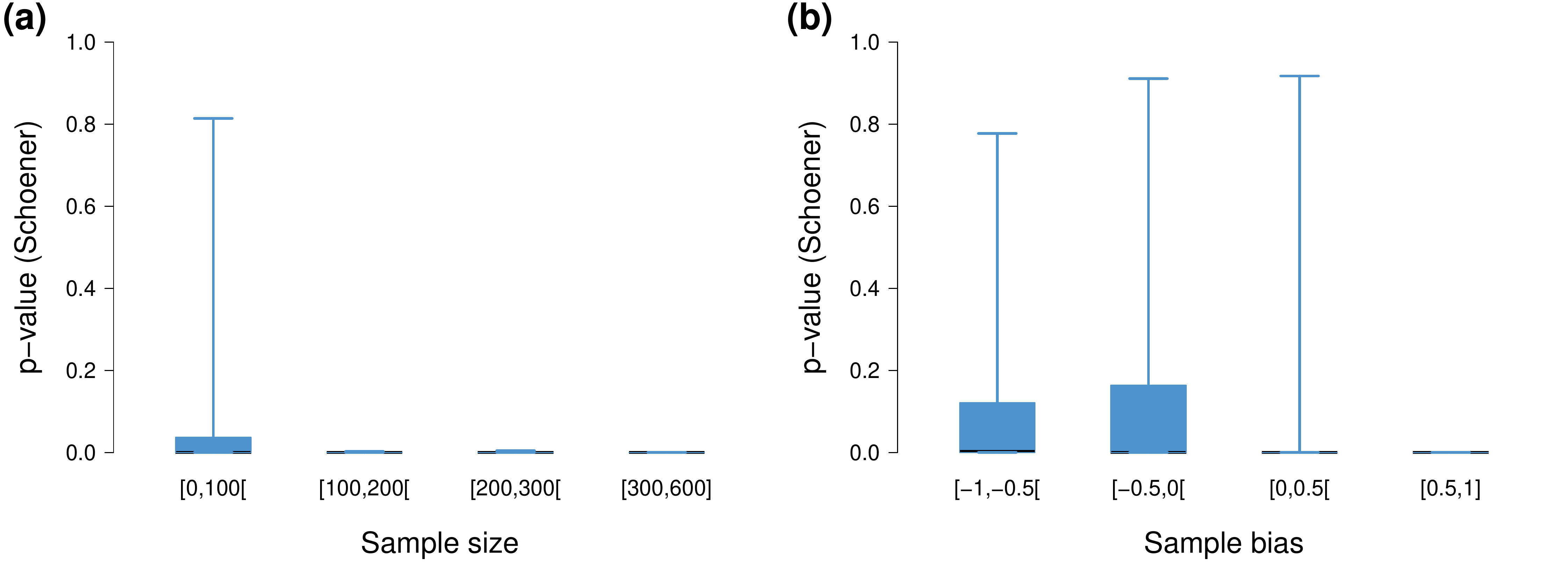}
		\caption{\sf \textbf{Relationship between sample size, sample bias and t-test p-value.} (a) Boxplots of the t-test p-value (difference between mean Schoener'D overlap between corrected and uncorrected groups and mean Schoener's D overlap between corrected model runs) according to the sample size (divided into four categories). (b) Boxplots of the t-test p-value (difference between mean Schoener's D overlap between corrected and uncorrected and mean Schoener's D overlap between corrected model runs) according to the sample bias (divided into four categories). Each boxplot is composed of the first decile, the first quartile, the median, the third quartile and the ninth decile.\label{FigS26}}
	\end{center}
\end{figure}

For a given species, we also investigated the influence of the modelling techniques on the effect of correction. For each species, we divided the modelling techniques into two groups: the techniques showing than $\bar{D}_{0}$ is significantly greater than $\bar{D}$ for the Schoener's D, (p-value lower than 0.05) and that with a p-value higher than 0.05. This allowed us to compute the fraction of modelling techniques in the majority groups for a given species, and the fraction of species for which a modelling technique was in the majority group. In Figure S27a, which plots the fraction of modelling techniques in the majority group, we observe that most of the modelling techniques are in agreement in Thau and Trondheim (100\% of the techniques are in agreement for half of the species). This is not the case for Grote Nete, where for 75\% of the species, less than 70\% of the modelling techniques are in agreement. Figure S27b shows the fraction of species for which a modelling technique was in the majority group. We see that the generalised boosting model (GBM) was in the majority group for all species in Thau and Trondheim. It is important to note that some predictions failed for the GBM and were therefore excluded from the computation of the overlap metrics (as shown in Table S2 and S3 in Appendix). The surface range envelope (SRE), random forest (RF), and generalised linear models (GLM) were in the majority group for 75\% of the species across the three study sites. The four remaining modelling techniques (ANN, GAM, FDA and CTA) were in the majority group for 50\% of the species across the three study sites.

\begin{figure}[!h]
	\begin{center}
		\includegraphics[width=\linewidth]{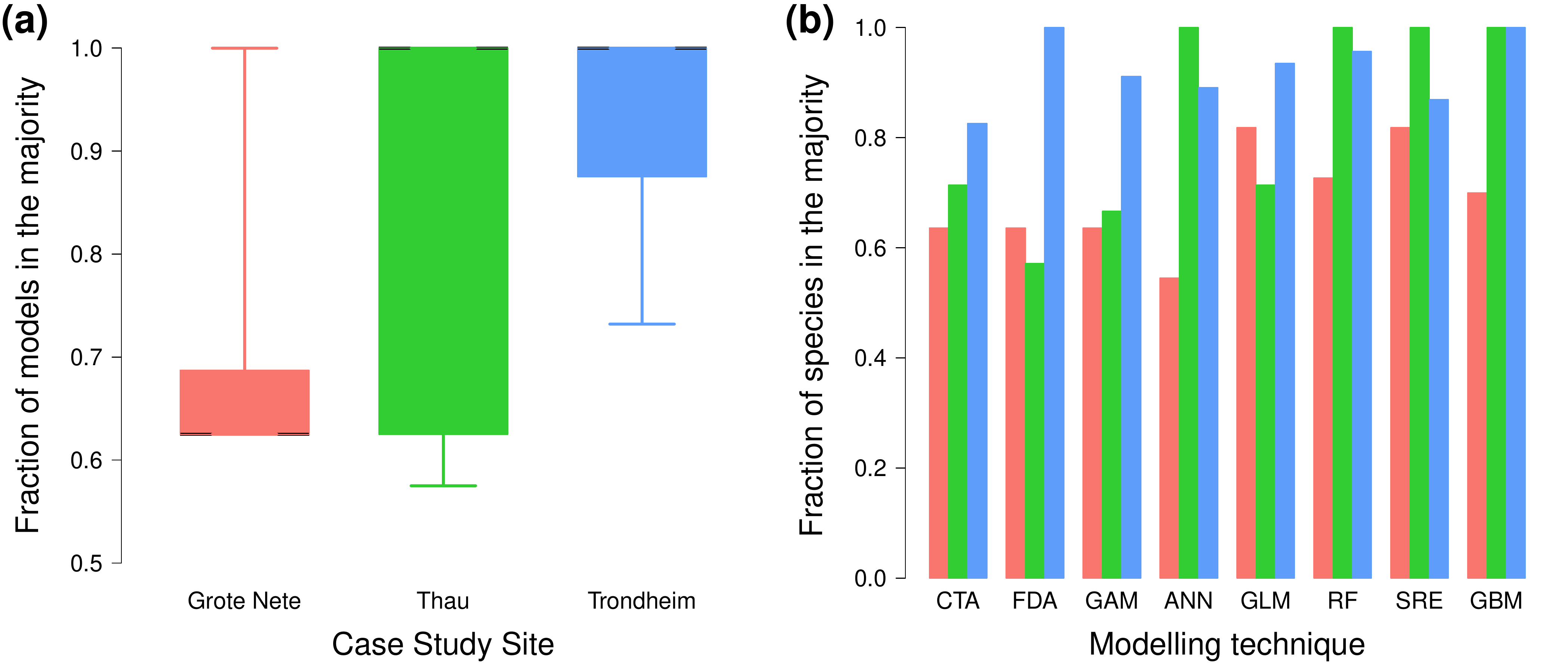}
		\caption{\sf \textbf{Variation of the effect of correction across modelling techniques.} (a) Boxplots of the fraction of modelling techniques in the majority group for each species according to the case study site. For each species, this ratio is based on the number of modelling techniques exhibiting a p-value (Schoener's D) lower than 0.05 (if the majority of modelling techniques exhibits a p-value lower than 0.05) or higher than 0.05 (if the majority of modelling techniques exhibits a p-value higher 0.05). (b) Barplots of the fraction of species for which a modelling technique was in the majority group (p-value higher or lower than 0.05 as the case may be) according to the case study site. \label{FigS27}}
	\end{center}
\end{figure}

\newpage
\clearpage
\newpage
\subsection*{Supplementary Figures}

\begin{figure}[!h]
	\begin{center}
		\includegraphics[width=\linewidth]{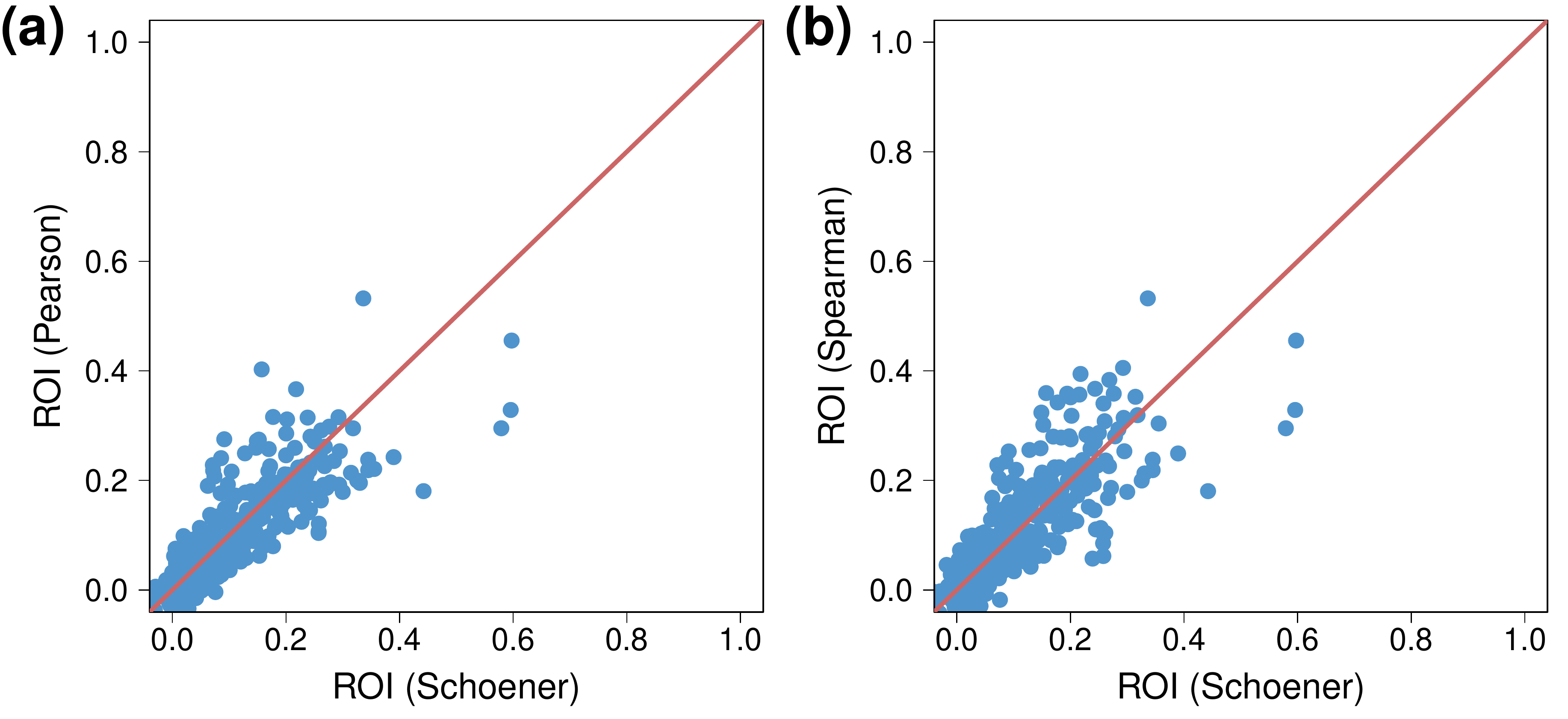}
		\caption{\sf \textbf{Relationship between the ROI obtained with Schoener'D and the ones obtained with Pearson's and Spearman's coefficients.} \label{FigS28}}
	\end{center}
\end{figure}

\begin{figure}[!h]
	\begin{center}
		\includegraphics[width=11cm]{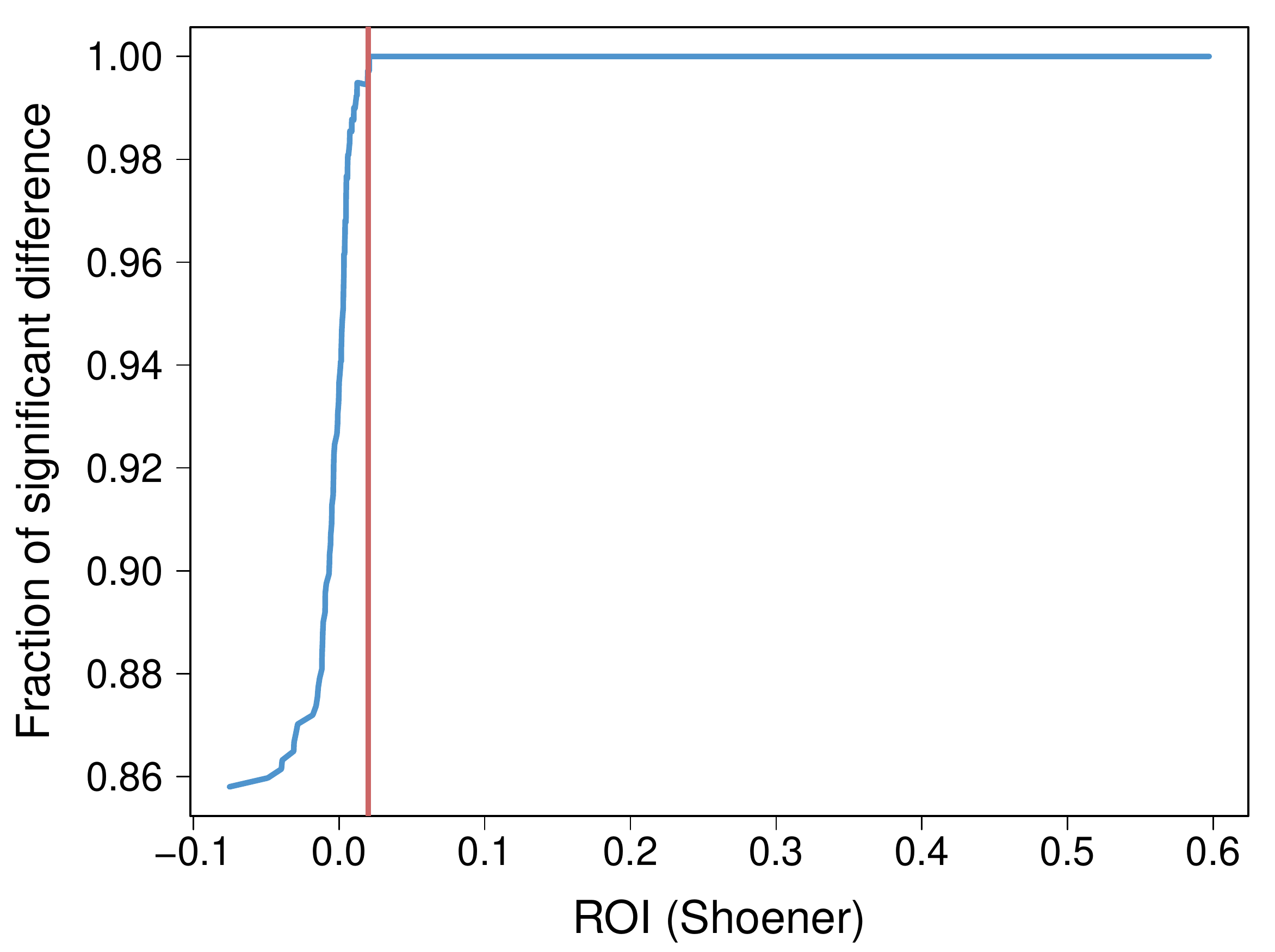}
		\caption{\sf \textbf{Relationship between ROI and t-test p-value.} Fraction of t-test p-value (difference between mean Schoener'D overlap between corrected and uncorrected and mean Schoener'D overlap between corrected model runs) lower than 0.05 as a function of the ROI obtained with the Schoener's D. The vertical grey line represents the ROI threshold 0.02. This means that all the p-value associated with ROI strictly higher than 0.02 are strictly lower than 0.05.\label{FigS29}}
	\end{center}
\end{figure}

\begin{figure}[!h]
	\begin{center}
		\includegraphics[width=\linewidth]{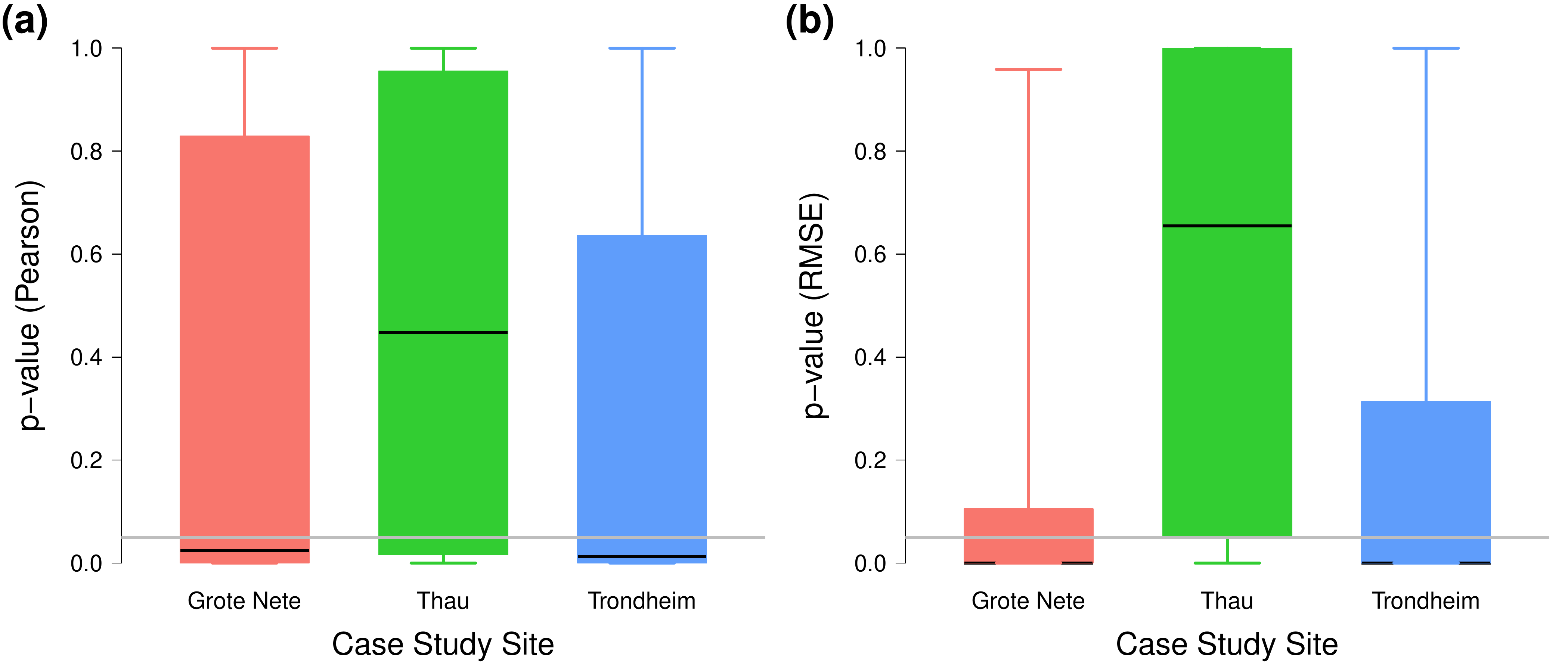}
		\caption{\sf \textbf{Comparison with the \enquote{true} probability of occurrence (virtual species) for the uncorrected and the corrected groups at three sites.} (a) Boxplots of the p-value obtained with a one-sided Student t-test to evaluate whether or not the Pearson's correlation coefficient with the \enquote{true} distribution is significantly greater for the corrected group than the uncorrected group for a given species and modelling technique. (b) Boxplots of the p-value obtained with a one-sided Student t-test to evaluate whether or not the RMSE with the \enquote{true} distribution is significantly lower for the corrected group than the uncorrected group for a given species and modelling technique.\label{FigS30}}
	\end{center}
\end{figure}

\newpage
\clearpage
\newpage
\subsection*{Supplementary Tables}

\vspace*{-0.5cm}
\begin{table}[!h]
	\fontsize{6}{6}\selectfont
	\caption{\textbf{Summary information about the 64 real species (species name, site, sample size (after thinning/resampling) and sample bias).} Sample bias was estimated from Boyce indices based on occurrence points and accessibility maps.}
	\label{TabS1}
	\centering
	\vspace*{0.25cm}
	\begin{tabular}{llrr}
		\hline
		\textbf{Species} & \textbf{Site} & \textbf{Sample size} & \textbf{Sample bias} \\ 
		\hline
		
		Alcedo atthis & Grote Nete & 193 & -0.95\\
		Anthus trivialis & Grote Nete & 99 & -0.94\\
		Castor fiber & Grote Nete & 41 & -0.66\\
		Dryocopus martius & Grote Nete & 246 & -0.94\\
		Lullula arborea & Grote Nete & 83 & -0.94\\
		Luscinia megarhynchos & Grote Nete & 70 & -0.89\\
		Luscinia svecica & Grote Nete & 55 & -0.73\\
		Oriolus oriolus & Grote Nete & 65 & -0.86\\
		Phoenicurus phoenicurus & Grote Nete & 63 & -0.59\\
		Poecile montanus & Grote Nete & 107 & -0.8\\
		Sciurus vulgaris & Grote Nete & 192 & 0.1\\
		Epidalea calamita & Thau & 58 & 0.2\\
		Natrix natrix & Thau & 14 & -0.46\\
		Plecotus austriacus & Thau & 50 & 0.05\\
		Rhinolophus ferrumequinum & Thau & 35 & -0.24\\
		Tarentola mauritanica & Thau & 93 & 0.74\\
		Timon lepidus & Thau & 51 & 0.55\\
		Triturus marmoratus & Thau & 20 & 0.3\\
		Accipiter gentilis & Trondheim & 200 & 0.83\\
		Accipiter nisus & Trondheim & 309 & 0.94\\
		Actitis hypoleucos & Trondheim & 165 & 0.74\\
		Aegithalos caudatus & Trondheim & 146 & 0.36\\
		Alauda arvensis & Trondheim & 42 & 0.78\\
		Anas penelope & Trondheim & 81 & 0.74\\
		Anthus pratensis & Trondheim & 169 & 0.89\\
		Anthus trivialis & Trondheim & 150 & 0.48\\
		Asio flammeus & Trondheim & 16 & 0.1\\
		Aythya fuligula & Trondheim & 113 & 0.29\\
		Bombycilla garrulus & Trondheim & 445 & 0.89\\
		Bucephala clangula & Trondheim & 277 & 0.69\\
		Calidris pugnax & Trondheim & 23 & 0.15\\
		Carduelis flammea & Trondheim & 366 & 0.89\\
		Certhia familiaris & Trondheim & 216 & 0.88\\
		Charadrius hiaticula & Trondheim & 45 & 0.84\\
		Cinclus cinclus & Trondheim & 135 & 0.73\\
		Coccothraustes coccothraustes & Trondheim & 150 & 0.5\\
		Cygnus cygnus & Trondheim & 194 & 0.94\\
		Delichon urbicum & Trondheim & 109 & 0.84\\
		Dendrocopos major & Trondheim & 485 & 0.83\\
		Dendrocopos minor & Trondheim & 62 & 0.74\\
		Dryocopus martius & Trondheim & 195 & 0.71\\
		Emberiza citrinella & Trondheim & 447 & 0.86\\
		Emberiza schoeniclus & Trondheim & 55 & 0.73\\
		Eptesicus nilssonii & Trondheim & 13 & 0.23\\
		Erithacus rubecula & Trondheim & 599 & 0.92\\
		Ficedula hypoleuca & Trondheim & 185 & 0.84\\
		Gallinago gallinago & Trondheim & 70 & 0.84\\
		Garrulus glandarius & Trondheim & 230 & 0.89\\
		Gavia arctica & Trondheim & 49 & 0.06\\
		Gavia stellata & Trondheim & 99 & 0.41\\
		Glaucidium passerinum & Trondheim & 40 & -0.29\\
		Grus grus & Trondheim & 171 & 0.66\\
		Haematopus ostralegus & Trondheim & 123 & 0.51\\
		Hippolais icterina & Trondheim & 133 & 0.78\\
		Hirundo rustica & Trondheim & 279 & 0.8\\
		Lanius excubitor & Trondheim & 55 & 0.81\\
		Lissotriton vulgaris & Trondheim & 16 & -0.06\\
		Loxia curvirostra & Trondheim & 171 & 0.89\\
		Loxia pytyopsittacus & Trondheim & 34 & -0.59\\
		Luscinia svecica & Trondheim & 11 & -0.14\\
		Mergus merganser & Trondheim & 155 & 0.58\\
		Mergus serrator & Trondheim & 104 & 0.17\\
		Motacilla alba & Trondheim & 372 & 0.88\\
		Rana temporaria & Trondheim & 22 & 0.34\\
		
		\hline
	\end{tabular}
\end{table}

\begin{table}[!h]
	\fontsize{8}{8}\selectfont
	\caption{\textbf{Percentage of failed predictions by real species according to the model for the uncorrected group.}}
	\label{TabS2}
	\centering
	\vspace*{0.25cm}
	\begin{tabular}{lcccccccc}
		\hline
		\textbf{Species (Site)} & \textbf{ANN} & \textbf{CTA} & \textbf{FDA} & \textbf{GAM} & \textbf{GBM} & \textbf{GLM} & \textbf{RF} & \textbf{SRE} \\ 
		\hline
		
		Accipiter gentilis (Trondheim) & 0 & 0 & 0 & 0 & 0 & 0 & 0 & 0\\
		Accipiter nisus (Trondheim) & 0 & 0 & 0 & 0 & 0 & 0 & 0 & 0\\
		Actitis hypoleucos (Trondheim) & 0 & 0 & 0 & 0 & 0 & 0 & 0 & 0\\
		Aegithalos caudatus (Trondheim) & 0 & 0 & 0 & 0 & 0 & 0 & 0 & 0\\
		Alauda arvensis (Trondheim) & 0 & 0 & 0 & 0 & 0 & 0 & 0 & 0\\
		Alcedo atthis (Grote Nete) & 0 & 0 & 0 & 3 & 0 & 0 & 0 & 0\\
		Anas penelope (Trondheim) & 0 & 0 & 0 & 0 & 0 & 0 & 0 & 0\\
		Anthus pratensis (Trondheim) & 0 & 0 & 0 & 0 & 0 & 0 & 0 & 0\\
		Anthus trivialis (Grote Nete) & 0 & 0 & 0 & 0 & 0 & 0 & 0 & 0\\
		Anthus trivialis (Trondheim) & 0 & 0 & 0 & 0 & 0 & 0 & 0 & 0\\
		Asio flammeus (Trondheim) & 0 & 0 & 7 & 0 & 100 & 0 & 0 & 0\\
		Aythya fuligula (Trondheim) & 0 & 0 & 0 & 0 & 0 & 0 & 0 & 0\\
		Bombycilla garrulus (Trondheim) & 0 & 0 & 0 & 0 & 0 & 0 & 0 & 0\\
		Bucephala clangula (Trondheim) & 0 & 0 & 0 & 0 & 0 & 0 & 0 & 0\\
		Calidris pugnax (Trondheim) & 0 & 0 & 0 & 0 & 100 & 0 & 0 & 0\\
		Carduelis flammea (Trondheim) & 0 & 0 & 0 & 0 & 0 & 0 & 0 & 0\\
		Castor fiber (Grote Nete) & 0 & 0 & 0 & 0 & 0 & 0 & 0 & 0\\
		Certhia familiaris (Trondheim) & 0 & 0 & 0 & 0 & 0 & 0 & 0 & 0\\
		Charadrius hiaticula (Trondheim) & 0 & 0 & 0 & 0 & 0 & 0 & 0 & 0\\
		Cinclus cinclus (Trondheim) & 0 & 0 & 0 & 0 & 0 & 0 & 0 & 0\\
		Coccothraustes coccothraustes (Trondheim) & 0 & 0 & 0 & 0 & 0 & 0 & 0 & 0\\
		Cygnus cygnus (Trondheim) & 0 & 0 & 0 & 0 & 0 & 0 & 0 & 0\\
		Delichon urbicum (Trondheim) & 0 & 0 & 0 & 0 & 0 & 0 & 0 & 0\\
		Dendrocopos major (Trondheim) & 3 & 0 & 0 & 0 & 0 & 0 & 0 & 0\\
		Dendrocopos minor (Trondheim) & 0 & 0 & 0 & 0 & 0 & 0 & 0 & 0\\
		Dryocopus martius (Grote Nete) & 0 & 0 & 0 & 0 & 0 & 0 & 0 & 0\\
		Dryocopus martius (Trondheim) & 0 & 0 & 0 & 0 & 0 & 0 & 0 & 0\\
		Emberiza citrinella (Trondheim) & 0 & 0 & 0 & 0 & 0 & 0 & 0 & 0\\
		Emberiza schoeniclus (Trondheim) & 0 & 0 & 0 & 0 & 0 & 0 & 0 & 0\\
		Epidalea calamita (Thau) & 0 & 0 & 0 & 0 & 0 & 3 & 0 & 0\\
		Eptesicus nilssonii (Trondheim) & 0 & 0 & 27 & 0 & 100 & 17 & 0 & 0\\
		Erithacus rubecula (Trondheim) & 0 & 0 & 0 & 0 & 0 & 0 & 0 & 0\\
		Ficedula hypoleuca (Trondheim) & 0 & 0 & 0 & 0 & 0 & 0 & 0 & 0\\
		Gallinago gallinago (Trondheim) & 0 & 0 & 0 & 0 & 0 & 0 & 0 & 0\\
		Garrulus glandarius (Trondheim) & 0 & 0 & 0 & 0 & 0 & 0 & 0 & 0\\
		Gavia arctica (Trondheim) & 0 & 0 & 0 & 0 & 0 & 0 & 0 & 0\\
		Gavia stellata (Trondheim) & 0 & 0 & 0 & 0 & 0 & 0 & 0 & 0\\
		Glaucidium passerinum (Trondheim) & 0 & 0 & 0 & 0 & 0 & 10 & 0 & 0\\
		Grus grus (Trondheim) & 0 & 0 & 0 & 0 & 0 & 0 & 0 & 0\\
		Haematopus ostralegus (Trondheim) & 0 & 0 & 0 & 0 & 0 & 0 & 0 & 0\\
		Hippolais icterina (Trondheim) & 3 & 0 & 0 & 0 & 0 & 0 & 0 & 0\\
		Hirundo rustica (Trondheim) & 0 & 0 & 0 & 0 & 0 & 0 & 0 & 0\\
		Lanius excubitor (Trondheim) & 0 & 0 & 0 & 0 & 0 & 0 & 0 & 0\\
		Lissotriton vulgaris (Trondheim) & 0 & 0 & 10 & 100 & 100 & 3 & 0 & 0\\
		Loxia curvirostra (Trondheim) & 0 & 0 & 0 & 0 & 0 & 0 & 0 & 0\\
		Loxia pytyopsittacus (Trondheim) & 0 & 0 & 37 & 0 & 7 & 77 & 0 & 0\\
		Lullula arborea (Grote Nete) & 0 & 0 & 0 & 0 & 0 & 0 & 0 & 0\\
		Luscinia megarhynchos (Grote Nete) & 0 & 0 & 0 & 0 & 0 & 0 & 0 & 0\\
		Luscinia svecica (Grote Nete) & 0 & 0 & 0 & 0 & 0 & 0 & 0 & 0\\
		Luscinia svecica (Trondheim) & 0 & 0 & 27 & 0 & 100 & 3 & 0 & 0\\
		Mergus merganser (Trondheim) & 0 & 0 & 0 & 0 & 0 & 0 & 0 & 0\\
		Mergus serrator (Trondheim) & 0 & 0 & 0 & 0 & 0 & 0 & 0 & 0\\
		Motacilla alba (Trondheim) & 0 & 0 & 0 & 0 & 0 & 0 & 0 & 0\\
		Natrix natrix (Thau) & 0 & 0 & 33 & 100 & 100 & 7 & 0 & 0\\
		Oriolus oriolus (Grote Nete) & 0 & 0 & 0 & 0 & 0 & 0 & 0 & 0\\
		Phoenicurus phoenicurus (Grote Nete) & 0 & 0 & 0 & 0 & 0 & 0 & 0 & 0\\
		Plecotus austriacus (Thau) & 0 & 0 & 0 & 0 & 0 & 0 & 0 & 0\\
		Poecile montanus (Grote Nete) & 3 & 0 & 0 & 0 & 0 & 0 & 0 & 0\\
		Rana temporaria (Trondheim) & 0 & 0 & 23 & 0 & 100 & 7 & 0 & 0\\
		Rhinolophus ferrumequinum (Thau) & 0 & 0 & 3 & 0 & 0 & 0 & 0 & 0\\
		Sciurus vulgaris (Grote Nete) & 0 & 0 & 0 & 0 & 0 & 0 & 0 & 0\\
		Tarentola mauritanica (Thau) & 0 & 0 & 0 & 0 & 0 & 0 & 0 & 0\\
		Timon lepidus (Thau) & 0 & 0 & 0 & 0 & 0 & 0 & 0 & 0\\
		Triturus marmoratus (Thau) & 0 & 0 & 10 & 0 & 100 & 3 & 0 & 0\\
		
		\hline
	\end{tabular}
\end{table}

\begin{table}[!h]
	\fontsize{8}{8}\selectfont
	\caption{\textbf{Percentage of failed predictions by real species according to the model for the corrected group.}}
	\label{TabS3}
	\centering
	\vspace*{0.25cm}
	\begin{tabular}{lcccccccc}
		\hline
		\textbf{Species (Site)} & \textbf{ANN} & \textbf{CTA} & \textbf{FDA} & \textbf{GAM} & \textbf{GBM} & \textbf{GLM} & \textbf{RF} & \textbf{SRE} \\ 
		\hline
		
		Accipiter gentilis (Trondheim) & 0 & 0 & 0 & 0 & 0 & 0 & 0 & 0\\
		Accipiter nisus (Trondheim) & 0 & 0 & 0 & 0 & 0 & 0 & 0 & 0\\
		Actitis hypoleucos (Trondheim) & 0 & 0 & 0 & 0 & 0 & 0 & 0 & 0\\
		Aegithalos caudatus (Trondheim) & 0 & 0 & 0 & 0 & 0 & 0 & 0 & 0\\
		Alauda arvensis (Trondheim) & 0 & 0 & 0 & 0 & 0 & 0 & 0 & 0\\
		Alcedo atthis (Grote Nete) & 0 & 0 & 0 & 3 & 0 & 0 & 0 & 0\\
		Anas penelope (Trondheim) & 0 & 0 & 0 & 0 & 0 & 0 & 0 & 0\\
		Anthus pratensis (Trondheim) & 0 & 0 & 0 & 0 & 0 & 0 & 0 & 0\\
		Anthus trivialis (Grote Nete) & 0 & 0 & 0 & 0 & 0 & 0 & 0 & 0\\
		Anthus trivialis (Trondheim) & 0 & 0 & 0 & 0 & 0 & 0 & 0 & 0\\
		Asio flammeus (Trondheim) & 0 & 0 & 10 & 0 & 100 & 0 & 0 & 0\\
		Aythya fuligula (Trondheim) & 0 & 0 & 0 & 0 & 0 & 0 & 0 & 0\\
		Bombycilla garrulus (Trondheim) & 0 & 0 & 0 & 0 & 0 & 0 & 0 & 0\\
		Bucephala clangula (Trondheim) & 0 & 0 & 0 & 0 & 0 & 0 & 0 & 0\\
		Calidris pugnax (Trondheim) & 0 & 0 & 0 & 0 & 0 & 0 & 0 & 0\\
		Carduelis flammea (Trondheim) & 0 & 0 & 0 & 0 & 0 & 0 & 0 & 0\\
		Castor fiber (Grote Nete) & 0 & 0 & 0 & 0 & 100 & 0 & 0 & 0\\
		Certhia familiaris (Trondheim) & 0 & 0 & 0 & 0 & 0 & 0 & 0 & 0\\
		Charadrius hiaticula (Trondheim) & 0 & 0 & 0 & 0 & 0 & 0 & 0 & 0\\
		Cinclus cinclus (Trondheim) & 0 & 0 & 0 & 0 & 0 & 0 & 0 & 0\\
		Coccothraustes coccothraustes (Trondheim) & 0 & 0 & 0 & 0 & 0 & 0 & 0 & 0\\
		Cygnus cygnus (Trondheim) & 0 & 0 & 0 & 0 & 0 & 0 & 0 & 0\\
		Delichon urbicum (Trondheim) & 0 & 0 & 0 & 0 & 0 & 0 & 0 & 0\\
		Dendrocopos major (Trondheim) & 0 & 0 & 0 & 0 & 0 & 0 & 0 & 0\\
		Dendrocopos minor (Trondheim) & 0 & 0 & 0 & 0 & 0 & 0 & 0 & 0\\
		Dryocopus martius (Grote Nete) & 0 & 0 & 0 & 0 & 0 & 0 & 0 & 0\\
		Dryocopus martius (Trondheim) & 0 & 0 & 0 & 0 & 0 & 0 & 0 & 0\\
		Emberiza citrinella (Trondheim) & 0 & 0 & 0 & 0 & 0 & 0 & 0 & 0\\
		Emberiza schoeniclus (Trondheim) & 0 & 0 & 0 & 0 & 0 & 0 & 0 & 0\\
		Epidalea calamita (Thau) & 0 & 0 & 0 & 0 & 0 & 0 & 0 & 0\\
		Eptesicus nilssonii (Trondheim) & 0 & 0 & 30 & 0 & 100 & 20 & 0 & 0\\
		Erithacus rubecula (Trondheim) & 0 & 0 & 0 & 0 & 0 & 0 & 0 & 0\\
		Ficedula hypoleuca (Trondheim) & 0 & 0 & 0 & 0 & 0 & 0 & 0 & 0\\
		Gallinago gallinago (Trondheim) & 0 & 0 & 3 & 0 & 0 & 0 & 0 & 0\\
		Garrulus glandarius (Trondheim) & 0 & 0 & 0 & 0 & 0 & 0 & 0 & 0\\
		Gavia arctica (Trondheim) & 0 & 0 & 0 & 0 & 0 & 0 & 0 & 0\\
		Gavia stellata (Trondheim) & 0 & 0 & 0 & 0 & 0 & 0 & 0 & 0\\
		Glaucidium passerinum (Trondheim) & 0 & 0 & 0 & 0 & 0 & 10 & 0 & 0\\
		Grus grus (Trondheim) & 0 & 0 & 0 & 0 & 0 & 0 & 0 & 0\\
		Haematopus ostralegus (Trondheim) & 0 & 0 & 0 & 0 & 0 & 0 & 0 & 0\\
		Hippolais icterina (Trondheim) & 3 & 0 & 0 & 0 & 0 & 0 & 0 & 0\\
		Hirundo rustica (Trondheim) & 0 & 0 & 0 & 0 & 0 & 0 & 0 & 0\\
		Lanius excubitor (Trondheim) & 0 & 0 & 0 & 0 & 0 & 0 & 0 & 0\\
		Lissotriton vulgaris (Trondheim) & 0 & 0 & 40 & 100 & 100 & 30 & 0 & 0\\
		Loxia curvirostra (Trondheim) & 0 & 0 & 0 & 0 & 0 & 0 & 0 & 0\\
		Loxia pytyopsittacus (Trondheim) & 0 & 0 & 60 & 0 & 100 & 77 & 0 & 0\\
		Lullula arborea (Grote Nete) & 0 & 0 & 0 & 0 & 0 & 0 & 0 & 0\\
		Luscinia megarhynchos (Grote Nete) & 0 & 0 & 0 & 0 & 0 & 0 & 0 & 0\\
		Luscinia svecica (Grote Nete) & 0 & 0 & 0 & 0 & 0 & 0 & 0 & 0\\
		Luscinia svecica (Trondheim) & 7 & 0 & 33 & 0 & 100 & 3 & 0 & 0\\
		Mergus merganser (Trondheim) & 0 & 0 & 0 & 0 & 0 & 0 & 0 & 0\\
		Mergus serrator (Trondheim) & 0 & 0 & 0 & 0 & 0 & 0 & 0 & 0\\
		Motacilla alba (Trondheim) & 0 & 0 & 0 & 0 & 0 & 0 & 0 & 0\\
		Natrix natrix (Thau) & 0 & 0 & 13 & 100 & 100 & 10 & 0 & 0\\
		Oriolus oriolus (Grote Nete) & 0 & 0 & 0 & 0 & 0 & 0 & 0 & 0\\
		Phoenicurus phoenicurus (Grote Nete) & 0 & 0 & 0 & 0 & 0 & 0 & 0 & 0\\
		Plecotus austriacus (Thau) & 0 & 0 & 0 & 0 & 0 & 0 & 0 & 0\\
		Poecile montanus (Grote Nete) & 0 & 0 & 0 & 0 & 0 & 0 & 0 & 0\\
		Rana temporaria (Trondheim) & 0 & 0 & 7 & 0 & 0 & 0 & 0 & 0\\
		Rhinolophus ferrumequinum (Thau) & 0 & 0 & 0 & 0 & 0 & 0 & 0 & 0\\
		Sciurus vulgaris (Grote Nete) & 3 & 0 & 0 & 0 & 0 & 0 & 0 & 0\\
		Tarentola mauritanica (Thau) & 0 & 0 & 0 & 0 & 0 & 0 & 0 & 0\\
		Timon lepidus (Thau) & 0 & 0 & 0 & 0 & 0 & 0 & 0 & 0\\
		Triturus marmoratus (Thau) & 0 & 0 & 3 & 0 & 100 & 0 & 0 & 0\\
		
		\hline
	\end{tabular}
\end{table}

\begin{table}[!h]
	\fontsize{10}{10}\selectfont
	\caption{\textbf{Summary information about the 21 virtual species (species name, site, sample size and sample bias).} Sample bias was estimated from Boyce indices based on occurrence points and accessibility maps.}
	\label{TabS4}
	\vspace*{0.25cm}
	\centering
	\begin{tabular}{llrr}
		\hline
		\textbf{Species} & \textbf{Site} & \textbf{Sample size} & \textbf{Sample bias} \\ 
		\hline
		
		Alteragris belgiumensis & Grote Nete & 100 & 0.32\\
		Fugiagris belgiumensis & Grote Nete & 300 & 0.89\\
		Philocourdogenos belgiumensis & Grote Nete & 300 & 0.96\\
		Philocourdos belgiumensis & Grote Nete & 300 & 0.76\\
		Philograss belgiumensis & Grote Nete & 100 & 0.62\\
		Philoherbagenos belgiumensis & Grote Nete & 80 & 0.63\\
		Philoherbas belgiumensis & Grote Nete & 80 & -0.87\\
		Alteragris thauensis & Thau & 100 & 0.82\\
		Fugiagris thauensis & Thau & 300 & 0.98\\
		Philocourdogenos thauensis & Thau & 300 & 0.99\\
		Philocourdos thauensis & Thau & 300 & 0.98\\
		Philograss thauensis & Thau & 100 & 0.38\\
		Philoherbagenos thauensis & Thau & 80 & 0.63\\
		Philoherbas thauensis & Thau & 100 & -0.35\\
		Alteragris norwayensis & Trondheim & 100 & 0.94\\
		Fugiagris norwayensis & Trondheim & 300 & -0.15\\
		Philocourdogenos norwayensis & Trondheim & 300 & 0.81\\
		Philocourdos norwayensis & Trondheim & 300 & 0.85\\
		Philograss norwayensis & Trondheim & 100 & 0.96\\
		Philoherbagenos norwayensis & Trondheim & 80 & 0.78\\
		Philoherbas norwayensis & Trondheim & 100 & 0.76\\
		
		\hline
	\end{tabular}
\end{table}

\begin{table}[!h]
	\fontsize{10}{10}\selectfont
	\caption{\textbf{Percentage of failed predictions by virtual species according to the model for the uncorrected group.}}
	\label{TabS5}
	\vspace*{0.25cm}
	\centering
	\begin{tabular}{lcccccccc}
		\hline
		\textbf{Species (Site)} & \textbf{ANN} & \textbf{CTA} & \textbf{FDA} & \textbf{GAM} & \textbf{GBM} & \textbf{GLM} & \textbf{RF} & \textbf{SRE} \\ 
		\hline
		
		Alteragris belgiumensis (Grote Nete) & 0 & 0 & 0 & 0 & 100 & 0 & 0 & 0\\
		Alteragris norwayensis (Trondheim) & 0 & 0 & 0 & 0 & 100 & 0 & 0 & 0\\
		Alteragris thauensis (Thau) & 0 & 0 & 0 & 0 & 0 & 0 & 0 & 0\\
		Fugiagris belgiumensis (Grote Nete) & 0 & 0 & 0 & 0 & 100 & 0 & 0 & 0\\
		Fugiagris norwayensis (Trondheim) & 0 & 0 & 0 & 0 & 0 & 0 & 0 & 0\\
		Fugiagris thauensis (Thau) & 0 & 0 & 0 & 0 & 0 & 0 & 0 & 0\\
		Philocourdogenos belgiumensis (Grote Nete) & 0 & 0 & 0 & 0 & 0 & 0 & 0 & 0\\
		Philocourdogenos norwayensis (Trondheim) & 0 & 0 & 0 & 0 & 0 & 0 & 0 & 0\\
		Philocourdogenos thauensis (Thau) & 0 & 0 & 0 & 0 & 100 & 0 & 0 & 0\\
		Philocourdos belgiumensis (Grote Nete) & 0 & 0 & 0 & 0 & 100 & 0 & 0 & 0\\
		Philocourdos norwayensis (Trondheim) & 0 & 0 & 0 & 0 & 0 & 0 & 0 & 0\\
		Philocourdos thauensis (Thau) & 0 & 0 & 0 & 0 & 100 & 0 & 0 & 0\\
		Philograss belgiumensis (Grote Nete) & 0 & 0 & 0 & 63 & 0 & 0 & 0 & 0\\
		Philograss norwayensis (Trondheim) & 0 & 0 & 0 & 0 & 100 & 0 & 0 & 0\\
		Philograss thauensis (Thau) & 0 & 0 & 0 & 0 & 100 & 0 & 0 & 0\\
		Philoherbagenos belgiumensis (Grote Nete) & 0 & 0 & 0 & 0 & 100 & 0 & 0 & 0\\
		Philoherbagenos norwayensis (Trondheim) & 0 & 0 & 0 & 0 & 0 & 0 & 0 & 0\\
		Philoherbagenos thauensis (Thau) & 0 & 0 & 3 & 0 & 0 & 0 & 0 & 0\\
		Philoherbas belgiumensis (Grote Nete) & 0 & 0 & 0 & 0 & 0 & 0 & 0 & 0\\
		Philoherbas norwayensis (Trondheim) & 0 & 0 & 0 & 0 & 0 & 0 & 0 & 0\\
		Philoherbas thauensis (Thau) & 0 & 0 & 0 & 0 & 0 & 0 & 0 & 0\\
		
		\hline
	\end{tabular}
\end{table}

\begin{table}[!h]
	\fontsize{10}{10}\selectfont
	\caption{\textbf{Percentage of failed predictions by virtual species according to the model for the corrected group.}}
	\label{TabS6}
	\vspace*{0.5cm}
	\centering
	\begin{tabular}{lcccccccc}
		\hline
		\textbf{Species (Site)} & \textbf{ANN} & \textbf{CTA} & \textbf{FDA} & \textbf{GAM} & \textbf{GBM} & \textbf{GLM} & \textbf{RF} & \textbf{SRE} \\ 
		\hline
		
		Alteragris belgiumensis (Grote Nete) & 0 & 0 & 0 & 0 & 100 & 0 & 0 & 0\\
		Alteragris norwayensis (Trondheim) & 0 & 0 & 0 & 0 & 100 & 0 & 0 & 0\\
		Alteragris thauensis (Thau) & 3 & 0 & 0 & 3 & 0 & 3 & 0 & 0\\
		Fugiagris belgiumensis (Grote Nete) & 0 & 0 & 0 & 0 & 100 & 0 & 0 & 0\\
		Fugiagris norwayensis (Trondheim) & 0 & 0 & 0 & 0 & 0 & 0 & 0 & 0\\
		Fugiagris thauensis (Thau) & 0 & 0 & 0 & 0 & 100 & 0 & 0 & 0\\
		Philocourdogenos belgiumensis (Grote Nete) & 0 & 0 & 0 & 0 & 100 & 0 & 0 & 0\\
		Philocourdogenos norwayensis (Trondheim) & 0 & 0 & 0 & 0 & 100 & 0 & 0 & 0\\
		Philocourdogenos thauensis (Thau) & 0 & 0 & 0 & 0 & 100 & 0 & 0 & 0\\
		Philocourdos belgiumensis (Grote Nete) & 0 & 0 & 0 & 0 & 100 & 0 & 0 & 0\\
		Philocourdos norwayensis (Trondheim) & 0 & 0 & 0 & 0 & 100 & 0 & 0 & 0\\
		Philocourdos thauensis (Thau) & 0 & 0 & 0 & 0 & 100 & 0 & 0 & 0\\
		Philograss belgiumensis (Grote Nete) & 0 & 0 & 0 & 60 & 100 & 0 & 0 & 0\\
		Philograss norwayensis (Trondheim) & 0 & 0 & 0 & 0 & 100 & 0 & 0 & 0\\
		Philograss thauensis (Thau) & 0 & 0 & 0 & 0 & 100 & 0 & 0 & 0\\
		Philoherbagenos belgiumensis (Grote Nete) & 0 & 0 & 0 & 0 & 100 & 0 & 0 & 0\\
		Philoherbagenos norwayensis (Trondheim) & 0 & 0 & 0 & 0 & 100 & 0 & 0 & 0\\
		Philoherbagenos thauensis (Thau) & 3 & 0 & 0 & 0 & 0 & 0 & 0 & 0\\
		Philoherbas belgiumensis (Grote Nete) & 0 & 0 & 0 & 0 & 0 & 0 & 0 & 0\\
		Philoherbas norwayensis (Trondheim) & 0 & 0 & 0 & 0 & 0 & 0 & 0 & 0\\
		Philoherbas thauensis (Thau) & 0 & 0 & 0 & 0 & 0 & 0 & 0 & 0\\
		
		\hline
	\end{tabular}
\end{table}

\begin{table}[!h]
	\caption{\textbf{Tables of confusion associated with a significance threshold $\bm{\delta}$ = 0.05}. We obtained the following accuracy values: Schoener's D (ACC=0.65), Boyce (ACC=0.46), cAUC (ACC=0.54), AUC (ACC=0.55) and TSS (ACC=0.54).}
	\label{TabS7}
	\vspace*{0.5cm}
	\centering 
	\includegraphics[width=14cm]{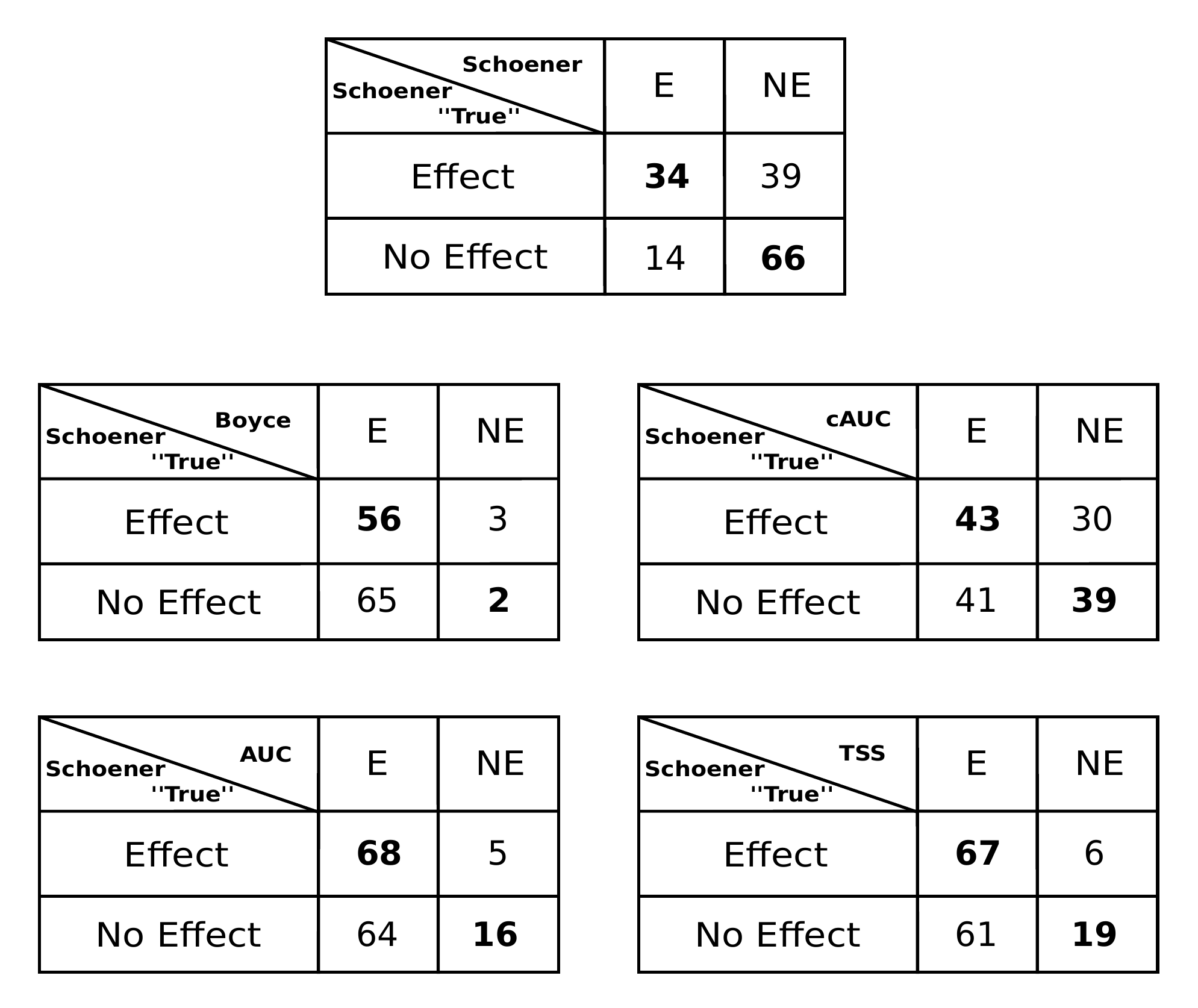}
\end{table}

\end{document}